\documentclass[11pt,letterpaper,english,final,journal,onecolumn]{IEEEtran}
\usepackage{color}
\usepackage{array}
\usepackage{multirow}
\usepackage{amsthm}
\usepackage{amsmath}
\usepackage{amssymb}
\usepackage{graphicx}
\usepackage[norelsize,algo2e,linesnumbered,longend,noline]{algorithm2e}
\SetAlFnt{\small}  
\usepackage{verbatim}

\usepackage[noadjust]{cite}
\usepackage{psfrag}

\renewcommand{\vec}[1]{\mbox{\boldmath$#1$}}

\newcommand{\beq}{\begin{equation}}
\newcommand{\eeq}{\end{equation}}
\newcommand{\eq}[1]{(\ref{#1})}

\newcommand{\bup}{\begin{upshape}}
\newcommand{\eup}{\end{upshape}}
\newcommand{\mc}{\mathcal}
\newcommand{\abs}[1]{\lvert#1\rvert}
\newcommand{\fig}[1]{Fig.~\ref{#1}}

\newcommand{\beqsub}{\begin{subequations}}
\newcommand{\eeqsub}{\end{subequations}}

\newcommand{\nulls}{\varnothing}

\newtheorem{lemma}{Lemma}
\newtheorem{corollary}{Corollary}

\newtheorem{theorem}{Theorem}
\newtheorem{definition}{Definition}

\newtheorem{fact}{Fact}
\newtheorem{example}{Example}

\makeatletter
\newcommand{\removelatexerror}{\let\@latex@error\@gobble}
\makeatother

\begin{document}
\IEEEoverridecommandlockouts

\title{Stable XOR-based Policies for the Broadcast Erasure Channel with Feedback}

\author{Sophia Athanasiadou, Marios Gatzianas,~\IEEEmembership{Member,~IEEE}, \\ Leonidas Georgiadis,~\IEEEmembership{Senior~Member,~IEEE} and Leandros Tassiulas,~\IEEEmembership{Fellow,~IEEE}%
\thanks{This research has been co-financed by the European Union (European Social Fund -- ESF) and Greek national funds through the Operational Program ``Education and Lifelong Learning'' of the National Strategic Reference Framework (NSRF) -- Research Funding Program: Thales. Investing in knowledge society through the European Social Fund. M.~Gatzianas was supported by the ERC Starting Project Grant NOWIRE ERC-2009-StG-240317.}%
\thanks{S.~Athanasiadou and L.~Georgiadis are with the Department of Electrical and Computer Engineering, Division of Telecommunications, Aristotle University of Thessaloniki, Thessaloniki, 54 124, Greece. M.~Gatzianas was with the School of Computer and Communication Sciences, EPFL, Switzerland. L.~Tassiulas is with the Computer Engineering and Telecommunications Department, University of Thessaly, Volos 38 221, Greece.}} 
\maketitle

\begin{abstract}
In this paper we describe a network coding scheme for the Broadcast
Erasure Channel with multiple unicast stochastic flows, in the case of
a single source transmitting packets to $N$ users, where per-slot
feedback is fed back to the transmitter in the form of ACK/NACK
messages. This scheme performs only binary (XOR) operations and
involves a network of queues, along with special rules for coding and
moving packets among the queues, that ensure instantaneous
decodability. The system under consideration belongs to a class of
networks whose stability properties have been analyzed in earlier work, which
is used to provide a stabilizing policy employing the currently
proposed coding scheme. Finally, we show the optimality
of the proposed policy for $N=4$ and i.i.d. erasure events, in the sense that the 
policy's stability region matches a derived outer bound (which coincides with 
the system's information-theoretic capacity region), even when a restricted set
of coding rules is used.
\end{abstract}

\section{Introduction} \label{sec:Introduction}

The information-theoretic capacity region of the
Broadcast Erasure Channel (BEC) in the case of one transmitter and
$N$ unicast sessions has been recently studied in \cite{GGT11}
and \cite{Wang_Allerton}. Both of these papers propose coding algorithms
based on transmission of linear combinations of packets. These algorithms
are shown to achieve capacity in the following settings: 1) $N\leq3$
and arbitrary channel statistics, and 2) arbitrary $N$ and channel
statistics which satisfy certain assumptions (i.e.~symmetric channels
and one-sided fair channels). However, these schemes are characterized 
by high complexity (as operations take place in a sufficiently large sized 
finite field) and decoding delay, since a sufficient number of linear combinations 
has to be received until a packet is decoded. In \cite{AGG}, we proposed a
network coding scheme that overcomes these obstacles by using only
XOR operations, generalizing the 2-user network coding scheme in \cite{Leo_2user}
to the case of 3 users. Thus, two low complexity algorithms were proposed,
namely \texttt{XOR1} and \texttt{XOR2}, which additionally had the advantageous property 
of ``instantaneous decodability''. By this term, it is meant that a receiver is able
to decode packet $p$ destined for it as soon as it receives an XOR
combination of packets containing $p$. Algorithm \texttt{XOR2}
was proved to achieve capacity for the case of i.i.d. channels as
well as spatially independent channels with erasure probabilities
that do not exceed 8/9.

However, the system considered in \cite{AGG} is a saturated system,
where a predefined number of packets needs to be transmitted to each
user. This model is not frequently encountered in practice. Moreover,
algorithms \texttt{XOR1} and \texttt{XOR2} cannot be easily generalized
to more than 3 users. This happens because, at each time slot, coding
choices have to be determined a priori so that each transmission is
optimally exploited in terms of allowing multiple users to simultaneously
decode their packets as well as create favorable future coding opportunities.
However, for $N>3$, the number of coding choices increases dramatically so 
that there is no clear intuition on the optimal choice (this will become 
apparent once the model and queue structure is described).

In the current work, we propose a general network coding scheme for
the case of a single transmitter sending packets to $N$ users through
the BEC with feedback, generalizing the scheme proposed in \cite{AGG}.
Any packet arriving to the transmitter is initially placed in one of
$N$ queues.  Depending on the received feedback, these packets (or XOR
combinations of them) may travel through a network of queues, before
they reach their destination, in order to exploit the overhearing
benefit of the broadcast channel. Coding and packet movement rules are
imposed in order to ensure instantaneous decodability of packets and
better exploitation of coding opportunities.

While in \cite{AGG} we examined a saturated system, in this paper
we consider a stochastic model where packets may arrive randomly
at the transmitter at any time slot. Additionally, we use a backpressure
type online algorithm that makes each coding choice based on instantaneous
quantities, such as queue sizes, without requiring knowledge of future
events. Therefore, we do not need to predefine the coding choices (as in 
\cite{AGG}), and the proposed network coding scheme can be applied to an 
arbitrary number of users. For the specific case of 4 users and i.i.d. erasure
events, we present a stabilizing policy that uses
only a subset of all possible coding choices and prove that the policy
stability region coincides with the information theoretic capacity
region of the standard BEC with feedback. This result is quite intriguing,
considering the restrictions imposed on the policy (XOR operations
only, instantaneous decodability, reduced set of coding choices).

The network stability of single hop broadcast erasure channels with
feedback has also been examined in \cite{SagEph09}, which considered
broadcast traffic only and investigated the stability regions of plain
retransmission and linear network coding schemes (parameterized over
the field size) as opposed to a proposed dynamic virtual queue-based
policy. The latter policy was shown to be optimal for 2 users while,
for $N>2$ and i.i.d. erasures, it achieved a stable rate that differs
from the cut-set bound by a factor of $O(\epsilon^{m+1})$, where
$m$ is the number of queue ``levels'' that participate in the
coding decision (see \cite{SagEph09} for more details and definitions;
$m$ can be loosely regarded as a measure of the encoding complexity)
and $\epsilon$ is the erasure probability. Although the structure
of the virtual queues and coding rules are inspired by similar concepts
as in our work, the actual rules for moving packets between the queues
are much more involved in our work since we are interested in achieving
the optimal stability region for all values of $\epsilon$ instead
of only asymptotic optimality as $\epsilon\to0$ (these notions of
optimality ignore any overhead). An additional cause for rule complexity
in our work is the fact that multiple unicast sessions are much more
difficult to handle (due to the inherent competition between different
sessions) than a single broadcast session. Furthermore, there is no
guarantee in \cite{SagEph09}, for the general case of $N$ users,
regarding instantaneous decodability.

The work in \cite{TMS09} studied a network which is described by
an underlying complete graph where each edge is modeled as a Markov
chain ON/OFF channel (i.e.~a generalization of the memoryless erasure
channel), while there also exists a special ``relay'' node with
XOR coding capabilities which can overhear all transmissions. Any
transmissions to/from the relay are error-free. The work considers
multiple unicast flows, originating in all nodes except for the relay,
and explicitly accounts for instantaneous decodability by mapping
this constraint into a specially constructed conflict graph (a similar
graph structure is used in \cite{SoVa10} to model the same constraint).
It proposes an online backpressure policy that requires computing
in each slot the maximum weight independent set of the time-varying
conflict graph. Although the work bears similarities to our paper
in terms of mathematical techniques and the optimization problem that
results, the model is quite different. Hence, the proposed coding
policies are quite different and the results in \cite{TMS09} cannot
be used to show one of our main results, namely that the proposed
scheduling and coding policies achieve channel capacity for BEC with
i.i.d. erasures. In particular, the broadcast channel at the relay
(which is the only node that can perform XOR coding) is error-free
in \cite{TMS09}, while we are interested in broadcast erasure channels.

In summary, the contribution of this paper is as follows:
\begin{enumerate}
\item We develop a systematic network-coding-based framework
for constructing instantaneously decodable feedback-based XOR coding
schemes for the BEC with multiple unicast sessions and an arbitrary
number of users. This requires a (highly non-trivial and quite involved)
generalization of the rules in \cite{AGG} and the replacement of the 
algorithmic core in \cite{AGG} with a backpressure-type online algorithm
proposed in \cite{pasxos2}, which makes each coding choice based on
instantaneous quantities instead of a predefined set of ordered actions.
The new policy, which cannot possibly be constructed from \cite{AGG} through
any obvious procedure, is elegant and conceptually simple, considering its
general applicability.

\item We derive an outer bound, for arbitrary $N$, on the 
stability region of the network through an elegant flow argument and relate
this to a bound on the information-theoretic capacity region of the ``extended''
BEC channel (where idle slots are allowed).

\item Finally, for the special case of $N=4$ and 
i.i.d. erasures across users, we carefully restrict the allowable coding 
choices and present a stabilizing policy on top of the previous network 
coding scheme whose stability region is essentially identical to the capacity 
region of the 4-user system (whereas in 2. above we only relate outer bounds). Hence, we 
show that XOR combining achieves both
instantaneous decodability and throughput optimality in this setting. Considering
that the proposed policy uses only a subset of all possible coding
choices and only XOR operations, while guaranteeing instantaneous
decodability, this result is quite unexpected.
\end{enumerate}
The rest of the paper is organized as follows: in
Section \ref{sec:System-model-and}, the system model is introduced
along with some useful notation. In Section \ref{sec:Network-coding-algorithms},
the proposed network coding scheme is described, while in Section
\ref{sec:Stability-region} the applied stabilizing policy is presented.
In Section \ref{sec:UpperBound}, an outer bound on the stability
region of the system under study is derived. In Section \ref{sec:4users},
we prove, for the case of 4 users and i.i.d. erasure events, that
the stability region of such a system coincides with the capacity
outer bound of the standard broadcast erasure channel with feedback.
In Section \ref{sec:Implementation-Issues} we examine some implementation
issues while Section \ref{sec:Conclusions} concludes the paper. Some
technical proofs are contained in the Appendix.

\section{System model and notation} \label{sec:System-model-and}

We describe some notation that will be used in the following. Sets are
denoted by calligraphic letters, e.g.~$\mc{M}$, and the empty set by
$\emptyset$. The cardinality of set $\mc{M}$ is denoted by
$\left|\mc{M}\right|$ and we write $M=\left|\mc{M}\right|$.  Random
variables are denoted by capital letters and their values by small
case letters. Vectors are denoted by bold letters,
e.g.~$\boldsymbol{A}=\left(A_{1}, \ldots,A_{n}\right)$. The expected
value of a random vector is the vector consisting of the expected
values of its components, i.e.,
$\mathbb{E}\left[\boldsymbol{A}\right]=
\left(\mathbb{E}\left[A_{1}\right],\ldots,\mathbb{E}\left[A_{n}\right]\right)$.

We consider a time-slotted system where slot $t=0,1,\ldots$
corresponds to the time interval $[t,\: t+1).$ The system consists of
  a base station $B$ and a set $\mc{N}=\left\{ 1,2,\ldots,N\right\} $
  of receivers (users). At the beginning of slot $t$, $A_{i}(t)$ data
  packets arrive at $B$ with an average rate of
  $\lambda_{i}=\mathbb{E}\left[A_{i}(t)\right]$; these packets must be
  delivered to receiver $i$ and are referred to as ``flow $i$''
  packets, where we denote
  $\boldsymbol{A}(t)=\left(A_{1}(t),\ldots,A_{N}
  \left(t\right)\right)$. All packets consist of $L$
  bits, and the transmission time of each packet is $1$ slot. A packet
  transmitted by $B$ may be either correctly received or completely
  erased by any receiver (broadcast medium). After each transmission,
  the receivers send feedback to $B$ (through an error-free zero-delay
  channel) informing whether the transmitted packet has been correctly
  received or not (ACK/NACK feedback). We also assume that if no
  packet is transmitted in a slot (say, because all queues are empty),
  then all receivers realize that the slot is idle.

Packet arrivals are assumed to be independent and identically
distributed across time, but arbitrarily correlated across users.
That is, the process $\left\{ \boldsymbol{A}\left(t\right)\right\}
_{t=0}^{\infty}$ consists of i.i.d. random vectors, while the
components of each vector $\boldsymbol{A}\left(t\right)$ may be
arbitrarily correlated. Similarly, packet erasures are i.i.d across
time and are initially assumed to be arbitrarily correlated across
users (we later concentrate on the special case of spatially
i.i.d. erasures). The packet arrival and erasure processes are
independent. For subsets $\mc{S},\mc{G}\subseteq\mc{N}$ with
$\mc{S}\cap\mc{G}= \emptyset$, we denote by $P_{\mc{G},\mc{S}}$ the
probability that a transmitted packet is erased at \textit{all}
receivers in $\mc{G}$ and received by \textit{all} receivers in $S$
(no condition is imposed on packet reception or erasure for receivers
in $\mc{N}-\left(\mc{S} \cup \mc{G}\right)$).  We also denote by
$\epsilon_{\mc{G}}$ the probability that a transmitted packet is
erased by all receivers in $\mc{G}$, i.e.,
$\epsilon_{\mc{G}}=P_{\mc{G},\varnothing}.$ For simplicity, we
slightly abuse the notation and write $\epsilon_{i}$ or
$\epsilon_{ij}$ instead of $\epsilon_{\left\{ i\right\} }$ or
$\epsilon_{\left\{ i,j\right\} }$, respectively.

\section{Network coding scheme description}  \label{sec:Network-coding-algorithms}

\subsection{Definitions}  \label{sub:Definitions}

Exogenous packets arriving at $B$ and being intended for user
$i\in\mc{N}$ are called ``native packets for $i$''. A packet is simply
termed ``native'' if it is a native packet for some user (due to the
unicast traffic, a packet is native for exactly one user). According
to the policies to be described below, all transmitted packets are
either native, or XOR combinations of native packets. In other words,
any transmitted packet $p$ can be written as $p=\bigoplus_{l=1}^n
s_{l}$ (where $\oplus$ denotes the XOR operation), where $s_{l}$ are native
packets, and we say that ``$p$ contains $s_{l}$'' or ``$s_{l}$ is
contained in $p$'', or ``$s_l$ is a constituent packet of $p$''. 
As will be seen, it is possible, and actually
beneficial, for $p$ to contain native packets for more than one
user. To shorten the description in the following, we say that a
packet $p$ is an XOR combination of native packets even when $p$
consists of a single native packet. Also, a native packet $q$ for user
$i$ is \textit{unknown} to $i$ at a given time if it has not been
decoded by $i$ by that time. The following definitions, which are
introduced in earlier work \cite{AGG}, will be crucial in the
subsequent analysis.
\begin{definition}
\label{listener}User $i$ is a \emph{Listener} of a packet $p$ iff
both of the following conditions are true: 
\begin{enumerate}
\item $p$ is an XOR combination of packets, not necessarily native, that
$i$ has correctly received. 

\item $p$ contains no native packet for $i$ that is unknown to $i$. Equivalently,
if $p$ contains a native packet $s$ for user $i$, then the packet $s$ is known to 
(i.e.~has already been decoded by) $i$. 
\end{enumerate}
\end{definition}

\begin{definition} \label{desti}
User $i$ is a \emph{Destination} of a packet $p$ iff
either $p$ is a native packet for user $i$ that
is unknown to $i,$ or $p$ can be decomposed as an XOR combination
of the form $p=q\oplus c$ where 
\begin{enumerate}
\item $q$ is a native packet for $i$ and unknown to $i,$ and 

\item $i$ is a Listener of $c$. 
\end{enumerate}
\end{definition}
We hereafter use the terms Listener, Destination to exclusively refer
to the above technical definitions. The decomposition of a packet
$p=q\oplus c$ with Destination $i$ alluded to in Definition
\ref{desti} is unique, since $c$ cannot itself contain an unknown
native packet for $i$, due to the second condition of Definition
\ref{listener} (since $i$ is also a Listener of $c$).  Hence, a packet
$p$ for which user $i$ is a Destination can contain exactly one
unknown native packet $q$ for $i$, which we denote as $q=p(i)$ (we
call $p(i)$ the ``unknown native packet'' of $i$ in
$p$ \label{enu:the-unknown-native}). On the other hand, notice that
the second condition of the Listener definition does not assert that
$p$ always contains a native packet $s$ for user $i$, only that the
existence of such a packet implies that $s$ is known to
$i$. Furthermore, the properties of Destination and Listener are
time-dependent since they depend on notions such as ``packets known to
user $i$'', which are inherently time-dependent. Clearly, the Listener
property is absorbing, in the sense that if user $i$ is a Listener for
packet $p$ at slot $t$, it remains a Listener for $p$ for all slots
$\tau>t$.

To better understand the previous definitions and some of their fine
points, we offer the following illustrative examples:
\begin{itemize}
\item Denote all native packets for users $i,j$ with $\tilde{r},\tilde{s}$,
respectively; we will use indices $\tilde{r}_{1},\tilde{r}_{2},\ldots,$
and $\tilde{s}_{1},\tilde{s}_{2},\ldots,$ to refer to different native
packets for the same user. Suppose $p=\tilde{r}\oplus\tilde{s}$ is
transmitted, where $\tilde{r},\tilde{s}$ are unknown to $i$ and
$j$, respectively, and have been previously received by $j,i$, respectively.
Then, according to Definition \ref{desti}, both $i$ and $j$ are
Destinations for $p$. If $p$ is only received by a
third user $k$, then $k$ becomes a Listener for $p$ (since $\tilde{r},\tilde{s}$
are not native packets for $k$).
If $i$ receives $p$ in the future, then $i$ instantly decodes its
native packet $\tilde{r}$, ceases to be a Destination for $p$, and
becomes a Listener for $p$, as $p$ no longer contains a native packet
of $i$ that is unknown to $i$.

\item Suppose that $p=\tilde{r}_{l}\oplus\tilde{s}_{l}$ is transmitted
and received by $i$, where neither $\tilde{r}_{l}$ nor $\tilde{s}_{l}$
has been decoded by $i$ in the past. Then, according to Definition
\ref{listener}, $i$ is not a Listener of $p$ (since $p$ contains
an unknown native packet $\tilde{r}_{l}$ for $i$), even though it
knows $p$. In juxtaposition to the previous example, we note the
following subtle point: although a user can only become a Listener
of a packet after receiving an XOR combination containing the packet,
the previous example shows that it is not always true that every successful
reception of a packet by a user automatically makes the user a Listener
for the received packet. To take that example one step further, suppose
now that $\tilde{p}=\tilde{r}_{m}\oplus p$ is transmitted immediately
after $p$ and received by $i$. Then, $i$ is not a Destination for
$\tilde{p}$ (since Definition \ref{desti} would require $i$ to
be a Listener of $p$ at the time of $\tilde{p}$'s transmission)
even though $i$ is able to decode $\tilde{r}_{m}$. Since $\tilde{p}$
is an Innovative packet\footnote{since each transmitted packet $p$ is an XOR 
combination of native packets, we can write $p$ as $p=\bigoplus_{n} a_{n,p}^{(i)}
\tilde{r}_{n}^{(i)}\oplus d_{p}$, where $\tilde{r}_{n}^{(i)}$ are all native packets 
for user $i$, the (composite) packet $d_{p}$ contains no native packet for $i$
and $a_{n,p}^{(i)}\in GF(2)$ are suitable coefficients. Hence, for
each transmitted packet $p$ and each user $i$, we can associate
a vector $\mathbf{a}_{p}^{(i)}=(a_{n,p}^{(i)})$ over the field $GF(2)$
and consider the space spanned by the vectors $\mathbf{a}^{(i)}$
that correspond to all packets previously received by user $i$. Packet
$p$ is defined in \cite{Lun04} to be Innovative
for user $i$ if the $\mathbf{a}_{p}^{(i)}$ vector is linearly independent
w.r.t. the $\mathbf{a}^{(i)}$ vectors of all previously received
packets by $i$. Hence, an Innovative packet essentially brings ``fresh''
information to a user.} for $i$, we conclude that the notion of ``$i$ 
is a Destination for $\tilde{p}$'' is a stronger notion than ``$\tilde{p}$ is
Innovative for $i$''. As will be seen, the proposed policies ensure
that this scenario never occurs; it is mentioned here only to illustrate
the Innovative/Destination distinction. 
\end{itemize}

As will be seen, transmitted packets may have several receivers as
Destinations or Listeners. The next fact follows from Definition \ref{desti}. 
\begin{fact}  \label{instant-decoding}
\textup{If user $i$ is a Destination for a packet $p$ and $i$ receives $p$, 
then $i$ is able to immediately (i.e.~instantly) decode the unknown native
packet intended for it that is contained in $p$.}
\end{fact}
Hence, one way of guaranteeing instant decodability in the proposed
scheme would be to guarantee that whenever a transmitted packet $p$
contains an unknown native packet for some user $i$, then $i$ is
a Destination for $p$. This desirable property will be eventually
proved once the coding scheme is fully described.

\subsection{Queue management and coding choices}  \label{sub:Queue-management-and}

Under the proposed policies, packets may be placed in various queues
at the transmitter side, based on the received feedback. A general
queue $Q_{\mc{D}}^{\mc{L}}$ is characterized by two index sets
$\mc{L},\mc{D}$ satisfying the following criteria: \newline
\textbf{Compatibility criteria (CC) for sets $\mc{L},\mc{D}$}
\begin{enumerate}
\item $\mc{L},\mc{D}\subseteq\mc{N},$ 
\item $\mc{L}\cap\mc{D}=\emptyset,$ 
\item $\mc{D}\neq\emptyset,$ 
\item $\mc{L}=\emptyset$ only if $\left|\mc{D}\right|=1$. 
\end{enumerate}
For simplicity, we will denote queue $Q_{\{i,j\}}^{\{k\}}$ by
$Q_{ij}^{k},$ and queue $Q_{\{i\}}^{\emptyset}$ by $Q_{i}.$ Also, we
use the notation $p_{\mc{D}}^{\mc{L}}$ to denote a packet that is
stored in queue $Q_{\mc{D}}^{\mc{L}}$ and denote with
$\abs{Q^{\mc{L}}_{\mc{D}}}$ the number of packets stored in
$Q^{\mc{L}}_{\mc{D}}$. We hereafter assume that all sets $\mc{L}$,
$\mc{D}$ for queues $Q^{\mc{L}}_{\mc{D}}$ satisfy the CC and will not
state this explicitly.

In addition to the above network of queues, it will be helpful to
introduce a network of ``virtual'' queues $V^{\mc{L}}_{\mc{D}}(i)$,
for all $\mc{L},\mc{D}$ and $i\in \mc{D}$ as follows: each
$V^{\mc{L}}_{\mc{D}}(i)$ exclusively contains ``tokens'' identifying
native packets, namely the unknown native packets for user $i\in
\mc{D}$ which are contained in packets stored in
$Q^{\mc{L}}_{\mc{D}}$. We refer to these tokens as ``virtual packets''
and write $p^{\mc{L}}_{\mc{D}}(i)$ to refer both to a token stored in
$V^{\mc{L}}_{\mc{D}}(i)$ as well as to the native packet identified by
this token. In the following, we will use the term ``packet movement''
between virtual queues to actually refer to token movement (tokens are
atomic entities so they cannot be further decomposed: each token moves
as a unit). Hence, queues $V^{\mc{L}}_{\mc{D}}(i)$ do not really exist
at the transmitter side and should only be examined at a conceptual
level, since they will be useful in Sections \ref{sec:UpperBound},
\ref{sec:4users}. In contrast to the ``virtual'' network, the queues
$Q^{\mc{L}}_{\mc{D}}$ and the packets stored in them will be referred
to as ``real''.

We also associate with each queue $Q^{\mc{L}}_{\mc{D}}$ a group of
non-negative integer counters $K^{\mc{L}}_{\mc{D}}(i)$, for each $i\in
\mc{D}$, which are interpreted as the number of unknown native packets
for user $i$ contained in packets stored in $Q^{\mc{L}}_{\mc{D}}$
(equivalently, the number of tokens for user $i$ in
$Q^{\mc{L}}_{\mc{D}}$), i.e.~it holds by definition
$K^{\mc{L}}_{\mc{D}}(i)= \abs{V^{\mc{L}}_{\mc{D}}(i)}$. We will later
prove the important property $K^{\mc{L}}_{\mc{D}}(i)=
\abs{Q^{\mc{L}}_{\mc{D}}}$ for all $i\in \mc{D}$. Initially, all
queues are empty and all counters set to 0.

\begin{figure}[t]
\centering
\psfrag{Level1}[][][0.8]{Level 1}
\psfrag{Level2}[][][0.8]{Level 2}
\psfrag{Level3}[][][0.8]{Level 3}

\psfrag{Q1}[][][0.8]{$Q_1$}
\psfrag{Q2}[][][0.8]{$Q_2$}
\psfrag{Q3}[][][0.8]{$Q_3$}
\psfrag{Q2_1}[][][0.8]{$Q^2_1$}
\psfrag{Q1_2}[][][0.8]{$Q^1_2$}
\psfrag{Q3_1}[][][0.8]{$Q^3_1$}
\psfrag{Q1_3}[][][0.8]{$Q^1_3$}
\psfrag{Q3_2}[][][0.8]{$Q^3_2$}
\psfrag{Q2_3}[][][0.8]{$Q^2_3$}
\psfrag{Q3_12}[][][0.8]{$Q^3_{12}$}
\psfrag{Q2_13}[][][0.8]{$Q^2_{13}$}
\psfrag{Q1_23}[][][0.8]{$Q^1_{23}$}
\psfrag{Q23_1}[][][0.8]{$Q^{23}_1$}
\psfrag{Q13_2}[][][0.8]{$Q^{13}_2$}
\psfrag{Q12_3}[][][0.8]{$Q^{12}_3$}
\includegraphics[scale=0.5]{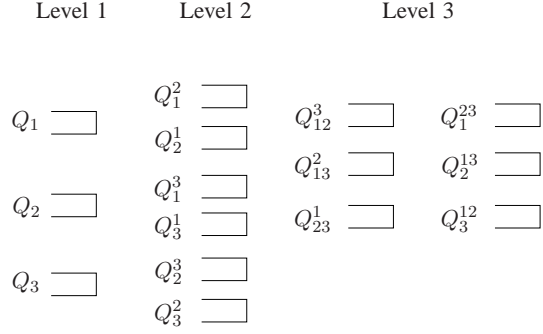}
\caption{Network of queues for $N=3$ (virtual queues are not shown since they are not used by the transmitter).}
\label{queuefig} 
\end{figure}

We classify queues into $N$ levels, where level $w\in\left\{
1,\ldots,N\right\}$ contains all queues $Q_{\mc{D}}^{\mc{L}}$ such
that $\left|\mc{L}\right|+\left|\mc{D}\right|=w$. Moreover, we
classify queues of level $w\geq3$ into sublevels, where sublevel $w.u$
includes queues of level $w$ with $\left|\mc{L}\right|=u$,
$u\in\left\{ 1,\ldots,w-1\right\}$. In Figure \ref{queuefig}, we give
an example of the queue network when $N=3$.
Under the proposed scheme, XOR combinations of packets are
transmitted, which contain at most one packet from each of the queues
$Q_{\mc{D}}^{\mc{L}}.$ While the specific choice of packets depends on
the received feedback and the specific algorithm that is employed, the
following rule always holds.

\textbf{Basic Coding Rule (BCR)}
A set $\mc{P}=\left\{
p_{\mc{D}_{1}}^{\mc{L}_{1}},\ldots,p_{\mc{D}_{\nu}}^{\mc{L}_{\nu}}\right\}$
of $\nu$ packets, \emph{one} from each of the \emph{different} queues
$\left\{
Q_{\mc{D}_{1}}^{\mc{L}_{1}},\ldots,Q_{\mc{D}_{\nu}}^{\mc{L}_{\nu}}\right\}
$, can be combined (by XORing) into a single coded packet only if
\begin{equation}
\mc{D}_{n}\subseteq\mc{L}_{r},\;\forall\, r\neq n,\; n,r\in\left\{ 
  1,\ldots,\nu\right\} .\label{eq:Basic Coding Rule}
\end{equation}
Note that the Basic Coding Rule implies that
$\mc{D}_{n}\cap\mc{D}_{r}=\emptyset$, for all $r\neq n$, $n,\:
r\in\left\{ 1,\ldots,\nu\right\} $. Indeed, $i\in\mc{D}_{n}$ implies,
through (\ref{eq:Basic Coding Rule}), that $i\in\mc{L}_{r}$ and, since
according to CC it holds $\mc{D}_{r}\cap\mc{L}_{r}=\emptyset,$ it
follows that $i\notin\mc{D}_{r}$.

We have not yet fully specified the criterion according to which a
packet is stored in a queue. It will be convenient for packets stored
in the same queue to have some common characteristics or
properties. Since the notions of Destination/Listener are crucial for
keeping track of the packet's history, we use these two notions as the
basis for the packet storage rules. Specifically, we require the
following properties to hold:

\textbf{Basic Properties (BP) of packets stored in queues $Q_{\mc{D}}^{\mc{L}}$}:
\begin{enumerate}
\item Each packet $p_{\mc{D}}^{\mc{L}}\in Q_{\mc{D}}^{\mc{L}}$
is an XOR combination of native packets (including the special case
of a single native packet), not necessarily for the same user. \label{enu:xor_native}

\item For each packet $p_{\mc{D}}^{\mc{L}}\in Q_{\mc{D}}^{\mc{L}}$,
the set of Destinations for $p^{\mc{L}}_{\mc{D}}$ is $\mc{D}$ and all $i\in\mc{L}$
are Listeners for $p_{\mc{D}}^{\mc{L}}$.  \label{enu:LD}

\item For each packet $p_{\mc{D}}^{\mc{L}}\in Q_{\mc{D}}^{\mc{L}}$,
if $p_{\mc{D}}^{\mc{L}}$ contains an unknown native packet
$q$ for some user $i$, then $i$ is a Destination for $p_{\mc{D}}^{\mc{L}}$.
Hence, taking BP\ref{enu:LD} into account, it follows that $i\in\mc{D}$. \label{enu:onlyD}

\item For each native packet $q$ for user $i$ that
has not been decoded by $i$ yet, there exists exactly one packet 
$p_{\mc{D}}^{\mc{L}}\in Q_{\mc{D}}^{\mc{L}}$ (for some sets $\mc{L},\mc{D}$) 
such that $q=p_{\mc{D}}^{\mc{L}}(i)$, i.e.~$p_{\mc{D}}^{\mc{L}}$ is a composite 
packet that contains $q$. \label{enu:nolost}
\end{enumerate}
We should stress the following subtle difference
in terms of reference between BP\ref{enu:xor_native}--BP\ref{enu:onlyD}
and BP\ref{enu:nolost}: BP\ref{enu:xor_native}--BP\ref{enu:onlyD}
describe properties of packets stored in any queue $Q_{\mc{D}}^{\mc{L}}$,
while BP\ref{enu:nolost} is an existence statement that essentially
describes properties of native packets, which are then related to
some queue $Q_{\mc{D}}^{\mc{L}}$.

In retrospect, the Basic Properties justify the
Compatibility Criteria imposed on $\mc{D},\mc{L}$. Specifically,
the fact that $\mc{D},\mc{L}$ contain Destinations and
Listeners, respectively, for a packet $p$ implies that $\mc{L}\cap\mc{D}=\emptyset$,
since $p$ cannot contain any packet that is unknown to a Listener
user, due to condition 2 of Definition \ref{listener} (hence, a
Listener can never be a Destination, although a Destination for a
packet becomes a Listener upon reception of the packet).
The condition $\mc{D}\neq\emptyset$ captures the fact that a
packet need only be stored in the queues for as long as it contains
an unknown native packet for at least one user. Finally, before any
transmissions occur, each native packet has a singleton Destination
set and an empty Listener set.

The next result follows immediately from BP.
\begin{lemma} \label{lem:allsame}
For all $\mc{L},\mc{D}$ that satisfy CC, BP implies that
$K^{\mc{L}}_{\mc{D}}(i)=\abs{Q^{\mc{L}}_{\mc{D}}}$ for all $i\in
\mc{D}$.
\end{lemma}

\begin{IEEEproof}
We slightly abuse notation and use $Q^{\mc{L}}_{\mc{D}}$ to refer to
the queue indexed by $\mc{L},\mc{D}$ as well as the set of packets
stored in the queue. We also denote with $\mc{P}_i$ the set of unknown
native packets for user $i$ that are contained in packets stored in
$Q^{\mc{L}}_{\mc{D}}$. By definition, it holds
$K^{\mc{L}}_{\mc{D}}(i)=\abs{\mc{P}_i}$, so that it suffices to show
$\abs{\mc{P}_i}=\abs{Q^{\mc{L}}_{\mc{D}}}$.  Consider any $i\in
\mc{D}$; by BP\ref{enu:nolost}, any unknown native packet for user $i$
in $\mc{P}_i$ is contained in exactly one packet stored in
$Q^{\mc{L}}_{\mc{D}}$, which implies $\abs{\mc{P}_i} \leq
\abs{Q^{\mc{L}}_{\mc{D}}}$.  Also, by BP\ref{enu:LD}, any packet
$p^{\mc{L}}_{\mc{D}} \in Q^{\mc{L}}_{\mc{D}}$ contains exactly one
unknown native packet for user $i$ (since $i\in \mc{D}$ is a
Destination for $p^{\mc{L}}_{\mc{D}}$) and, by BP\ref{enu:nolost}, no
two distinct packets in $Q^{\mc{L}}_{\mc{D}}$ can contain the same
unknown native packet for $i$, which implies
$\abs{Q^{\mc{L}}_{\mc{D}}} \leq \abs{\mc{P}_i}$. This completes the
proof.
\end{IEEEproof}

The significance of the BP (apart from a systematic way of storing
packets in queues) lies in the fact that, combined with BCR, they
guarantee the desired instantaneous decodability property, as described
in the next result.
\begin{lemma}  \label{lem:instant_decode}
If BP holds at the beginning of slot $t$
and the transmitted packet $p$ at slot $t$ is created according
to BCR, the following statement is true: if $p$ contains an unknown
native packet for some user $i$, then $i$ is a Destination for $p$.
Hence, by Fact \ref{instant-decoding}, any user for which $p$ contains
an unknown native packet can instantly decode it upon reception of $p$.
\end{lemma}

\begin{IEEEproof}
Let the transmitted packet $p=\bigoplus_{k=1}^{\nu}p_{\mc{D}_{k}}^{\mc{L}_{k}}$,
formed according to BCR, contain some unknown native packet $q$ for
user $i$. Then, $q$ must be contained in one of the $p_{\mc{D}_{k}}^{\mc{L}_{k}}$
packets that comprise $p$, say $p_{\mc{D}_{k^{\ast}}}^{\mc{L}_{k^{\ast}}}$.
BP\ref{enu:onlyD} now implies that, since $q$ is unknown to $i$,
$i$ is a Destination for $p_{\mc{D}_{k^{\ast}}}^{\mc{L}_{k^{\ast}}}$
so that, by BP\ref{enu:LD}, it holds $i\in\mc{D}_{k^{\ast}}$.
Hence, we can write $p_{\mc{D}_{k^{\ast}}}^{\mc{L}_{k^{\ast}}}=q\oplus c$,
where $i$ is a Listener for $c$. Furthermore, the BCR implies that
$i\in\mc{L}_{r}$ for all $r\neq k^{\ast}$, since it holds $i\in\mc{D}_{k^{\ast}}$,
so that we can write $p=q\oplus c\oplus\bigoplus_{r\neq k^{\ast}}p_{\mc{D}_{r}}^{\mc{L}_{r}}$.
By BP\ref{enu:LD} again, $i$ is a Listener for each $p_{\mc{D}_{r}}^{\mc{L}_{r}}$
(since $i\in\mc{L}_{r})$, whence it follows that $i$ is a Destination
for $p$. Fact \ref{instant-decoding} now implies that $i$ can instantly
decode $q$ upon reception of $p$.
\end{IEEEproof}

Notice that Lemma \ref{lem:instant_decode} proves
a property which is essentially identical to BP\ref{enu:onlyD}, albeit
for the transmitted packet $p$ only (whereas BP\ref{enu:onlyD} holds
for all packets stored in queues $Q_{\mc{D}}^{\mc{L}}$).
In fact, the previous lemma can be strengthened into the following
statement, which specifies the users that can potentially
instantly decode unknown native packets after reception of $p$. This
corollary will be crucially used in the proof of subsequent results.
\begin{corollary}  \label{cor:Dforall}
If BP holds at the beginning of slot $t$ and
the transmitted packet $p$ is created according to BCR, then $p$
contains unknown native packets for all users in $\cup_{k=1}^{\nu}\mc{D}_{k}$,
and only for them (in fact, $\cup_{k=1}^{\nu}\mc{D}_{k}$
is the set of Destinations for $p$ at the beginning of slot $t$).
Also, only the users in $\mc{S}\cap\left(\cup_{k=1}^{\nu}\mc{D}_{k}\right)$,
where $\mc{S}$ is the set of users that receive $p$, can decode
any unknown native packets contained in $p$.
\end{corollary}

\begin{IEEEproof}
We have already shown in the proof of Lemma \ref{lem:instant_decode}
that if $p$ contains an unknown native packet for some user $i$,
then there exists some $k^{\ast}$ such that $i\in\mc{D}_{k^{\ast}}$,
which implies that $i\in\cup_{k=1}^{\nu}\mc{D}_{k}$. For the
converse, consider any user $i\in\cup_{k=1}^{\nu}\mc{D}_{k}$.
Then, there exists some $k^{\ast}\in\left\{ 1,\ldots,\nu\right\}$
such that $i\in\mc{D}_{k^{\ast}}$ and, repeating the argument
in the proof of Lemma \ref{lem:instant_decode}, we conclude that
$i$ is a Destination for $p$. Hence, the set of Destinations for $p$ 
at the beginning of slot $t$ is $\cup_{k=1}^{\nu}\mc{D}_{k}$.
Finally, it is obvious that a user $i$ can only decode an unknown
native packet $q$ (intended for $i$) after successful reception
of a packet $p$ that contains $q$. Hence, only the Destinations
of $p$ that receive it, i.e.~the users in $\mc{S}\cap\left(\cup_{k=1}^{\nu}
\mc{D}_{k}\right)$ can decode unknown native packets at the end of slot $t$.
\end{IEEEproof}

Notice that we have not yet proved the BP but only stated them as
desirable properties that the proposed scheme should possess. The
proof of BP, by induction on time, will be given after the full description
of the scheme. It still remains to examine how feedback can be efficiently
used to update our knowledge about the Listeners and Destinations
of a packet. This is performed in the next subsection.

\subsection{Packet movement}  \label{sub:Packet-movement}

We now describe how packets are moved between queues
$Q_{\mc{D}}^{\mc{L}}$ based on the received feedback. 
The next result is necessary here and follows immediately from BCR.
\begin{lemma}  \label{lem:Consider-packet-,}
Consider a packet $p=p_{\mc{D}_{1}}^{\mc{L}_{1}}\oplus\ldots\oplus
p_{\mc{D}_{\nu}}^{\mc{L}_{\nu}}$ formed according to BCR, where
$\left|\mc{D}_{i}\right|+\left|\mc{L}_{i}\right|\leq k$, for some
$i\in\left\{ 1,\ldots,\nu\right\} $. Then, it holds
$\nu\leq\left|\cup_{r=1}^{\nu}\mc{D}_{r}\right|=
\sum_{r=1}^{\nu}\left|\mc{D}_{r}\right|\leq k$.
\end{lemma}

\begin{IEEEproof}
Assume w.l.o.g. that $\abs{\mc{D}_{1}}+\abs{\mc{L}_{1}} \leq k$. The
BCR dictates $\mc{D}_{r}\subseteq\mc{L}_{1},\;\forall r\in\left\{
2,\ldots,\nu\right\} $, which implies $\bigcup_{r=2}^\nu \mc{D}_r
\subseteq \mc{L}_{1}$ and $\bigcup_{r=1}^\nu \mc{D}_r \subseteq
\mc{D}_1 \cup \mc{L}_1$.  Since all $\mc{D}_{r}$ sets are disjoint and
$\mc{L}_{1}\cap\mc{D}_{1}=\emptyset$, it holds $\sum_{r=1}^{\nu}
\abs{\mc{D}_{r}}=\abs{\bigcup_{r=1}^{\nu}\mc{D}_{r}} \leq \abs{
  \mc{D}_{1} \cup \mc{L}_{1}}=\abs{\mc{D}_{1}}+\abs{\mc{L}_{1}} \leq
k$.  Since $\mc{D}_{r}\neq\emptyset$ for all $r$
(i.e.~$\left|\mc{D}_{r}\right|\geq1$), it also holds
$\sum_{r=1}^{\nu}\left|\mc{D}_{r}\right|\geq\nu$, which completes the
desired inequality.
\end{IEEEproof}

As previously mentioned, we wish to always satisfy BP, since they
guarantee instantaneous decodability through Lemma
\ref{lem:instant_decode}.  Hence, the rationale behind the rules for
packet movement can be broadly stated as follows: ``after transmission
occurs at slot $t$ and feedback is gathered, packets may be placed in
new queues such that the BP are satisfied at the end of slot $t$
(equivalently, beginning of slot $t+1$). The role of feedback is to help the 
transmitter update its knowledge of the
Destinations and Listeners for each packet''. The following example will
serve to illustrate this point. In this example, we also describe how
the virtual packets (i.e.~tokens) are moved among the virtual
queues. Although the latter movement is purely virtual, this
description will be crucial in the ensuing analysis.

\begin{example}  \label{multiple_choice} \begin{upshape}
We consider the case of 3 users and, assuming BP holds at the
beginning of slot $t$, packet $p=p_{12}^{3}\oplus p_{3}^{12}$ is
transmitted at slot $t$ (this combination satisfies the Basic Coding
Rule). We assume that only user $2$ receives the packet; since, by
Corollary \ref{cor:Dforall}, user 2 is a Destination for $p$, it can
decode the unknown native packet $p_{12}^{3}(2)$ contained in
$p_{12}^{3}$, so that $K^3_{12}(2)$ is reduced by 1. For the other
packet movements, two choices are consistent with BP:
\begin{enumerate}
\item Packet $p_{12}^{3}$ is moved to queue $Q_{1}^{23}$ and packet
  $p_{3}^{12}$ is not moved; hence, regarding the virtual queues, only
  token $p^3_{12}(1)$ is (virtually) moved to $V^{23}_1(1)$ and
  $K^3_{12}(1)$ is reduced by 1 while $K^{23}_1(1)$ is increased by 1
  while all other counters are unaffected. This is consistent with BP
  since, after receiving $p$, receiver 2 becomes a Listener for
  $p_{12}^{3}=p\oplus p_{3}^{12}$ at the end of slot $t$, while
  receiver 3 is already a Listener for $p_{12}^{3}$ (due to
  BP\ref{enu:LD} at beginning of $t$) and remains so due to the
  absorbing property of Listener.

\item Packet $p$ is moved to queue $Q_{13}^{2}$ and packets
  $p_{12}^{3}$, $p_{3}^{12}$ are removed from queues $Q_{12}^{3}$,
  $Q_{3}^{12}$ respectively; hence, token $p^3_{12}(1)$ is moved to
  $V^2_{13}(1)$ and $p^{12}_3(3)$ is moved to
  $V^2_{13}(3)$. Additionally, counters $K^3_{12}(1)$, $K^{12}_3(3)$
  are reduced by 1 while $K^2_{13}(1)$, $K^2_{13}(3)$ are increased by
  1.  This is also consistent with BP since, after receiving $p$,
  receiver 2 becomes a Listener of $p$. Furthermore, by Corollary
  \ref{cor:Dforall}, users 1, 3 are Destinations for $p$ at the
  beginning of slot $t$ and, since no user received $p$, the unknown
  native packets for 1,3 contained in $p$ (at the beginning of slot
  $t$) remain unknown at the end of slot $t$. Hence, users 1,3 are
  still Destinations at the end of slot $t$.
\end{enumerate}

\end{upshape} \end{example}
Intuition at this point tells us that the higher the level of a queue
in which a packet $p$ is stored, the better are the chances of sending
multiple unknown native packets with a single transmission.
Specifically, by combining packets of queues in level $w$, we can send
up to $w$ unknown native packets per transmission, as stated in Lemma
\ref{lem:Consider-packet-,}. For example, $p=p_{1}^{2}\oplus
p_{2}^{1}$ contains two unknown native packets, one for user $1$ and
one for user $2$. To provide a more general example of a BCR-formed
packet that contains the maximum allowable number of unknown packets
for the given level queues, consider sets $\mc{L}_{i},\mc{D}_{i}$ for
$i=1,\ldots,\nu$ such that $\mc{L}_{i}\cup\mc{D}_{i}=\mc{W}$ for all
$i$ and $\cup_{i=1}^{\nu}\mc{D}_{i}=\mc{W}$, where
$\left|\mc{W}\right|=w$. It is now easy to show that packet
$p=\overset{\nu}{\underset{i=1}{\bigoplus}}p_{\mc{D}_{i}}^{\mc{W}-\mc{D}_{i}}$satisfies
the BCR, where all $p_{\mc{D}_{i}}^{\mc{W}-\mc{D}_{i}}$ are at level
$w$, and contains exactly $\left|\cup_{i=1}^{\nu}\mc{D}_{i}\right|=w$
unknown packets. For example, within queues of level $w=2$ and user
set $\mc{W}=\left\{ 1,2\right\} $ the most beneficial combination is
$p_{1}^{2}\oplus p_{2}^{1}$ which results in transmitting 2 unknown
native packets with a single transmission, while within queues of
level $w=3$ and user set $\mc{W}=\left\{ 1,2,3\right\} $ the most
beneficial combinations are any of the following types:
$p_{23}^{1}\oplus p_{1}^{23}$, $p_{13}^{2}\oplus p_{2}^{13}$,
$p_{12}^{3}\oplus p_{3}^{12}$ and $p_{1}^{23}\oplus p_{2}^{13}\oplus
p_{3}^{12}$. All these types result in 3 unknown native packets
transmitted simultaneously.

Additionally, among queues of a given level, packets at higher
sublevel queues can be combined with other packets in more ways than
packets of queues at lower sublevels. For example, $p_{12}^{3}$ can
only be combined with $p_{3}^{12}$ while $p_{3}^{12}$ can be combined
with 1) $p_{12}^{3}$, 2) $p_{1}^{23}$, 3) $p_{2}^{13}$ and 4)
$p_{1}^{23}\oplus p_{2}^{13}$. The benefit of having more available coding choices 
for a higher sublevel packet is that the probability of ``wasting'' a slot is reduced, 
as the following specific example illustrates for $N=3$: assume that the transmitter can 
either send a packet $p=p^3_{12}$ or a packet $p=p^{13}_2\oplus p^2_1$. Both choices have the 
same number of Destinations. In the first case, the slot is ``wasted'' (i.e.~no decoding or packet 
movement takes place) with probability $\epsilon_{12}$ (i.e.~iff $p$ is erased by users 1,2). However, 
in the second case, even if $p$ is erased by both of its Destinations (i.e.~users 1,2) and 
received by user 3, we can move $p^2_1=p\oplus p^{13}_2$ to $Q^{23}_1$ (since $p$ is known to 3);
as a result, the slot is ``wasted'' with a lower probability $\epsilon_{123}$, which
corresponds to the case that $p$ is erased by all users.

Of course, one can argue instead that if the only non-empty queues were $Q^3_{12}$ and $Q^{13}_2$, then
(applying an argument similar to that of the previous paragraph) it would be better to transmit
$p^3_{12}$ instead of $p^{13}_2$, since the former packet ``wastes'' a slot with probability 
$\epsilon_{12}$ and the latter with a higher probability $\epsilon_2$. Nevertheless, we have to 
consider that in a ``loaded'' system (i.e.~when the exogenous arrivals are close to the boundary of 
the stability region), most of the queues will be non-empty so that this scenario (where it is preferable 
to transmit a lower sublevel packet) is unlikely to occur. Hence, we intuitively expect that the scenario 
described in the previous paragraph will dominate performance-wise and this why, when multiple choices 
for packet movement arise (all of which satisfy the BP after movement), we select
the one that ensures that all packets involved in a transmission are
placed in a higher level and, within the same level, higher sublevels,
(else they are not moved at all). Thus, in Example \ref{multiple_choice}
above, we choose the first option, since $p_{12}^{3}$ is moved from
sublevel $3.1$ to $3.2$ and $p_{3}^{12}$ is not moved, while in the
second option $p_{3}^{12}$ descends from sublevel $3.2$ to $3.1$.

\begin{figure}[ht]
\includegraphics{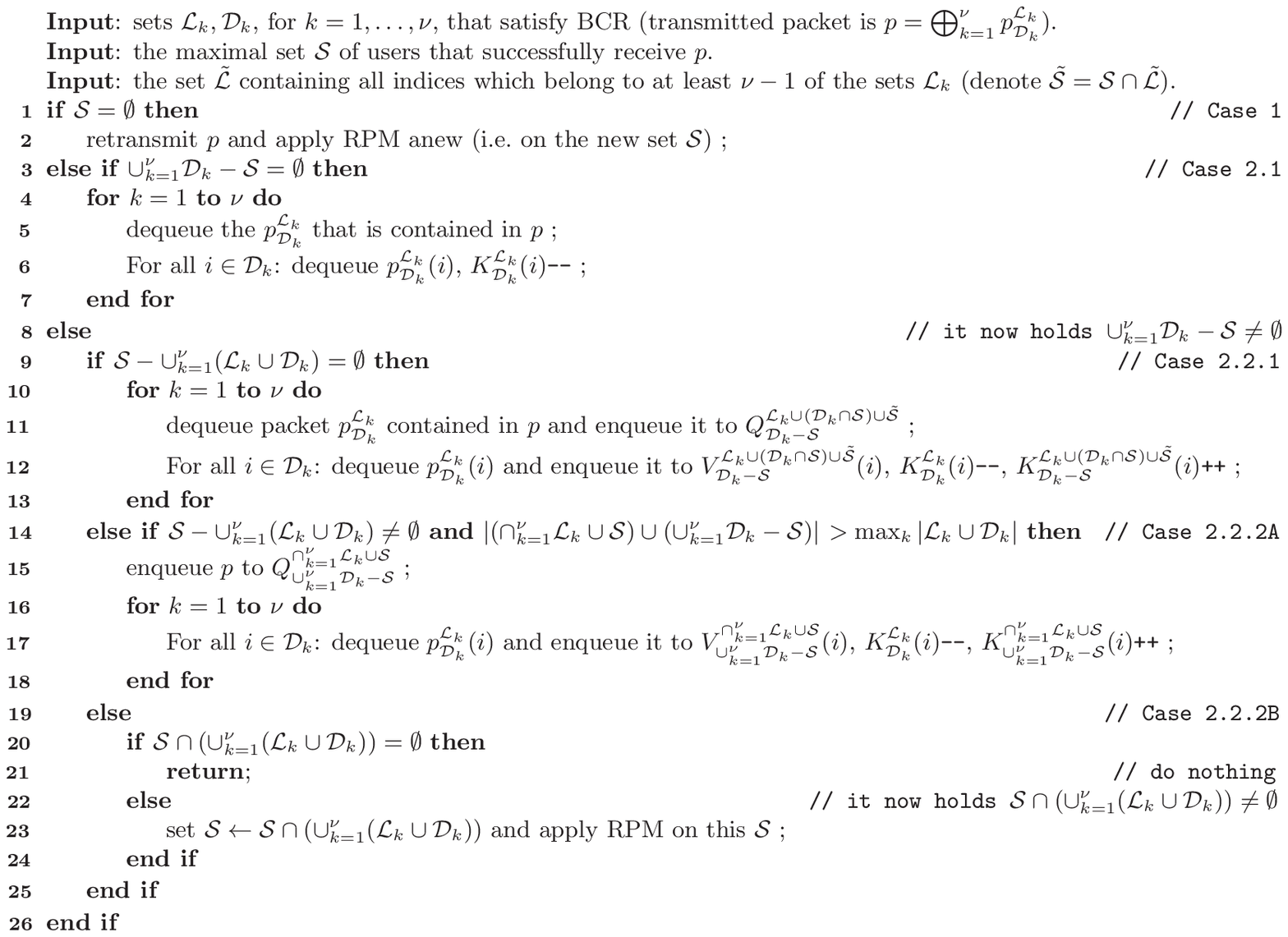}
\caption{Pseudocode representation for the Rules for Packet Movement.}
\label{fig:pseudo}
\end{figure}

The following specific rules for packet movement (shown in pseudocode form
in \fig{fig:pseudo}) have been devised according to the above rationale 
i.e.~assuming, for now, that BP holds at the beginning of slot $t$, we should 
move the packets in such a way that BP also holds at the end of slot $t$. For
the reader's benefit, we provide a high level description of the algorithmic
logic for each case and we use a mnemonic name in parentheses to easily
distinguish the cases.

\textbf{Rules for Packet Movement (RPM)}: Let packet $p$ of the form
$p=\bigoplus_{k=1}^{\nu} p_{\mc{D}_{k}}^{\mc{L}_{k}}$ satisfying the
Basic Coding Rule (BCR) be chosen for transmission at slot $t$, and
let $\mc{S}$ be the maximal set of users that receive $p$ (i.e.~the
packet is erased by all users in $\mc{S}^{c}$).  We define the set
$\tilde{\mc{L}}$ as follows: $i\in\tilde{\mc{L}}$ iff $i$ belongs to
at least $\nu-1$ of the sets $\mc{L}_{k}$, for
$k=1,\ldots,\nu$. Hence, before transmission of $p$, user
$i\in\tilde{\mc{L}}$ is a Listener for all but at most one of the
packets $p_{\mc{D}_{k}}^{\mc{L}_{k}}$, with $k=1,\ldots,\nu$. We also
denote with $\tilde{\mc{S}}=\mc{S}\cap\tilde{\mc{L}}$ the set of users
in $\tilde{\mc{L}}$ that received $p$. Note that it is quite possible
for $\tilde{\mc{S}}$ to be empty even though $\mc{S}\neq\emptyset$
(e.g.~$p=p_{2}^{1}\oplus p_{1}^{2}$, which satisfies BCR, with
$\mc{S}=\left\{ 3\right\}$ and $\tilde{\mc{L}}=\{1,2\}$). The
following rules are now checked and the corresponding actions are
performed (if applicable). Although only the real packets and queues
are handled by the transmitter, we also consider (at a conceptual
level) the virtual network and describe how it would be affected in
each case.

\begin{enumerate}
\item (\textit{$p$ is erased by all users}): If $\mc{S}=\emptyset$, then the transmitted packet is erased by all users.
Hence, no new information is gained by the users and the Destination/Listener sets for each packet in the network
remains unaffected (the current slot essentially being ``wasted''), which implies that no packet movement occurs 
and $p$ is retransmitted in the next slot.

\item Otherwise, it holds $\mc{S}\neq\emptyset$. In this case, by Corollary
  \ref{cor:Dforall} and Fact \ref{instant-decoding}, all users in
  $\cup_{k=1}^{\nu}\mc{D}_{k}$ (i.e.~the Destinations of packet $p$)
  that receive $p$ can instantly decode their unknown native packet,
  i.e.~for all $k\in\left\{ 1,\ldots,\nu\right\} $ and
  $i\in\mc{D}_{k}\cap\mc{S}$, packet $p_{\mc{D}_{k}}^{\mc{L}_{k}}(i)$
  is decoded by $i$ and its corresponding token is removed from the
  virtual network (as a result, $K^{\mc{L}_k}_{\mc{D}_k}(i)$ is
  reduced by 1).  Notice also that any $i\in\mc{D}_{k}\cap\mc{S}$
  becomes a Listener for $p_{\mc{D}_{k}}^{\mc{L}_{k}}$ after receiving
  $p$. Regarding the potential packet movements and counter changes:
\begin{enumerate}
\item (\textit{all Destinations of $p$ receive $p$}): 
If $\cup_{k=1}^{\nu}\mc{D}_{k}\subseteq\mc{S}$
(i.e.~$\cup_{k=1}^{\nu}\mc{D}_{k}-\mc{S}=\emptyset$), then all 
native packets $p^{\mathcal{L}_k}_{\mathcal{D}_k}(i)$, for $k=1,\ldots,\nu$
and $i\in \mathcal{D}_k$, are instantly decoded by their intended destinations
and their tokens are removed from the virtual network (as explained above),
since the corresponding native packets are no longer useful, having been decoded
by their intended users. For the same reason, for $k=1,\ldots,\nu$, all packets
$p^{\mathcal{L}_k}_{\mathcal{D}_k}$ that comprise $p$ are removed from the
respective queue $Q^{\mathcal{L}_k}_{\mathcal{D}_k}$ and no other packet/token
movement takes place.  \label{enu:RPM_2.1}

\item Otherwise, it holds $\cup_{k=1}^{\nu}\mc{D}_{k}-\mc{S}\neq\emptyset$
and we distinguish the following cases:  \label{enu:RPM_2.2}

\begin{enumerate}
\item \label{enu:case1} (\textit{only Destinations/Listeners of constituent
packets of $p$ receive $p$}): It holds $\mc{S}\subseteq \cup_{k=1}^{\nu}
(\mc{L}_{k}\cup\mc{D}_{k})$, equivalently
$\hat{\mc{S}}=S-\cup_{k=1}^{\nu}
(\mc{L}_{k}\cup\mc{D}_{k})=\emptyset$.  Notice that, for $\nu>1$, the
latter condition is equivalent, by the BCR, to
$\mc{S}\subseteq\cup_{k=1}^{\nu}\mc{L}_{k}$, while for $\nu=1$ it
reduces to $\mc{S}\subseteq\mc{L}_{1}\cup\mc{D}_{1}$.  In both cases,
and for each $k\in\left\{ 1,\ldots,\nu\right\} $, packet
$p_{\mc{D}_{k}}^{\mc{L}_{k}},\:\mc{{\rm
    where\:}D}_{k}-\mc{S}\neq\emptyset,$ is moved to queue
$Q_{\mc{D}_{k}-\mc{S}}^{\mc{L}_{k}\cup\left(\mc{D}_{k}\cap\mc{S}\right)\cup\tilde{\mc{S}}}$
and, for each $i\in \mc{D}_k-\mc{S}$, token
$p^{\mc{L}_k}_{\mc{D}_k}(i)$ is moved to $V^{\mc{L}_k\cup
  (\mc{D}_k\cap \mc{S})\cup
  \tilde{\mc{S}}}_{\mc{D}_k-\mc{S}}(i)$. Hence, counter
$K^{\mc{L}_k}_{\mc{D}_k}(i)$ is reduced by 1 while $K^{\mc{L}_k\cup
  (\mc{D}_k\cap \mc{S})\cup \tilde{\mc{S}}}_{\mc{D}_k-\mc{S}}(i)$ is
increased by 1. If, for some $k$, it holds
$\mc{D}_{k}-\mc{S}=\emptyset$, then all $p_{\mc{D}_{k}}^{\mc{L}_{k}}$
are removed from the respective queues. The two cases
$\mc{D}_{k}-\mc{S}\stackrel{\neq}{=}\emptyset$ can be jointly handled
following the convention that whenever a packet is moved to a queue
$Q_{\mc{D}}^{\mc{L}}$ with $\mc{D}=\emptyset$, it actually leaves the
network.  This will be systematically used below to avoid
repetition. The consistency of these packet movements with Basic
Properties is subsequently proved in Lemma \ref{lem:movok}.  Hence,
according to this rule, packet $p_{\mc{D}_{k}}^{\mc{L}_{k}}$ is either
not moved at all (if
$\tilde{\mc{S}}\cup\left(\mc{D}_{k}\cap\mc{S}\right)=\emptyset$), or
is moved to a higher level (or within the same level but higher
sublevel) queue, or exits the network completely (if
$\mc{D}_{k}-\mc{S}=\emptyset$). Also notive that, as intuitively
expected based on Definitions \ref{listener}, \ref{desti}, the current case 
guarantees that the Destination set (resp.~Listenerr set) of a packet cannot 
decrease(resp.~increase after a packet movement).

\item \label{enu:case2} It holds
$\hat{\mc{S}}=S-\cup_{k=1}^{\nu}(\mc{L}_{k}\cup\mc{D}_{k})\neq\emptyset$/
Again, this condition is equivalent to
$\hat{\mc{S}}=\mc{S}-\cup_{k=1}^{\nu}\mc{L}_{k}\neq\emptyset$, for
$\nu>1$, and
$\hat{\mc{S}}=\mc{S}-(\mc{L}_{1}\cup\mc{D}_{1})\neq\emptyset$ for
$\nu=1$. We further distinguish two subcases:

\begin{enumerate}
\item (\textit{received feedback creates a combined Listener/Destination set
in a level higher than that of all constituent packets of $p$}):
If $\abs{\left(\cap_{k=1}^{\nu} \mc{L}_{k} \cup\mc{S} \right)\cup
  \left(\cup_{k=1}^{\nu} \mc{D}_{k}-\mc{S}\right)}>
\max_{k=1,\ldots,\nu}\abs{\mc{L}_{k}\cup\mc{D}_{k}}$,\footnote{it is
  easy to verify that this inequality is always true for $\nu=1$.}
then packet $p$ is moved to
$Q_{\cup_{k=1}^{\nu}\mc{D}_{k}-\mc{S}}^{\cap_{k=1}^{\nu}\mc{L}_{k}\cup\mc{S}}$
and packets $p_{\mc{D}_{k}}^{\mc{L}_{k}}$ are removed from queues
$Q_{\mc{D}_{k}}^{\mc{L}_{k}}$. In the virtual network, for each $i\in
\mc{D}_{k}-\mc{S}$, token $p^{\mc{L}_k}_{\mc{D}_k}(i)$ is moved from
$V^{\mc{L}_k}_{\mc{D}_k}(i)$ to $V^{\cap_{k=1}^\nu \mc{L}_k\cup
  \mc{S}}_{\cup_{k=1}^\nu \mc{D}_k-\mc{S}}(i)$ (so that counters
$K^{\mc{L}_k}_{\mc{D}_k}(i)$ and
$K_{\cup_{k=1}^{\nu}\mc{D}_{k}-\mc{S}}^{\cap_{k=1}^{\nu}\mc{L}_{k}
  \cup\mc{S}}\left(i\right)$ are reduced by 1 and increased by 1,
respectively). Lemma \ref{lem:movok} shows again that this packet
movement is consistent with Basic Properties and the packets are moved
only to higher level or sublevel queues (or exit the network).

\item (\textit{no higher level Listener/Destination set, relative to constituent packets of $p$,
can be created based on received feedback}): 
If $\abs{\left(\cap_{k=1}^{\nu}\mc{L}_{k} \cup \mc{S}\right)
  \cup \left( \cup_{k=1}^{\nu} \mc{D}_{k}-\mc{S}\right)} \leq
  \max_{k=1,\ldots,\nu} \abs{\mc{L}_{k}\cup\mc{D}_{k}}$ then
\begin{itemize}
\item if $\mc{S}\cap\left(\cup_{k=1}^{\nu}(\mc{L}_{k}\cup\mc{D}_{k})\right)=\emptyset$,
no further action is taken. 

\item else, set
$\mc{S}\leftarrow\mc{S}\cap\left(\cup_{k=1}^{\nu}(\mc{L}_{k}\cup\mc{D}_{k})\right)$
and apply the above rules again for the new $\mc{S}$. Notice that Case
2.2.1 is now applicable for the new $\mc{S}$.
\end{itemize}
\end{enumerate}

\end{enumerate}
\end{enumerate}
\end{enumerate}

As previously mentioned, the validity of the above actions is proved
in the following result, which in turn guarantees the instant decodability
property. Induction on time then shows that BP is true for all slots
$t$ if BCR and RPM are applied in each slot.
\begin{lemma}  \label{lem:movok} 
Assuming that the Basic Properties are satisfied
at the beginning of slot $t$, then the application of the Basic Coding
Rule and Rules for Packet Movement to the packet transmitted at slot
$t$ satisfies the Basic Properties at the beginning of slot $t+1$. 
\end{lemma}

\begin{IEEEproof}
See Appendix \ref{sub:Proof-of-Lemma}.
\end{IEEEproof}
Since the Rules for Packet Movement have a complicated logical structure,
we provide the following concrete example for clarification. 
\begin{example}  \label{big_example} \begin{upshape}
Suppose packet $p=p_{1}^{2346}\oplus p_{24}^{135}\oplus p_{3}^{1246}$
is transmitted, so $\nu=3$ and $\mc{D}_{1}=\left\{ 1\right\} ,\mc{D}_{2}=\left\{ 2,4\right\} ,\mc{D}_{3}=\left\{ 3\right\} ,$
$\mc{L}_{1}=\left\{ 2,3,4,6\right\} ,$ $\mc{L}_{2}=\left\{ 1,3,5\right\} ,$
$\mc{L}_{3}=\left\{ 1,2,4,6\right\} $. Hence, $\cup_{k=1}^{3}\mc{D}_{k}=\left\{ 1,2,3,4\right\}$. 

\begin{itemize}
\item Suppose $p$ is received by users $2,5$ and $6$, so $\mc{S}=\left\{ 2,5,6\right\}$.
It holds $\cup_{k=1}^{3}\mc{D}_{k}-\mc{S}=\left\{ 1,3,4\right\} \neq\emptyset$
and $\mc{\hat{\mc{S}}=S}-\cup_{k=1}^{3}(\mc{L}_{k}\cup\mc{D}_{k})=\left\{ 2,5,6\right\} -\left\{ 1,2,3,4,5,6\right\} =\emptyset$,
so we are in case 2.2.1. We have $\mc{S}\cap\left(\cup_{k=1}^{3}(\mc{L}_{k}\cup\mc{D}_{k})\right)=\mc{S}=\left\{ 2,5,6\right\} $
and $\tilde{\mc{S}}=\left\{ 2,6\right\}$, because user $5$
does not belong to $\nu-1=2$ sets $\mc{L}_{k}$ but only to
set $\mc{L}_{2}$. The 3 packets are moved as follows:

\begin{itemize}
\item packet $p_{1}^{2346}$ is not moved because $\mc{D}_{1}\cap\mc{S}=\left\{ 1\right\} \cap\left\{ 2,5,6\right\}=\emptyset$
(equivalently, it is moved to $Q_{\mc{D}_{1}-\mc{S}}^{\mc{L}_{1}\cup\left(\mc{D}_{1}\cap\mc{S}\right)\cup\tilde{\mc{S}}}$,
i.e.~$Q_{\left\{ 1\right\} }^{\left\{ 2,3,4,6\right\} \cup\emptyset\cup\left\{ 2,6\right\} }=Q_{1}^{2346}$, 
which is where it is currently stored). 

\item packet $p_{24}^{135}$ is moved to $Q_{\mc{D}_{2}-\mc{S}}^{\mc{L}_{2}\cup\left(\mc{D}_{2}\cap\mc{S}\right)\cup\tilde{\mc{S}}}$,
i.e.~$Q_{\left\{ 2,4\right\} -\left\{ 2,5,6\right\} }^{\left\{ 1,3,5\right\} \cup\left(\left\{ 2,4\right\} \cap\left\{ 2,5,6\right\} 
\right)\cup\left\{ 2,6\right\} }=Q_{4}^{12356}$. 

\item packet $p_{3}^{1246}$ is not moved because $\mc{D}_{3}\cap\mc{S}=\left\{ 3\right\} \cap\left\{ 2,5,6\right\} =\emptyset$
(equivalently, it is moved to $Q_{\mc{D}_{3}-\mc{S}}^{\mc{L}_{3}\cup\left(\mc{D}_{3}\cap\mc{S}\right)\cup\tilde{\mc{S}}}$,
i.e.~$Q_{\left\{ 3\right\} }^{\left\{ 1,2,4,6\right\} \cup\emptyset\cup\left\{ 2,6\right\} }=Q_{3}^{1246}$). 
\end{itemize}

\item Suppose now that $p$ is received by users $7$ and $8$, so $\mc{S}=\left\{ 7,8\right\} $.
It holds $\cup_{k=1}^{3}\mc{D}_{k}-\mc{S}=\left\{ 1,2,3,4\right\}$
and $\mc{\hat{\mc{S}}=S}-\cup_{k=1}^{3}(\mc{L}_{k}\cup\mc{D}_{k})=\left\{ 7,8\right\} -\left\{ 1,2,3,4,5,6\right\} 
=\left\{ 7,8\right\} \neq\emptyset$, so we are in case 2.2.2. We have
\begin{align*}
& \left|\left(\cap_{k=1}^{3}\mc{L}_{k}\cup\mc{S}\right)\cup\left(\cup_{k=1}^{3}\mc{D}_{k}-\mc{S}\right)\right|\\
& =\left|\left(\left(\left\{ 2,3,4,6\right\} \cap\left\{ 1,3,5\right\} \cap\left\{ 1,2,4,6\right\} \right)\cup\left\{ 7,8\right\} \right)
    \cup\left(\left(\left\{ 1\right\} \cup\left\{ 2,4\right\} \cup\left\{ 3\right\} \right)-\left\{ 7,8\right\} \right)\right|\\
& =\left|\left\{ 1,2,3,4,7,8\right\} \right|=6.
\end{align*}
 We also have 
\[
\max_{k=1,\ldots,3}\left|\mc{L}_{k}\cup\mc{D}_{k}\right|=\max\left\{ \left|\left\{ 1,2,3,4,6\right\} \right|,
  \left|\left\{ 1,2,3,4,5\right\} \right|,\left|\left\{ 1,2,3,4,6\right\} \right|\right\} =5.
\]
Therefore, we are in subcase 2.2.2A, and $p$ is moved to $Q_{\cup_{k=1}^{3}\mc{D}_{k}-\mc{S}}^{\cap_{k=1}^{3}\mc{L}_{k}\cup\mc{S}}$,
i.e.~$Q_{1234}^{78}$. 

\item If $p$ is received by user $7$, then $\mc{S}=\left\{ 7\right\} $.
It holds $\cup_{k=1}^{3}\mc{D}_{k}-\mc{S}\neq\emptyset$ and $\mc{\hat{\mc{S}}=S}-\cup_{k=1}^{3}(\mc{L}_{k}\cup\mc{D}_{k})=
\left\{ 7\right\} -\left\{ 1,2,3,4,5,6\right\} =\left\{ 7\right\} \neq\emptyset$, so we are in case 2.2.2. We have 
\[
\left|\left(\cap_{k=1}^{3}\mc{L}_{k}\cup\mc{S}\right)\cup\left(\cup_{k=1}^{3}\mc{D}_{k}-\mc{S}\right)\right|=
  \left|\left\{ 1,2,3,4,7\right\} \right|=5,
\]
and $\max_{k=1,\ldots,3}\left|\mc{L}_{k}\cup\mc{D}_{k}\right|=5$.
We also have $\mc{S}\cap\left(\cup_{k=1}^{3}(\mc{L}_{k}\cup\mc{D}_{k})\right)=\left\{ 7\right\} \cap\left\{ 1,2,3,4,5,6\right\} =\emptyset$,
therefore we are in the first case of 2.2.2B and no packets are moved. 

\item If $p$ is received by users $2$ and $7$, then $\mc{S}=\left\{ 2,7\right\}$.
We have $\cup_{k=1}^{3}\mc{D}_{k}-\mc{S}\neq\emptyset$ and $\mc{\hat{\mc{S}}=S}-\cup_{k=1}^{3}(\mc{L}_{k}\cup\mc{D}_{k})=
\left\{ 2,7\right\} -\left\{ 1,2,3,4,5,6\right\} =\left\{ 7\right\} \neq\emptyset$, so we are in case 2.2.2. We have 
\[
\left|\left(\cap_{k=1}^{3}\mc{L}_{k}\cup\mc{S}\right)\cup\left(\cup_{k=1}^{3}\mc{D}_{k}-\mc{S}\right)\right|=
  \left|\left\{ 1,2,3,4,7\right\} \right|=5,
\]
and $\max_{k=1,\ldots,3}\left|\mc{L}_{k}\cup\mc{D}_{k}\right|=5$. We also have $\mc{S}\cap\left(\cup_{k=1}^{3}(\mc{L}_{k}\cup\mc{D}_{k})\right)
=\left\{ 2,7\right\} \cap\left\{ 1,2,3,4,5,6\right\} =\left\{ 2\right\} \neq\emptyset$, therefore we are in the second case of 2.2.2B. 
Next, we set $\mc{S}\leftarrow\mc{S}\cap\left(\cup_{k=1}^{3}(\mc{L}_{k}\cup\mc{D}_{k})\right)$, i.e.~$\mc{S}\leftarrow\left\{ 2\right\}$, 
and apply the same rules to the new $\mc{S}$, which brings us to case 2.2.1. We have $\tilde{\mc{S}}=\left\{ 2\right\} $ and the 3 packets are
moved as follows:
\begin{itemize}
\item packet $p_{1}^{2346}$ is not moved because $\mc{D}_{1}\cap\mc{S}=\left\{ 1\right\} \cap\left\{ 2\right\} =\emptyset$
(equivalently, it is moved to $Q_{\mc{D}_{1}-\mc{S}}^{\mc{L}_{1}\cup\left(\mc{D}_{1}\cap\mc{S}\right)\cup\tilde{\mc{S}}}$,
i.e.~$Q_{\left\{ 1\right\} }^{\left\{ 2,3,4,6\right\} \cup\emptyset\cup\left\{ 2\right\} }=Q_{1}^{2346}$). 
\item packet $p_{24}^{135}$ is moved to $Q_{\mc{D}_{2}-\mc{S}}^{\mc{L}_{2}\cup\left(\mc{D}_{2}\cap\mc{S}\right)\cup\tilde{\mc{S}}}$,
i.e.~$Q_{\left\{ 2,4\right\} -\left\{ 2\right\} }^{\left\{ 1,3,5\right\} \cup\left(\left\{ 2,4\right\} \cap\left\{ 2\right\} \right)
\cup\left\{ 2\right\} }=Q_{4}^{1235}$. 

\item packet $p_{3}^{1246}$ is not moved because $\mc{D}_{3}\cap\mc{S}=\left\{ 3\right\} \cap\left\{ 2\right\} =\emptyset$
(equivalently, it is moved to $Q_{\mc{D}_{3}-\mc{S}}^{\mc{L}_{3}\cup\left(\mc{D}_{3}\cap\mc{S}\right)\cup\tilde{\mc{S}}}$,
i.e.~$Q_{\left\{ 3\right\} }^{\left\{ 1,2,4,6\right\} \cup\emptyset\cup\left\{ 2\right\} }=Q_{3}^{1246}$). 
\end{itemize}
\end{itemize}

\end{upshape} \end{example}

The above choice of the Rules for Packet Movement allows for potential
feedback information loss, regarding which user knows which packet.
This is best illustrated in the third case of Example \ref{big_example}
where, although user 7 becomes a Listener for packet $p$ at the end
of slot $t$, this information is actually discarded. As explained,
this choice is made on intuitive grounds in order to keep the system
manageable and amenable to analysis. However, as will be seen in the
next Section, for $N=4$ even a more restrictive choice of rules suffices
to implement a policy with asymptotically (as packet length increases)
maximal stability region when the channel erasure probabilities are
i.i.d.

\subsection{Comparison between the Rules for Packet Movement and the rules in \cite{AGG}}

The reader who is familiar with the work in \cite{AGG} will notice that the current RPM constitute
an involved extension and \textit{strict} generalization of the rules in \cite{AGG}, i.e.~all
allowable packet movements in \cite{AGG} are still allowable in this work (and additional movements, not possible
in \cite{AGG}, are now allowed). A proof of this fact entails a straightforward enumeration of all possible 
feedback and application of the relevant RPM case and is omitted. However, for the reader's benefit, we provide 
Tables~\ref{act1}--\ref{act5}, which summarize the packet movements for all phases in 
\cite{AGG} and show which RPM case applies to them.

\section{Stabilizing Scheduling Policy}  \label{sec:Stability-region}

In this Section, we investigate the design of policies that, under
the coding restrictions and packet movements described in Section
\ref{sec:Network-coding-algorithms}, stabilize the system whenever
possible. We first need some definitions. \newpage

\begin{table}[t]
\centering
\caption{Selecting $p_i$ for transmission in phase 1 of \texttt{XOR2} in \cite{AGG}.}
\renewcommand{\arraystretch}{1.5}
\begin{tabular}{|c|c|c|c|l|} \hline 
\multirow{2}{*}{user $i$} & \multirow{2}{*}{user $j$} & \multirow{2}{*}{user $k$} & 
\multirow{2}{*}{action performed in \cite{AGG}} & \multirow{2}{*}{Corresponding case in RPM (for arbitrary $N$)} \\ 
  &  &  &  & leading to identical action \\ \hline 
R  & R  & R  & dequeue $p_i$; user $i$ decodes & Case 2.1 \\ \hline
R  & R  & E  & dequeue $p_i$; user $i$ decodes & Case 2.1 \\ \hline
R  & E  & R  & dequeue $p_i$; user $i$ decodes & Case 2.1 \\ \hline
R  & E  & E  & dequeue $p_i$; user $i$ decodes & Case 2.1 \\ \hline
E  & R  & R  & dequeue $p_i$, move $p_i$ to $Q^{jk}_i$ & Case 2.2.2A \\ \hline 
E  & R  & E  & dequeue $p_i$, move $p_i$ to $Q^j_i$ & Case 2.2.2A \\ \hline 
E  & E  & R  & dequeue $p_i$, move $p_i$ to $Q^k_i$ & Case 2.2.2A \\ \hline 
E  & E  & E  & retransmit & Case 1 \\ \hline 
\end{tabular}
\label{act1} 
\end{table}

\begin{table}[h!]
\centering
\caption{Selecting $p^i_j\oplus p^j_i$ for transmission in phase 2 of \texttt{XOR2} in \cite{AGG}.}
\renewcommand{\arraystretch}{1.5}
\begin{tabular}{|c|c|c|c|l|} \hline 
\multirow{2}{*}{user $i$} & \multirow{2}{*}{user $j$} & \multirow{2}{*}{user $k$} & 
\multirow{2}{*}{action performed in \cite{AGG}} & \multirow{2}{*}{Corresponding case in RPM (for arbitrary $N$)} \\ 
  &  &  &  & leading to identical action \\ \hline 
R  & R  & R  & dequeue $p_i^j$, $p_j^i$; users $i,j$ decode & Case 2.1 \\ \hline 
R  & R  & E  & dequeue $p_i^j$, $p_j^i$; users $i,j$ decode & Case 2.1 \\ \hline 
R  & E  & R  & dequeue $p_i^j$, $p_j^i$, move $p$ to $Q_j^{ik}$; user
$i$ decodes & Case 2.2.2A \\ \hline 
R  & E  & E  & dequeue $p_i^j$; user $i$ decodes & Case 2.2.1 \\ \hline
E  & R  & R  & dequeue $p_i^j$, $p_j^i$, move $p$ to $Q_i^{jk}$; user
$j$ decodes & Case 2.2.2A \\ \hline 
E  & R  & E  & dequeue $p_j^i$; user $j$ decodes & Case 2.2.1 \\ \hline 
E  & E  & R  & dequeue $p_i^j$, $p_j^i$, move $p$ to $Q_{ij}^k$ & Case 2.2.2A \\ \hline 
E  & E  & E  & retransmit & Case 1 \\ \hline 
\end{tabular}
\label{act2} 
\end{table}

\begin{table}[h!]
\centering
\caption{Selecting $p^i_{jk}\oplus p^{jk}_i$ for transmission in phase 3 (part 1) of \texttt{XOR2} in \cite{AGG}.}
\renewcommand{\arraystretch}{1.5}
\begin{tabular}{|c|c|c|c|l|} \hline 
\multirow{2}{*}{user $i$} & \multirow{2}{*}{user $j$} & \multirow{2}{*}{user $k$} & 
\multirow{2}{*}{action performed in \cite{AGG}} & \multirow{2}{*}{Corresponding case in RPM (for arbitrary $N$)} \\ 
  &  &  &  & leading to identical action \\ \hline 
R  & R  & R  & dequeue $p_i^{jk},\: p_{jk}^i$; all 3 users decode & Case 2.1 \\ \hline 
R  & R  & E  & dequeue $p_i^{jk},\: p_{jk}^i$, move $p_{jk}^i$ to $Q_k^{ij}$;
users $i,j$ decode & Case 2.2.1 \\ \hline 
R  & E  & R  & dequeue $p_i^{jk},\: p_{jk}^i$, move $p_{jk}^i$ to $Q_j^{ik}$;
users $i,k$ decode & Case 2.2.1 \\ \hline 
R  & E  & E  & dequeue $p_i^{jk}$; user $i$ decodes & Case 2.2.1 \\ \hline 
E  & R  & R  & dequeue $p_{jk}^i$; users $j$, $k$ decode & Case 2.2.1 \\ \hline 
E  & R  & E  & dequeue $p_{jk}^i$, move $p_{jk}^i$ to $Q_k^{ij}$; user $j$
decodes & Case 2.2.1 \\ \hline 
E  & E  & R  & dequeue $p_{jk}^i$, move $p_{jk}^i$ to $Q_j^{ik}$; user $k$
decodes & Case 2.2.1 \\ \hline 
E  & E  & E  & retransmit & Case 1 \\ \hline 
\end{tabular}
\label{act3.1} 
\end{table}

\newpage 
 
\begin{table}[t]
\centering
\caption{Selecting $p^i_{jk}$ for transmission in phase 3 (part 2) of \texttt{XOR2} in \cite{AGG}.}
\renewcommand{\arraystretch}{1.5}
\begin{tabular}{|c|c|c|c|l|} \hline 
\multirow{2}{*}{user $i$} & \multirow{2}{*}{user $j$} & \multirow{2}{*}{user $k$} & 
\multirow{2}{*}{action performed in \cite{AGG}} & \multirow{2}{*}{Corresponding case in RPM (for arbitrary $N$)} \\ 
  &  &  &  & leading to identical action \\ \hline 
R  & R  & R  & dequeue $p^i_{jk}$; users $j,k$ decode & Case 2.1 \\ \hline 
R  & R  & E  & dequeue $p^i_{jk}$; move $p^i_{jk}$ to $Q^{ij}_k$; user $j$ decodes & Case 2.2.1 \\ \hline 
R  & E  & R  & dequeue $p^i_{jk}$, move $p^i_{jk}$ to $Q^{ik}_j$;
user $k$ decodes & Case 2.2.1 \\ \hline
R  & E  & E  & $p^i_{jk}$ remains in $Q^i_{jk}$ & Case 2.2.1 \\ \hline 
E  & R  & R  & dequeue $p^i_{jk}$; users $j,k$ decode & Case 2.1 \\ \hline 
E  & R  & E  & dequeue $p^i_{jk}$, move $p^i_{jk}$ to $Q^{ij}_k$; user $j$
decodes & Case 2.2.1 \\ \hline 
E  & E  & R  & dequeue $p^i_{jk}$, move $p^i_{jk}$ to $Q^{ik}_j$; user $k$
decodes & Case 2.2.1 \\ \hline 
E  & E  & E  & retransmit & Case 1 \\ \hline 
\end{tabular}
\label{act3.2} 
\end{table}

\begin{table}[h!]
\centering
\caption{Selecting $p^j_i\oplus p^{ik}_j$ for transmission in phase 4 (part 1) of \texttt{XOR2} in \cite{AGG}.}
\renewcommand{\arraystretch}{1.5}
\begin{tabular}{|c|c|c|c|l|} \hline 
\multirow{2}{*}{user $i$} & \multirow{2}{*}{user $j$} & \multirow{2}{*}{user $k$} & 
\multirow{2}{*}{action performed in \cite{AGG}} & \multirow{2}{*}{Corresponding case in RPM (for arbitrary $N$)} \\ 
  &  &  &  & leading to identical action \\ \hline 
R  & R  & R  & dequeue $p^j_i$, $p^{ik}_j$; users $i,j$ decode & Case 2.1 \\ \hline 
R  & R  & E  & dequeue $p^j_i$, $p^{ik}_j$; users $i,j$ decode & Case 2.1 \\ \hline 
R  & E  & R  & dequeue $p^j_i$; user $i$ decodes & Case 2.2.1 \\ \hline
R  & E  & E  & dequeue $p^j_i$; user $i$ decodes & Case 2.2.1 \\ \hline
E  & R  & R  & dequeue $p^j_i$, $p^{ik}_j$, move $p^j_i$ to $Q^{jk}_i$; user $j$ decodes & Case 2.2.1 \\ \hline 
E  & R  & E  & dequeue $p^{ik}_j$; user $j$ decodes & Case 2.2.1 \\ \hline
E  & E  & R  & dequeue $p^j_i$, move $p^j_i$ to $Q^{jk}_i$ & Case 2.2.1 \\ \hline
E  & E  & E  & retransmit & Case 1 \\ \hline 
\end{tabular}
\label{act4.1} 
\end{table}

\begin{table}[h!]
\centering
\caption{Selecting $p^j_i$ for transmission in phase 4 (part 2) of \texttt{XOR2} in \cite{AGG}.}
\renewcommand{\arraystretch}{1.5}
\begin{tabular}{|c|c|c|c|l|} \hline 
\multirow{2}{*}{user $i$} & \multirow{2}{*}{user $j$} & \multirow{2}{*}{user $k$} & 
\multirow{2}{*}{action performed in \cite{AGG}} & \multirow{2}{*}{Corresponding case in RPM (for arbitrary $N$)} \\ 
  &  &  &  & leading to identical action \\ \hline 
R  & R  & R  & dequeue $p^j_i$; user $i$ decodes & Case 2.1 \\ \hline
R  & R  & E  & dequeue $p^j_i$; user $i$ decodes & Case 2.1 \\ \hline 
R  & E  & R  & dequeue $p^j_i$; user $i$ decodes & Case 2.1 \\ \hline
R  & E  & E  & dequeue $p^j_i$; user $i$ decodes & Case 2.1 \\ \hline
E  & R  & R  & dequeue $p^j_i$, move $p^j_i$ to $Q^{jk}_i$ & Case 2.2.2A \\ \hline
E  & R  & E  & $p^j_i$ remains in $Q^j_i$ & Case 2.2.1 \\ \hline
E  & E  & R  & dequeue $p^j_i$, move $p^j_i$ to $Q^{jk}_i$ & Case 2.2.2A \\ \hline
E  & E  & E  & retransmit & Case 1 \\ \hline 
\end{tabular}
\label{act4.2} 
\end{table} 
 
\newpage

\subsection{System Stability and Stability Region}

Let $X\left(t\right),$ $t=0,1,\ldots$ be a stochastic process. 
\begin{definition}
[Stability] The process $X\left(t\right),$ $t=0,1,\ldots$ is stable iff 
\[
\lim_{q\rightarrow\infty}\underset{{\scriptstyle t\rightarrow\infty}}{\limsup}\Pr\left(X\left(t\right)>q\right)=0.
\]
\end{definition} 

\begin{table}[t]
\centering
\caption{Selecting $p^{jk}_i\oplus p^{ik}_j \oplus p^{ij}_k$ for transmission in phase 5 of \texttt{XOR2} in \cite{AGG}.}
\renewcommand{\arraystretch}{1.5}
\begin{tabular}{|c|c|c|c|l|} \hline 
\multirow{2}{*}{user $i$} & \multirow{2}{*}{user $j$} & \multirow{2}{*}{user $k$} & 
\multirow{2}{*}{action performed in \cite{AGG}} & \multirow{2}{*}{Corresponding case in RPM (for arbitrary $N$)} \\ 
  &  &  &  & leading to identical action \\ \hline 
R  & R  & R  & dequeue $p^{jk}_i$, $p^{ik}_j$, $p^{ij}_k$; users $i,j,k$ decode & Case 2.1 \\ \hline
R  & R  & E  & dequeue $p^{jk}_i$, $p^{ik}_j$; users $i,j$ decode & Case 2.2.1 \\ \hline
R  & E  & R  & dequeue $p^{jk}_i$, $p^{ij}_k$; users $i,k$ decode & Case 2.2.1 \\ \hline
R  & E  & E  & dequeue $p^{jk}_i$; user $i$ decodes & Case 2.2.1 \\ \hline
E  & R  & R  & dequeue $p^{ik}_j$, $p^{ij}_k$; users $j,k$ decode & Case 2.2.1 \\ \hline
E  & R  & E  & dequeue $p^{ik}_j$; user $j$ decodes & Case 2.2.1 \\ \hline
E  & E  & R  & dequeue $p^{ij}_k$; user $k$ decodes & Case 2.2.1 \\ \hline
E  & E  & E  & retransmit & Case 1 \\ \hline 
\end{tabular}
\label{act5} 
\end{table}

Consider next a time-slotted system $\mc{U}$. At the beginning of each
slot, a number of new packets belonging to a set $\mc{N}$ of ``flows''
arrive to the system. Newly arriving packets of flow $i\in\mc{N}$ are
placed at infinite size queues, i.e.~no incoming packets are ever
dropped. These packets are processed by a policy $\pi$ belonging to a
set $\Pi$ of admissible policies. We hereafter use the term ``policy''
to refer to a collection of rules for choosing which packets, stored
in a set of queues $\mc{Q}$, to combine through a XOR operation and
how to move packets between the queues in $\mc{Q}$ (the rules also
allow for a packet to exit the system). The exact rules will be stated
later. Let $A_{i}\left(t\right)$, $i\in\mc{N}$, be the number of flow
$i$ packets arriving at the system at the beginning of slot $t.$ For
the purposes of this paper, we assume that the process $\left\{
\boldsymbol{A}(t)\right\} _{t=0}^{\infty}$, where
$\boldsymbol{A}\left(t\right)=\left( A_i(t): \, i\in \mc{N} \right)$,
consists of i.i.d vectors with
$\mathbb{E}[\boldsymbol{A}(t)]=\boldsymbol{\lambda}\geq
\boldsymbol{0}$. We denote with $Q_{l}^{\pi}\left(t\right)$ the number
of packets in queue $Q_{l}\in\mc{Q}$ at time $t$ when policy
$\pi\in\Pi$ is applied, and define
$\hat{Q}^{\pi}(t)=\sum_{Q_{l}\in\mc{Q}}Q_{l}^{\pi}\left(t\right).$
\begin{definition}[System Stability] 
\hspace{1cm}
\begin{enumerate}
\item For a given arrival rate vector $\boldsymbol{\lambda}$, system $\mc{U}$
is stable under policy $\pi$ if the process $\hat{Q}^{\pi}(t)$ is stable.
 
\item The stability region $\mc{R}^{\pi}$ of a policy $\pi\in\Pi$ is the closure 
of the set of arrival rates for which $\mc{U}$ is stable under $\pi$.
 
\item The stability region $\mc{R}_{\Pi}$ of system $\mc{U}$
under the set of policies $\Pi$ is the closure of the set $\cup_{\pi\in\Pi}\mc{R}^{\pi}.$ 

\item A policy $\pi^{*}\in\Pi$ is stabilizing within $\Pi$ if $\mc{R}_{\Pi}=\mc{R}^{\pi^{*}}$. 
\end{enumerate}
\end{definition}

Consider now the system under study in the current work. At the
beginning of each slot, a decision must be made at the base station
concerning the combination of packets from the real queues that must
be XORed to form the packet
$p=p_{\mc{D}_{1}}^{\mc{L}_{1}}\oplus\ldots\oplus
p_{\mc{D}_{\nu}}^{\mc{L}_{\nu}}$ to be transmitted. Such a decision is
called a ``control''
$I_{\mc{D}_{1},\ldots,\mc{D}_{\nu}}^{\mc{L}_{1},\ldots,\mc{L}_{\nu}}$
and we denote the set of such controls by $\mc{I}$. Notice that, by
definition, a control is identified by the set $\left\{
\left(\mc{D}_{i},\mc{L}_{i}\right)\right\} _{i=1}^{\nu}$ and not by
the order of the elements in the set, i.e.~control
$I_{\mc{D}_{1},\ldots,\mc{D}_{\nu}}^{\mc{L}_{1},\ldots,\mc{L}_{\nu}}$
is identical to control
$I_{\mc{D}_{\sigma(1)},\ldots,\mc{D}_{\sigma(\nu)}}^{\mc{L}_{\sigma(1)},\ldots,\mc{L}_{\sigma(\nu)}}$
for any permutation $\sigma(i)$ of the indices on $\left\{
1,\ldots,\nu\right\}$.

We assume henceforth that the Basic Coding Rule is followed for the
formation of packet $p$. For this system, an admissible policy
consists of selecting, at the beginning of each time slot, one of the
available controls
$I_{\mc{D}_{1},\ldots,\mc{D}_{\nu}}^{\mc{L}_{1},\ldots,\mc{L}_{\nu}}$
to form a packet $p$ for transmission. After $p$ is transmitted,
packets are moved among the real queues
$Q_{\mc{D}_{k}}^{\mc{L}_{k}}(i)$ according to the Rules for Packet
Movement (RPM) described in Section
\ref{sec:Network-coding-algorithms}. We also consider the virtual
network, where a token for an exogenous native packet for user $i\in
\mc{N}$ is initially stored in $V_i(i)$ and then travels through the
virtual network according to the RPM (as it now applies to the virtual
queues only). Hence, there exist two different queue networks, a
``real network'' $\mc{Q}= \bigcup_{\mc{L},\mc{D}} \left\{
Q^{\mc{L}}_{\mc{D}} \right\}$ and a ``virtual network'' $\mc{V}=
\bigcup_{\mc{L},\mc{D}} \bigcup_{i\in \mc{D}} \left\{
V^{\mc{L}}_{\mc{D}}(i) \right\}$, although only the former is actually
present in the transmitter (the latter should be understood as part of
a thought experiment that facilitates the analysis).

\begin{figure}[t]
\centering
\psfrag{Q3_12(1)}[][][0.8]{$V^2_{12}(1)$}
\psfrag{Q3_12(2)}[][][0.8]{$V^3_{12}(2)$}
\psfrag{Q12_3(3)}[][][0.8]{$V^{12}_3(3)$}
\psfrag{Q23_1(1)}[][][0.8]{$V^{23}_1(1)$}
\psfrag{Q13_2(2)}[][][0.8]{$V^{13}_2(2)$}
\psfrag{d1}[][][0.8]{$d_1$} \psfrag{d2}[][][0.8]{$d_2$} \psfrag{d3}[][][0.8]{$d_3$}
\includegraphics[scale=0.45]{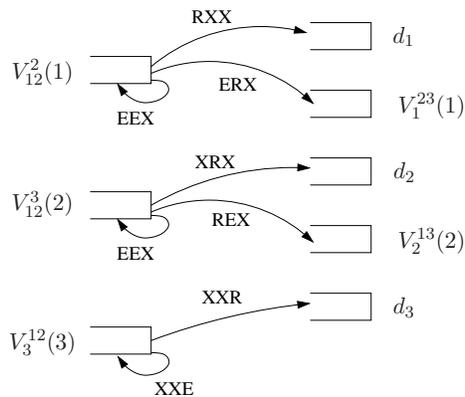}
\caption{Possible movements of tokens $p_{12}^{3}(1),\: p_{12}^{3}(2),\: p_{3}^{12}(3)$.
Destination of user $i$ is denoted as $d_{i}$. Received feedback
is denoted as $\left(u_{1},u_{2},u_{3}\right)$, where $u_{i}$ is
the feedback from user $i$, where R, E stand for received, erased,
respectively, while X denotes an unspecified value (either R or E).}
\label{transfig-1} 
\end{figure}

We now identify $\Pi$ as the set of admissible policies that select
transmitted packets according to the Basic Coding Rule and then move
packets based on the Rules for Packet Movement. A characteristic of
such movements is that the destination (i.e.~queue) of a packet
movement cannot be determined at the beginning of transmission since
it depends on the feedback received after packet transmission.  For
example, assume that $N=3$ and control $I_{12,3}^{3,12}$ is applied,
i.e.~packet $p=p_{12}^{3}\oplus p_{3}^{12}$ is transmitted.  The
tokens involved in this transmission are
$p_{12}^{3}(1),p_{12}^{3}(2),p_{3}^{12}(3)$. Figure \ref{transfig-1}
shows the possible movements of these tokens according to the received
feedback.

Under the above definition of $\Pi$, any policy $\pi\in \Pi$ can be
individually applied to the ``real'' and ``virtual'' network. Defining
$\hat{Q}^\pi(t)=\sum_{\mc{L},\mc{D}} \abs{\left(
  Q^{\mc{L}}_{\mc{D}}(t) \right)^\pi}$ and $\hat{V}^\pi(t)=
\sum_{\mc{L},\mc{D}} \sum_{i\in \mc{D}} \abs{\left(
  V^{\mc{L}}_{\mc{D}}(i)(t) \right)^\pi}$ as the total backlog at slot
$t$ in each network (and hereafter dropping the $\pi$ superscript in
the queues), we can use Lemma~\ref{lem:allsame} to write
\begin{equation}
\hat{Q}^\pi(t)= \sum_{\mc{L},\mc{D}} \abs{Q^{\mc{L}}_{\mc{D}}(t)} \leq \sum_{\mc{L},\mc{D}} \abs{\mc{D}} 
  \abs{Q^{\mc{L}}_{\mc{D}}(t)}= \hat{V}^\pi(t) \leq \sum_{\mc{L},\mc{D}} N \abs{Q^{\mc{L}}_{\mc{D}}(t)} ,
\end{equation}
since $\abs{\mc{D}}\leq N$, whence we conclude that
$\frac{\hat{V}^\pi(t)}{N} \leq \hat{Q}^\pi(t) \leq
\hat{V}^\pi(t)$. The last inequality implies that the real and virtual
networks have the same stability region. Surprisingly, it also implies
that the total number of packets stored in the real queues at any time
is generally less than the total number of unknown native packets at
that time.

Furthermore, it turns out that the virtual network falls in the class
of systems whose stability has been studied in \cite{pasxos2}. We next
summarize the formulation and main results in \cite{pasxos2} in a
manner that will be useful in the development that follows. Consider a
slotted-time network with a node set $\mc{M}\cup\left\{ d\right\} $,
where $d\not\in\mc{M}$, and directed edge (i.e.~link) set ${\cal E}$,
where the special node $d$ represents the destination of traffic
originated at the nodes in $\mc{M}$ (for now, assume there is a single
destination for all traffic).  Let
$\mc{E}_{out}^{m},\:\mc{E}_{in}^{m}$ denote, respectively, the set of
outgoing links and incoming links to node $m\in\mc{M}$ and assume that
$\mc{E}_{out}^{m}\neq\emptyset$ for all $m\in\mc{M}$.  We allow
self-loops in the network, i.e.~for node $m\in\mc{M}$, there may be a
link $(m,m),$ implying that the sets
$\mc{E}_{out}^{m},\:\mc{E}_{in}^{m}$ may both contain node $m.$ A
finite set of controls $\mc{I}$ is available. For each control
$I\in\mc{I},$ ``transmission'' takes place over the set of outgoing
links $\mc{E}_{out}^{m}$ of node $m\in\mc{M}$ in a random manner as
follows.
\begin{itemize}
\item If, at a given slot, control $I\in\mc{I}$ is applied, then,
for any node $m\in\mc{M}$, at most $\hat{\mu}_{m}(I)\in\left\{ 0,1\right\} $
packets may be transmitted ``over the set'' $\mc{E}_{out}^{m}$
in the following random manner: For each $I\in\mc{I}$, there
is a random sequence $R_{n}^{m}(I)$, with $n\geq1$, $m\in\mc{M}$,
where each $R_{n}^{m}\left(I\right)$ takes values in the set $\mc{E}_{out}^{m},$
with the following interpretation. A packet (if any) transmitted from
node $m$ over the set $\mc{E}_{out}^{m}$ when control $I$
is applied for the $n$-th time, is received \emph{only }by the recipient
of the link $R_{n}^{m}\left(I\right).$ Of course, if $R_{n}^{m}\left(I\right)=(m,m)$
then the packet is not received by any node in $\mc{E}_{out}^{m}$-$\left\{ m\right\} ,$
hence it remains at node $m$. 
\end{itemize}

For a given $n$ and $I$, the random variables $R_{n}^{m}\left(I\right),\ m\in\mc{M},$
may be arbitrarily correlated. Moreover, we assume that for each control
$I\in\mc{I},$ the random sequences $\left\{ R_{n}^{m}\left(I\right),\ m\in\mc{M}\right\} _{n=1}^{\infty}$
are i.i.d., independent of the arrival processes, and define $p_{e}^{m}\left(I\right)\triangleq
\Pr\left(R_{n}^{m}\left(I\right)=e\right)$ for $e\in\mc{E}_{out}^{m}$ so that 
\begin{equation}
\sum_{e\in\mc{E}_{out}^{m}}p_{e}^{m}(I)=1\quad\forall\, m\in\mc{M},\,\forall\, I\in\mc{I}.\label{eq:ProbSum}
\end{equation}
Strictly speaking, the description above is for nodes for which $\hat{\mu}_{m}\left(I\right)>0.$
In case $\hat{\mu}_{m}\left(I\right)=0$ for some $m\in\mc{M}$,
to avoid complicated notation, it is helpful to set $R_{n}^{m}\left(I\right)=e_{0}$
for some fixed $e_{0}\in\mc{E}_{out}^{m}$.

To describe the stability region $\mc{R}_{\Pi}$ of this network,
we need some preliminary definitions. For control $I\in\mc{I}$,
we define the set $\Gamma(I)$ of vectors $\boldsymbol{f}$ as 
\begin{equation} \label{eq:fows-control}
\Gamma\left(I\right)=\left\{ \boldsymbol{f}= (f_{e})_{e\in\mc{E}}:\: f_{e}=p_{e}^{m}(I)\mu_{m},\:0 \leq
  \mu_{m}\leq\hat{\mu}_{m}(I),\; m\in\mc{M},\, e\in\mc{E}_{out}^{m}\right\} ,
\end{equation}
and the convex hull $\mc{H}$ of the sets $\Gamma(I)$ as 
\begin{equation} \label{eq:ltservice}
\mc{H}=\mathrm{conv}\left(\Gamma\left(I\right),I\in\mc{I}\right).
\end{equation}

The stability region of the network $(\mc{M}\cup\{d\},\mc{E})$ is
described by the following Theorem.
\begin{theorem}  \label{thm:StabCond}
\cite{pasxos2} The stability region $\mc{R}_{\Pi}$
of the system is the set of arrival rates $\boldsymbol{\lambda}=\left\{ 
\lambda_{m}\right\}_{m\in\mc{M}}$, $\lambda_{m}\geq 0$, for which there 
exists a vector $\boldsymbol{f}\in\mc{H}$ such that for all nodes 
$m\in\mc{M}$ it holds
\begin{equation} \label{eq:mainStabCond}
\sum_{e\in\mc{E}_{in}^{m}}f_{e}+\lambda_{m}\leq\sum_{e\in\mc{E}_{out}^{m}}f_{e}.
\end{equation}
\end{theorem}

We will apply the formulation described above to the network
consisting of the virtual queues $V_{\mc{D}}^{\mc{L}}(i),\:
i\in\mc{D},$ i.e., we consider $\mc{M}=\left\{
V_{\mc{D}}^{\mc{L}}(i):\: i\in\mc{D}\right\} $ for all $\mc{L},\mc{D}$
that satisfy CC. For this network, since at most one virtual packet
(i.e.~token) is transmitted per slot from any queue $m$, we have
$\hat{\mu}_{m}(I)\in\left\{ 0,1\right\} $, $m\in\mc{M}$. Also, the
packet transition probabilities $p_{e}^{m}\left(I\right)$ for nodes
with $\hat{\mu}_{m}(I)=1$ can be easily calculated (an example is
given below). The only difference between the network
$(\mc{M}\cup\left\{ d\right\} ,\mc{E})$ and our model is that, in the
latter, there are $N$ token destinations, $d_{i},\: i\in\mc{N}$ (one
for each of the receivers) instead of a single one. However, we can
combine all these destinations to a single destination $d$, so that
any token arriving in $d_{i}$ is considered to arrive at $d$. This
affects neither the admissible policies, nor the queue sizes at the
various native queues at the base station. Hence, system stability is
not affected, provided that we are interested in the total queue size
at the base station.

\begin{example} \begin{upshape}
Consider the case $N=3$ and assume that control $I_{12,3}^{3,12}$
is chosen, hence a combination $p=p_{12}^{3}\oplus p_{3}^{12}$ is
transmitted, where $p_{12}^{3}=p_{12}^{3}\left(1\right)\oplus p_{12}^{3}\left(2\right)$
and $p_{3}^{12}=p_{3}^{12}\left(3\right)$ (recall
Section \ref{enu:the-unknown-native} for the interpretation of the
parentheses). The transition probabilities are then as follows:

\begin{itemize}
\item Token $p_{12}^{3}\left(1\right)$:

\begin{enumerate}
\item If $p$ is received by user $1$, $p_{12}^{3}\left(1\right)$ is removed
from $V_{12}^{3}\left(1\right)$ and delivered to $d_{1}$ (i.e.~to $d$
for the equivalent network). This event has probability
$P_{\emptyset,\left\{ 1\right\} }$.

\item If $p$ is erased at user 1 and received by user 2, packet $p_{12}^{3}$
is moved to queue $Q_{1}^{23}$ and token $p_{12}^{3}\left(1\right)$
is moved to $V_{1}^{23}\left(1\right)$. This event has probability
$P_{\left\{ 1\right\} ,\left\{ 2\right\} }$. 

\item If $p$ is erased at users 1 and 2, $p_{12}^{3}\left(1\right)$ remains
at $V_{12}^{3}\left(1\right)$. This event has probability $P_{\left\{ 1,2\right\} ,\emptyset}$. 
\end{enumerate}

\item Token $p_{12}^{3}\left(2\right)$: the transition probabilities are
determined as in the previous case, by interchanging the indices 1, 2. 

\item Token $p_{3}^{12}\left(3\right)$:
\begin{enumerate}
\item If $p$ is received by user 3, $p_{3}^{12}\left(3\right)$ is removed
from $V_{3}^{12}\left(3\right)$ and delivered to $d_{3}$. 
This event has probability $P_{\emptyset,\left\{ 3\right\} }$. 

\item If $p$ is erased at $3,$ $p_{3}^{12}\left(3\right)$ remains at
$V_{3}^{12}\left(3\right)$. This event has probability $P_{\left\{
  3\right\} ,\emptyset}$.
\end{enumerate}

\end{itemize}
\end{upshape} \end{example}

We now describe the stability region of Theorem \ref{thm:StabCond}
in a form that is more convenient for calculations. Any $\boldsymbol{f}$
in $\mc{H}$ can be written in the form 
\begin{equation} \label{eq:Probability}
\boldsymbol{f}=\sum_{I\in\mc{I}}\phi_{I}\boldsymbol{f}\left(I\right),\mbox{ for some }\left\{ \phi_{I}\right\} _{I\in\mc{I}}
  \mbox{ such that } \phi_{I}\geq0,\:\sum_{I\in\mc{I}}\phi_{I}\leq 1 ,
\end{equation}
where 
\[
\boldsymbol{f}\left(I\right)=\left(f_{e}\left(I\right)\right)_{e\in\mc{E}},
\]
\[
f_{e}\left(I\right)=p_{e}^{m}(I)\mu_{m}(I),\:0\leq\mu_{m}\left(I\right)\leq\hat{\mu}_{m}(I),\: m\in\mc{M},\,e\in\mc{E}_{out}^{m},
\]
and, for any control $I=I_{\mc{D}_{1},\ldots,\mc{D}_{\nu}}^{\mc{L}_{1},\ldots,\mc{L}_{\nu}}$,
we define the set $\mc{M}(I)=\bigcup_{r=1}^{\nu}\bigcup_{k\in\mc{D}_{r}}\left\{ V_{\mc{D}_{r}}^{\mc{L}_{r}}(k)\right\} $
so that 
\begin{equation}
\hat{\mu}_{m}(I)=\left\{ \begin{array}{cc}
1 & \text{if \ensuremath{m\in\mc{M}\left(I\right),}}\\
0 & \mbox{otherwise}.
\end{array} \right. \label{eq:help0}
\end{equation}
In words, $\hat{\mu}_m(I)$ indicates whether control $I$ involves the queue corresponding to node $m$ 
for creation of the transmitted packet according to BCR.

Hence it holds,
\begin{equation}
\sum_{e\in\mc{E}_{out}^{m}}f_{e}=\sum_{e\in\mc{E}_{out}^{m}}\sum_{I\in\mc{I}}\phi_{I}f_{e}\left(I\right)=
  \sum_{I\in\mc{I}}\phi_{I}\sum_{e\in\mc{E}_{out}^{m}}p_{e}^{m}(I)\mu_{m}\left(I\right),\label{eq:help1.0}
\end{equation}
and
\begin{equation}
\sum_{e\in\mc{E}_{in}^{m}}f_{e}=\sum_{e\in\mc{E}_{in}^{m}}\sum_{I\in\mc{I}}\phi_{I}f_{e}\left(I\right)=
  \sum_{I\in\mc{I}}\sum_{e=(l,m)\in\mc{E}_{in}^{m}}\phi_{I}\mu_{l}\left(I\right)p_{e}^{l}(I).\label{eq:help2.0}
\end{equation}
Since the tokens for new packet arrivals are always placed in queues
$V_{i}(i)$, $i\in\mc{N}$, we define
\begin{equation} \label{eq:InRates8}
\bar{\lambda}_{m}=\begin{cases}
1 & \mbox{if }m=V_{i}(i),\\
0 & \mbox{otherwise}.
\end{cases}
\end{equation}

Replacing (\ref{eq:help1.0}), (\ref{eq:help2.0}) in (\ref{eq:mainStabCond}), we have 
\begin{equation}
\sum_{I\in\mc{I}}\phi_{I}\left(\sum_{e=(l,m)\in\mc{E}_{in}^{m}}\mu_{l}\left(I\right)p_{e}^{l}(I)\right)+
  \bar{\lambda}_{m}\leq\sum_{I\in\mc{I}}\phi_{I}\left(\sum_{e\in\mc{E}_{out}^{m}}p_{e}^{m}(I)\mu_{m}\left(I\right)\right),
  \quad m\in\mc{M},\label{eq:param0}
\end{equation}
or equivalently, taking into account (\ref{eq:ProbSum}), 
\begin{align} \label{eq:param1}
\sum_{I\in\mc{I}}\phi_{I}\left(\sum_{\substack{e=(l,m)\in\mc{E}_{in}^{m}\\l\neq m}} \mu_{l}\left(I\right)p_{e}^{l}(I)\right)
  +\bar{\lambda}_{m} & \leq\sum_{I\in\mc{I}}\left(1-p_{(m,m)}^{m}(I)\right)\mu_{m}\left(I\right)\phi_{I},\quad m\in\mc{M},
\end{align}
Hence, the stability region $\mc{R}_{\Pi}$ of the system is described by either one of (\ref{eq:param0}), (\ref{eq:param1}),
combined with 
\begin{align}
0\leq \mu_{m}\left(I\right) & \leq\hat{\mu}_{m}(I), \label{eq:param2}\\
\phi_{I} & \geq 0, \label{eq:param3}\\
\sum_{I\in\mc{I}}\phi_{I} & \leq1, \label{eq:param6}
\end{align}
where $\hat{\mu}_{m}\left(I\right)$ is given by (\ref{eq:help0}). 

Two implementation issues are worth mentioning at this point. First,
there must exist a mechanism for the receivers to know the constituents
of the XOR combination of each received packet, in order to be able
to use this packet in the decoding process. The simplest way to implement
this is to use packet addresses to identify the native packets involved in
the XOR combination of the transmitted packet. These addresses can
be placed in the packet header. Reserving bits to describe packet
addresses implies some loss of throughput due to the introduced overhead.
To simplify the description, in the current and next Section we do
not take the overhead into account and address the issue of stability
in packets per slot. In Section \ref{sec:Implementation-Issues},
we discuss the number of addressed needed and loss of throughput due
to overhead.

The second issue is that, under the schemes described in Section \ref{sec:Network-coding-algorithms},
the receivers need to save received packets so that they can correctly
decode at a later time. The stability results
above consider only the queues at the base station.
Hence, if we are interested in taking the receiver
queues into consideration as well, we must ensure that the system
remains stable even if the sizes of these queues are added to the
total queue size at the base station. In fact, if the receivers are
never informed by the base station as to which of their received packets
will not be needed in the future, it is easy to devise scenarios where
the queue sizes at the receivers grow to infinity even though the
queues at the base station are stable. A simple way to deal with this
problem is described in Section \ref{sec:Implementation-Issues}.

\subsection{Stabilizing Policy}  \label{sub:Stabilizing-Policy}

Applying directly the results in \cite{pasxos2}, we obtain the
stabilizing policy described below. At the beginning of each time
slot, the policy chooses a control of the form
$I=I_{\mc{D}_{1},\ldots,\mc{D}_{\nu}}^{\mc{L}_{1},\ldots,\mc{L}_{\nu}}
\in\mc{I}$, where all counters $K^{\mc{L}_r}_{\mc{D}_r}(k)$, for
$r=1,\ldots,\nu$ and $k\in \mc{D}_r$, are non-zero\footnote{recall
  that $K^{\mc{L}_r}_{\mc{D}_r}(k)$ is defined as the number of tokens
  in virtual queue $V^{\mc{L}_r}_{\mc{D}_r}(k)$ and, by
  Lemma~\ref{lem:allsame}, can be deduced by information available in
  the real network. Hence, $K^{\mc{L}_r}_{\mc{D}_r}(k)>0$ is
  equivalent to saying that $V^{\mc{L}_r}_{\mc{D}_r}(k)$ is
  non-empty.} and forms the appropriate packet to be transmitted in
that slot, $p=\oplus_{r=1}^{\nu}p_{\mc{D}_{r}}^{\mc{L}_{r}}$,
according to the Basic Coding Rule. If control $I$ is chosen, one
token from each of the queues in the set
$\mc{M}\left(I\right)=\bigcup_{r=1}^\nu \bigcup_{k\in \mc{D}_r}
\left\{ V^{\mathcal{L}_r}_{\mathcal{D}_r}(k) \right\}$ may be moved to
another virtual queue inside the network, or may reach the destination
(thus, the native packet corresponding to the token exits the
network). No packets from any of the other queues are moved. The
algorithm for choosing the appropriate control is the following.

\textbf{Algorithm 1\label{Algorithm-1-At}}: At each decision slot: 
\begin{enumerate}
\item For each control $I=I_{\mc{D}_{1},\ldots,\mc{D}_{\nu}}^{\mc{L}_{1},\ldots,\mc{L}_{\nu}}\in\mc{I}$
that satisfies the BCR:

\begin{itemize}
\item Form the weights 
\[
c_{m}\left(I\right)=\max\left\{ K_{m}-\underset{{\scriptstyle e=(m,l)\in\mc{E}_{out}^{m}}}{\sum} 
  p_{e}^{m}\left(I\right)K_{l},\:0\right\} ,\; m\in\mc{M}\left(I\right),
\]
where $K_{m}$ is the length of the queue corresponding to node $m$
(corresponding to a queue in the virtual network, i.e.~if
$m=V^{\mc{L}}_{\mc{D}}(i)$ for some $\mc{L},\mc{D}$ and $i\in \mc{D}$,
then $K_m=K^{\mc{L}}_{\mc{D}}(i)$).

\item Form the reward under the given control, 
\[
C\left(I\right)=\underset{m\in\mc{M}\left(I\right)}{\sum}c_{m}\left(I\right).
\]
\end{itemize}

\item Find the control that maximizes the reward,
i.e.~$I^\ast=\arg\max{}_{I\in\mc{I}}C\left(I\right)$, transmit the
packet $p=\bigoplus_{k=1}^{\nu^\ast}
p^{\mathcal{L}^\ast_k}_{\mathcal{D}^\ast_k}$ that corresponds to
control
$I^\ast=I^{\mathcal{L}^\ast_1,\ldots,\mathcal{L}^\ast_{\nu^\ast}}_{\mathcal{D}^\ast_1,
  \ldots,\mathcal{D}^\ast_{\nu^\ast}}$ and apply the Rules for Packet
Movement after reception of feedback (including updating the $K$
counters).
\end{enumerate}

\begin{example} \begin{upshape}
Consider a network of $N=2$ users. The virtual queue network can be seen
in Figure \ref{fig_2_users}, where $d_{1}$ and $d_{2}$ are the two 
destination nodes. The set of all controls that obey the BCR is 
$\mc{I}=\left\{ I_{1},I_{2},I_{1}^{2},I_{2}^{1},I_{1,2}^{2,1}\right\} $.
Suppose all queues are non empty. At each decision slot: 

\begin{figure}[t]
\centering
\psfrag{Q1}[][][0.8]{$V_1(1)$}
\psfrag{Q2}[][][0.8]{$V_2(2)$}
\psfrag{Q2_1}[][][0.8]{$V^2_1(1)$}
\psfrag{Q1_2}[][][0.8]{$V^1_2(2)$}
\psfrag{d1}[][][0.8]{$d_1$}
\psfrag{d2}[][][0.8]{$d_2$}
\includegraphics[scale=0.6]{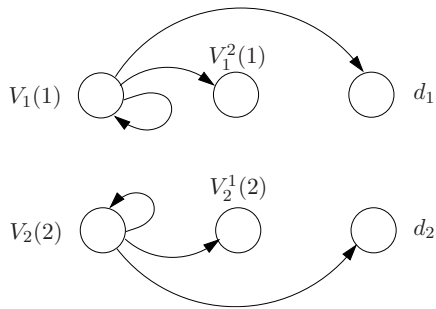}
\caption{Virtual queues in the case of $N=2$ users and possible movements of tokens.}
\label{fig_2_users} 
\end{figure}

\begin{enumerate}
\item For each control $I\in\mc{I}$: 

\begin{itemize}
\item The set $\mc{M}\left(I\right)$ is formed. Table \ref{table_set_of_queues} shows the set 
$\mc{M}(I)$ for each control. 

\item The next step is forming the weights $c_{m}\left(I\right)$ for every
$I$. For every node $m\in\mc{M}\left(I\right)$, all possible
outgoing edges $e=(m,l)$ in set $\mc{E}_{out}^{m}$, when applying
control $I$, or equivalently, all receiving nodes
$l$, must be determined. Table \ref{table_receiving_nodes}
shows all receiving nodes for each node $m$,
as well as the respective transition probabilities.

\item Next, for each node $m\in\mc{M}(I)$ and each control $I$ the
weight $c_{m}\left(I\right)$ is calculated, as can be seen in Table
\ref{table_weight}. 

\item Then, for each control $I$ the reward $C\left(I\right)$ is determined
(Table \ref{table_reward}). 
\end{itemize}
\item Finally, select the control that maximizes the reward 
\[
I^{*}=\arg\max{}_{I\in\mc{I}}C\left(I\right)=\arg\max\left\{ C\left(I_{1}\right),C\left(I_{2}\right),C\left(I_{1}^{2}\right),
  C\left(I_{2}^{1}\right),C\left(I_{1,2}^{2,1}\right)\right\} .
\]

\end{enumerate}
\end{upshape}
\end{example}

The previous example is simple enough that the stability region of the proposed algorithm can be
analytically determined as follows. For arrival rates $\lambda_1,\lambda_2$, we use the transition
probabilities in Table~\ref{table_receiving_nodes} and apply \eq{eq:param0}, \eq{eq:param2}--\eq{eq:param6}
to get the following set of inequalities (recall the notational shortcut at the end of 
Section~\ref{sec:System-model-and})
\begin{align}
V_1(1): \lambda_1 & \leq (1-\epsilon_{12}) \phi_1, \\
V_2(2): \lambda_2 & \leq (1-\epsilon_{12}) \phi_2, \\
V^2_1(1): (\epsilon_1-\epsilon_{12})\phi_1 & \leq (1-\epsilon_1) \left( \phi^2_1+\phi^{2,1}_{1,2} \right), \\
V^1_2(2): (\epsilon_2-\epsilon_{12})\phi_2 & \leq (1-\epsilon_2) \left( \phi^1_2+\phi^{2,1}_{1,2} \right),
\end{align}
with the additional constraint that $\phi_1,\phi_2,\phi^2_1,\phi^1_2,\phi^{2,1}_{1,2}$ are non-negative and
their sum is less than 1. Applying the Fourier-Motzkin algorithm to eliminate (i.e.~deparameterize) 
$\phi^{2,1}_{1,2},\phi^1_2,\phi^2_1,\phi_2,\phi_1$ in this order results, after some simple algebra 
(see Appendix~\ref{sec:Proof-FM}), in the set of inequalities $\left\{ \frac{\lambda_1}{1-\epsilon_1}+
\frac{\lambda_2}{1-\epsilon_{12}}\leq 1, \; \frac{\lambda_2}{1-\epsilon_2}+\frac{\lambda_1}{1-\epsilon_{12}} 
\leq 1 \right\}$, which matches the stability outer bound in \cite{Leo_2user} (this will be generalized to 
arbitrary $N$ in the next Section). This shows that the optimal policy derived in \cite{Leo_2user} for 
arbitrary erasures is a special case of the policy proposed in this paper.

\begin{table}[t]
\centering
\caption{Set of queues $\mc{M}\left(I\right)$ for each control $I$}
\renewcommand{\arraystretch}{1.25}  
\begin{tabular}{|c|c|c|c|c|c|} \hline
$I$ & $I_1$ & $I_2$ & $I^2_1$ & $I^1_2$ & $I^{2,1}_{1,2}$ \\ \hline
$\mc{M}\left(I\right)$ & $\left\{ V_1(1)\right\}$ & $\left\{ V_2\right(2)\}$ 
  & $\left\{ V^2_1(1)\right\}$ & $\left\{ V^1_2(2)\right\}$ 
  & $\left\{ V^2_1(1),V^1_2(2)\right\}$ \\ \hline
\end{tabular}
\label{table_set_of_queues}
\end{table}

\begin{table}[t]
\centering
\caption{Receiving nodes for each node $m$ and transition probabilities}
\renewcommand{\arraystretch}{1.25}  
\begin{tabular}{|c|c|c|c|c|c|c|c|} \hline
control & node $m$ & node $l$ & $p^m_{(m,l)}(I)$ & control & node $m$ & node $l$ & 
  $p^m_{(m,l)}(I)$ \\ \hline
\multirow{3}{*}{$I_1$} & \multirow{3}{*}{$V_1(1)$} & $V_1(1)$ & $P_{\{1,2\},\emptyset}$ & 
  \multirow{3}{*}{$I_2$} & \multirow{3}{*}{$V_2(2)$} & $V_2(2)$ & 
  $P_{\{1,2\},\emptyset}$ \\ \cline{3-4} \cline{7-8}
  & & $V^2_1(1)$ & $P_{\{1\},\{2\}}$ & & & $V^1_2(2)$ & $P_{\{2\},\{1\}}$ \\ \cline{3-4} \cline{7-8}
  & & $d_1$ & $P_{\emptyset,\{1\}}$ & & & $d_2$ & $P_{\emptyset,\{2\}}$ \\ \hline
\multirow{2}{*}{$I^2_1$} & \multirow{2}{*}{$V^2_1(1)$} & $V^2_1(1)$ & $P_{\{1\},\emptyset}$ & 
  \multirow{2}{*}{$I^1_2$} & \multirow{2}{*}{$V^1_2(2)$} & $V^1_2(2)$ & $P_{\{2\},\emptyset}$ \\
  \cline{3-4} \cline{7-8}
  & & $d_1$ & $P_{\emptyset,\{1\}}$ & & & $d_2$ & $P_{\emptyset,\{2\}}$ \\ \hline  
\multirow{2}{*}{$I^{2,1}_{1,2}$} & \multirow{2}{*}{$V^2_1(1)$} & $V^2_1(1)$ & $P_{\{1\},\emptyset}$ & 
  \multirow{2}{*}{$I^{2,1}_{1,2}$} & \multirow{2}{*}{$V^1_2(2)$} & $V^1_2(2)$ & 
  $P_{\{2\},\emptyset}$ \\ \cline{3-4} \cline{7-8}
  & & $d_1$ & $P_{\emptyset,\{1\}}$ & & & $d_2$ & $P_{\emptyset,\{2\}}$ \\ \hline
\end{tabular}
\label{table_receiving_nodes}
\end{table}

\begin{table}[h]
\centering
\caption{Weight $c_m(I)$ for each node $m$ and each control $I$}
\renewcommand{\arraystretch}{1.25}  
\begin{tabular}{|c|} \hline
$c_m\left(I\right)$ \\ \hline
$c_{V_1(1)}\left(I_1\right)=\max\left\{ K_{V_1(1)}-P_{\left\{ 12\right\},\emptyset} K_{V_1(1)}
  -P_{\left\{ 1\right\} ,\left\{ 2\right\}} K_{V^2_1(1)}-P_{\emptyset,\left\{ 1\right\} }
  K_{d_{1}},\:0\right\} $ \\ \hline
$c_{V_2(2)}\left(I_{2}\right)=\max\left\{ K_{V_2(2)}-P_{\left\{ 12\right\},\emptyset} K_{V_2(2)}
 -P_{\left\{ 2\right\} ,\left\{ 1\right\} } K_{V^1_2(2)}-P_{\emptyset,\left\{ 2\right\} }
  K_{d_{2}},\:0\right\} $ \\ \hline
$c_{V^2_1(1)}\left(I_{1}^{2}\right)=\max\left\{ K_{V^2_1(1)}-P_{\left\{ 1\right\} ,\emptyset}
  K_{V^2_1(1)}-P_{\emptyset,\left\{ 1\right\} } K_{d_{1}},\:0\right\} $ \\ \hline
$c_{V^1_2(2)}\left(I_{2}^{1}\right)=\max\left\{ K_{V^1_2(2)}-P_{\left\{ 2\right\} ,\emptyset}
  K_{V^1_2(2)}-P_{\emptyset,\left\{ 2\right\} } K_{d_{2}},\:0\right\} $ \\ \hline
$c_{V^2_1(1)}\left(I_{1,2}^{2,1}\right)=\max\left\{ K_{V^2_1(1)}-P_{\left\{ 1\right\} ,\emptyset} 
  K_{V^2_1(1)}-P_{\emptyset,\left\{ 1\right\} } K_{d_{1}},\:0\right\} $ \\ \hline
$c_{V^1_2(2)}\left(I_{1,2}^{2,1}\right)=\max\left\{ K_{V^1_2(2)}-P_{\left\{ 2\right\} ,\emptyset}
  K_{V^1_2(2)}-P_{\emptyset,\left\{ 2\right\} } K_{d_{2}},\:0\right\} $ \\ \hline
\end{tabular}
\label{table_weight}
\end{table}

\begin{table}[h]
\centering
\caption{Reward $C\left(I\right)$ for each control $I$}
\renewcommand{\arraystretch}{1.25}  
\begin{tabular}{|c|c|} \hline
$I$ & $C\left(I\right)$ \\ \hline
$I_1$ & $C\left(I_1\right)= \sum_{m\in \{V_1(1)\}} c_m(I)= c_{V_1(1)}(I_1)$ \\ \hline
$I_2$ & $C\left(I_2\right)= \sum_{m\in \{V_2(2)\}} c_m(I)= c_{V_2(2)}(I_2)$ \\ \hline
$I^2_1$ & $C\left(I^2_1\right)= \sum_{m\in \left\{ V^2_1(1)\right\}} c_m(I)=
  c_{V^2_1(1)}\left(I^2_1\right)$ \\ \hline
$I^1_2$ & $C\left(I^1_2\right)= \sum_{m\in\left\{ V^1_2(2)\right\} } c_m(I)=
  c_{V^1_2(2)}\left(I^1_2\right)$ \\ \hline
$I^{2,1}_{1,2}$ & $C\left(I_{1,2}^{2,1}\right)= \sum_{m\in\left\{ V^2_1(1),V^1_2(2)\right\} } c_m(I)=
  c_{V^2_1(1)} \left(I_{1,2}^{2,1}\right) +c_{V^1_2(2)}\left(I_{2,1}^{1,2}\right)$ \\ \hline
\end{tabular}
\label{table_reward}
\end{table}

\subsection{Comparison between Algorithm 1 for $N=3$ and the algorithm in \cite{AGG}}

It should be stated that, although the application of the RPM to the case $N=3$
yields the \textit{exact} same rules as in \cite{AGG}, the performance of Algorithm 1 is not identical to 
the algorithm \texttt{XOR2} in \cite{AGG}. In fact, although \texttt{XOR2} in \cite{AGG} (which assumed a fixed
\textit{a priori} number of packets and no new arrivals) can be suitably modified
so that it is applicable to the case of stochastic arrivals, the resulting policy will be no better than
Algorithm 1 in this paper, since the latter yields, by construction, a stabilizing policy over the class of
policies that apply BCR and RPM (and this includes the policy in \cite{AGG}).

A more intuitive reason for the performance difference is that \texttt{XOR2} in \cite{AGG} and the current work 
apply different procedures for selecting the XOR combination to be transmitted. Namely,
\cite{AGG} selects packets for transmission by combining queues in different levels in an order that is defined 
\textit{a priori}, while Algorithm 1 imposes no such fixed order and determines the packet for transmission by
maximizing a suitable backlog-weighted sum. Hence, Algorithm 1 is not burdened by any \textit{a priori} choices, 
which may actually be suboptimal.

\section{Outer Bound on the Stability Region}  \label{sec:UpperBound}

In this Section, we derive an outer bound on the stability region
of the system under study by deparameterizing (i.e.~eliminating the
flow variables $\vec{f}$ in) Theorem \ref{thm:StabCond}.
This bound is identical with the bound on the information-theoretic
capacity region of the BEC with feedback presented in \cite{GGT11,Wang_Allerton}.
Although it was shown in \cite{GPTL} that the capacity region of
the system under consideration is the same as the stability region
of the system, we cannot directly invoke this result to derive the
stability region outer bound via the capacity outer bound in \cite{GGT11,Wang_Allerton}.
The reason is that the latter capacity bound does not take into account
the case of slots without any packet transmission, i.e.~idle slots,
so that, in principle, coding algorithms may take advantage of idle
slots to increase capacity beyond the outer bound in \cite{GGT11,Wang_Allerton}.
To distinguish between the two channels, we call the BEC studied in
\cite{GGT11,Wang_Allerton} the ``standard'' BEC, and refer to
the channel under study in this paper (i.e.~the one containing idle
slots) as the ``extended'' BEC.

As will be seen, the capacity of the standard BEC, measured in information
bits per transmitted symbol, differs from the capacity of the extended
BEC by at most 1 bit; in fact, this difference decreases exponentially
w.r.t. the packet length $L$. Specifically, the following Theorem
is proved in the Appendix (we denote with $\epsilon_{\mc{S}}$
the probability that a transmitted packet is erased by all users in
set $\mc{S}$).
\begin{theorem} \label{thm:outer_bound_ext}
A capacity outer bound $\mc{C}_{out}$,
measured in packets per transmitted symbol, for the $N$-user ``extended''
BEC with feedback is given by (assuming that $\epsilon_{i}<1$ for
all $i\in\mc{N}$) 
\begin{equation}
\mc{C}_{out}=\left\{ \vec{R}:\max_{\sigma\in\mc{P}}\;\left(\sum_{k\in\mc{N}}
  \frac{R_{\sigma(k)}}{1-\epsilon_{\{\sigma(1),\ldots,\sigma(k)\}}}
  -2^{-L/A_{\sigma}}A_{\sigma}/L\right)\leq 1 \right\} ,
\end{equation}
where $\mc{P}$ is the set of permutations $\sigma$ on $\mc{N}$
and $A_{\sigma}=\sum_{k\in\mc{N}}\,\frac{1}{1-\epsilon_{\{\sigma(1),\ldots,\sigma(k)\}}}$. 
\end{theorem}

\begin{corollary}
Using the same notation as in Theorem~\ref{thm:outer_bound_ext} and measuring
rates in units of bits per transmitted symbol, a capacity outer bound $\mc{C}_{out}$ for the 
$N$-user ``extended'' BEC with feedback is given by (assuming that $\epsilon_{i}<1$ for
all $i\in\mc{N}$) 
\begin{equation}
\mc{C}_{out}=\left\{ \vec{R}:\max_{\sigma\in\mc{P}}\;\left(\sum_{k\in\mc{N}}
  \frac{R_{\sigma(k)}}{1-\epsilon_{\{\sigma(1),\ldots,\sigma(k)\}}}
  -2^{-L/A_{\sigma}}A_{\sigma} \right) \leq L \right\} .
\end{equation} 
\end{corollary}

The next Theorem, which is proved in Appendix \ref{sec:Proof-Of-Theorem},
describes the main result of this Section. 
\begin{theorem}  \label{thm:UpperBound}
The following relation holds 
\begin{equation} \label{eq:capacityReg}
\mc{R}_{\Pi}\subseteq\left\{ \boldsymbol{\lambda}:\:\max_{\sigma\in\mc{P}}\sum_{i\in\mc{N}}
  \frac{\lambda_{\sigma\left(i\right)}}{1-\epsilon_{\tilde{\mc{S}}(i)}}\leq1\right\} \triangleq \mc{C}_{u},
\end{equation}
where $\mc{P}$ is the set of permutations on $\mc{N}$
and $\tilde{\mc{S}}(i)=\left\{ \sigma\left(1\right),\ldots,\sigma(i)\right\}$.
\end{theorem}

Since $\mc{C}_{u}$ is identical to an outer bound on the capacity
region of the ``standard'' BEC (and the ``extended'' BEC capacity
region differs from this by at most 1 bit), it follows that any class
$\Pi$ of policies that achieves $\mc{C}_{u}$ (i.e.~$\mc{R}_{\Pi}=\mc{C}_{u})$
is essentially optimal. A special case where this occurs is examined
in the next Section.

\section{The Case of i.i.d. Channels: Stability Region for $4$ Users}  \label{sec:4users}

In this Section, we assume that the erasure events for all receivers
are i.i.d, and denote by $\epsilon$ the probability of such an event.
We also repeat the definition $P_{\mc{G},\mc{S}}=\epsilon^{\left|\mc{G}\right|}(1-\epsilon)^{\left|\mc{S}\right|}$.
We consider the case of a channel with 4 receivers and show that,
for all $0 \leq \epsilon<1$, if $\boldsymbol{\lambda}\in\mc{C}_{u}$,
then $\boldsymbol{\lambda}\in\mc{R}_{\Pi}$, i.e.~$\mc{R}_{\Pi}\supseteq\mc{C}_{u}$.
Hence, in this case we have $\mc{R}_{\Pi}=\mc{C}_{u}$ and
the stability region using only XOR operations coincides (barring
addressing overhead) with the capacity region of the standard broadcast
channel. Also, it is within one bit, and asymptotically
(as the packet length increases) equal to the stability region of
the extended BEC under general coding schemes.

To proceed, we restrict the set of available controls by allowing
only intra-level coding, i.e.~we only consider controls of the form
$I_{\mc{D}_{1},\ldots,\mc{D}_{\nu}}^{\mc{L}_{1},\ldots,\mc{L}_{\nu}}$
where $\left|\mc{L}_{r}\cup\mc{D}_{r}\right|=\left|\mc{L}_{s}\cup\mc{D}_{s}\right|$
for all $r,s\in\left\{ 1,\ldots,\nu\right\} $. This restriction simplifies
the calculations and shows that even a restricted set of controls suffices
to achieve the maximal stability region when channel erasure events
are i.i.d. We note however, that if channel statistics are non-i.i.d.,
the additional controls are helpful in increasing the stability region
of the policy. The set of permitted controls is described in Table
\ref{table1}, where $i,j,k,l\in\left\{ 1,2,3,4\right\} $ are distinct.

\begin{table}
\caption{Permitted controls for levels 1 to 4. }
\centering 
\renewcommand{\arraystretch}{1.5}
\begin{tabular}{|c|c|c|c|c|}  \hline 
  & Level 1 & Level 2 & Level 3 & Level 4 \\ \hline 
\multirow{10}{*}{Permitted controls} & Control & Control & Control & Control \\ \cline{2-5} 
  & $I_{i}$  & $I_{i,j}^{j,i}$  & $I_{i,jk}^{jk,i}$  & $I_{i,jkl}^{jkl,i}$ \\ \cline{2-5} 
  &  & $I_{i}^{j}$ & $I_{jk}^{i}$  & $I_{jkl}^{i}$ \\ \cline{2-5} 
  &  &  & $I_{i,j,k}^{jk,ik,ij}$  & $I_{ij,kl}^{kl,ij}$ \\ \cline{2-5} 
  &  &  & $I_{i,j}^{jk,ik}$  & $I_{ij,k,l}^{kl,ijl,ijk}$ \\ \cline{2-5} 
  &  &  & $I_{i}^{jk}$  & $I_{ij}^{kl}$ \\ \cline{2-5} 
  &  &  &  & $I_{i,j,k,l}^{jkl,ikl,ijl,ijk}$ \\ \cline{2-5} 
  &  &  &  & $I_{i,j,k}^{jkl,ikl,ijl}$ \\ \cline{2-5} 
  &  &  &  & $I_{i,j}^{jkl,ikl}$ \\ \cline{2-5} 
  &  &  &  & $I_{i}^{jkl}$ \\ \hline 
\end{tabular}
\label{table1} 
\end{table}

For the rest of this Section, we assume without loss of generality that
\begin{equation} \label{eq:order}
\lambda_{1} \geq \lambda_{2} \geq \lambda_{3} \geq\lambda_{4},
\end{equation}
which implies that 
\begin{equation} \label{eq:maxsl}
\max_{\sigma\in \mc{P}} \, \sum_{i=1}^4 \frac{\lambda_{\sigma(i)}}{1-\epsilon^i}= 
  \sum_{i=1}^4 \frac{\lambda_i}{1-\epsilon^i} .
\end{equation}
We will show that $\boldsymbol{\lambda}\in\mc{C}_{u}$ implies
$\boldsymbol{\lambda}\in\mc{R}_{\Pi}$, which, by combining \eq{eq:maxsl}, \eq{eq:capacityReg}, 
is equivalent to solving the following problem for any $0\leq\epsilon<1$.

\textbf{Problem}: If $\lambda_{1}\geq\lambda_{2}\geq\lambda_{3}\geq\lambda_{4}$
and $\sum_{i=1}^{4}\frac{\lambda_{i}}{1-\epsilon^{i}}\leq1,$ find
parameters $\phi_{I}$ satisfying (\ref{eq:param1})-(\ref{eq:param6}),
where $\mc{M}$ is the set of all queues $Q_{\mc{D}}^{\mc{L}}\left(i\right),\: i\in\mc{D},$
and $\mc{L},\mc{D}$ satisfy CC.

In the following, we will describe the procedure according to which
$\mu_{m}\left(I\right),\phi_{I}$, $m\in\mc{M},\: I\in\mc{I},$
are calculated. First, we set 
\begin{equation} \label{eq:setmu}
\mu_{m}\left(I\right)=\hat{\mu}_{m}\left(I\right),\: m\in\mc{M},\: I\in\mc{I},
\end{equation}
ensuring that (\ref{eq:param2}) is satisfied. It remains to determine
$\phi_{I},\: I\in\mc{I}$. Notice that, for any given value of
$\epsilon$, (\ref{eq:setmu}) transforms (\ref{eq:param1}), (\ref{eq:param3}),
(\ref{eq:param6}) into a linear program (LP) w.r.t $\phi_{I}$, so
that achievability of the rate $\boldsymbol{\lambda}$ is reduced
to LP feasibility (a similar LP-based approach is used to describe
an achievable scheme for a 2 user MIMO setting over broadcast erasure
channels in \cite{WL12}). However, since $\epsilon$ takes a continuum
of values, we cannot solve the resulting LP for each $\epsilon$ but
need to determine $\phi_{I}$ analytically.

To simplify the notation somewhat, for control $I=I_{\mc{D}_{1},\ldots,\mc{D}_{\nu}}^{\mc{L}_{1},\ldots,\mc{L}_{\nu}}$
we denote 
\[
\phi_{I}=\phi_{\mc{D}_{1},\ldots,\mc{D}_{\nu}}^{\mc{L}_{1},\ldots,\mc{L}_{\nu}}.
\]
An overview of the approach follows. We start from inequalities (\ref{eq:param1})
referring to queues at level 1, i.e.~$V_{i}\left(i\right)$, and determine
all $\phi_{i},$ ensuring that these inequalities are satisfied. In
general, having determined $\phi_{I}$ for all controls $I$ that
involve queues up to level $l,$ we consider
the inequalities (\ref{eq:param1}) referring to queues at level $l+1$
and determine $\phi_{I}$ for all controls that involve queues at
level $l+1,$ ensuring that these inequalities are satisfied. During
this process, it is ensured that (\ref{eq:param3}) is satisfied.
After all $\phi_{I}$ are computed, it is checked that (\ref{eq:param6})
is also satisfied.

We now proceed with the detailed description of the manner in which
$\phi_{I},\; I\in\mc{I}$, are determined. We will use the following
terminology in the description. If, under an allowable control $I$,
it is possible to have a token movement from virtual
queue $m$ to virtual queue $l$, 
we say that there is a ``flow from virtual queue $m$ to virtual queue $l$''
under control $I$ and we name $p_{(m,l)}\left(I\right)$, the
``probability of flow'' from $m$ to $l$ under control $I$.
We also say that there is ``flow from virtual queue $m$ to virtual queue $l$''
if it is possible to have a token movement from queue $m$ to queue
$l$ under some of the allowable controls. 

\textbf{Level 1}: At this level, there are 4 queues (equivalently,
nodes in $\mc{M}$) of the form $V_{i}\left(i\right),\; i\in\left\{ 1,\ldots,4\right\} $.
There are no incoming flows from other nodes to $V_{i}\left(i\right)$,
but there are new native packet arrivals (equivalently, token arrivals) of rate $\lambda_{i}$ at every
$V_{i}\left(i\right)$. The only control that may result in packets
leaving $V_{i}\left(i\right)$ is $I_{i}$, so inequality (\ref{eq:param1})
becomes 
\begin{equation}
\lambda_{i}\leq(1-\epsilon^{4})\cdot\phi_{i} .
\end{equation}
To satisfy this inequality, we set, for all $i\in \{1,\ldots,4\}$, 
\begin{equation}  \label{eq:level1}
\boxed{ \phi_{i}=\lambda_{i}/(1-\epsilon^{4}) } .
\end{equation}

\textbf{Level 2}: At level 2, there are 12 queues of the form $V_{i}^{j}\left(i\right),\: i,j\in\left\{ 1,\ldots,4\right\},
\: i\neq j$. The only incoming flow to each of these nodes is under control $I{}_{i}$,
with probability $\epsilon^{3}(1-\epsilon)$, while there are two
outgoing flows, under controls $I_{i,j}^{j,i}$ and $I_{i}^{j}$,
that result in packets leaving with probability $1-\epsilon^{3}$.
Hence, inequality (\ref{eq:param1}) becomes 
\begin{equation} \label{eq:level2-1}
\epsilon^{3}(1-\epsilon)\cdot\phi_{i}\leq(1-\epsilon^{3})\cdot\phi_{i,j}^{j,i}+(1-\epsilon^{3}) \cdot
   \phi_{i}^{j}.
\end{equation}
Similarly, for node $V_{j}^{i}\left(j\right)$ we have 
\begin{equation} \label{eq:level2-2}
\epsilon^{3}(1-\epsilon)\cdot\phi_{j}\leq(1-\epsilon^{3})\cdot\phi_{i,j}^{j,i}+(1-\epsilon^{3})
   \cdot\phi_{j}^{i}.
\end{equation}
Since $\phi_i,\,\phi_j$ have already been determined by (\ref{eq:level1}), the LHS of
(\ref{eq:level2-1}), (\ref{eq:level2-2}) are known. We select $\phi^j_i=\phi^i_j=0$, for all $i\neq j$, 
so that 
\begin{equation} \label{eq:first_pair_lvl2}
\phi^{j,i}_{i,j} \geq \frac{\epsilon^3 (1-\epsilon)}{1-\epsilon^3} \, \max( \phi_i,\phi_j) ,
\end{equation}
and we choose $\phi^{j,i}_{i,j}$ to satisfy (\ref{eq:first_pair_lvl2}) with equality.
Assuming w.l.o.g.~$i<j$ (so that $\lambda_i \geq \lambda_j$), it follows from (\ref{eq:level1}) that 
$\phi_i \geq \phi_j$, which implies 
\begin{equation}
\boxed{ \phi^{j,i}_{i,j}= \frac{\epsilon^3 (1-\epsilon)}{1-\epsilon^3} \, \phi_i } ,
\end{equation}
or
\begin{equation} \label{eq:level2-3}
\phi_{i,j}^{j,i}=\epsilon^{3}(1-\epsilon)\cdot \lambda_i/(1-\epsilon^{3})\cdot(1-\epsilon^{4}),\quad i< j.
\end{equation}

\textbf{Level 3}: At this level, there are 12 real queues of type
$Q_{ij}^{k}$ (corresponding to virtual queues $V_{ij}^{k}\left(i\right)$
and $V_{ij}^{k}\left(j\right)$) and 12 real queues of type $Q_{i}^{jk}$
(corresponding to virtual queues $V_{i}^{jk}(i)$), where $i,j,k\in\left\{ 1,\ldots,4\right\} $
with $i\neq j\neq k$.

\begin{itemize}
\item Incoming flow to $Q_{ij}^{k}$ (respectively, to both $V_{ij}^{k}\left(i\right)$
and $V_{ij}^{k}\left(j\right)$) occurs under control $I_{i,j}^{j,i}$
with probability $\epsilon^{3}(1-\epsilon)$. Outgoing flows from
nodes of this form occur under controls $I_{ij,k}^{k,ij}$ and $I_{ij}^{k}$,
with probability $1-\epsilon^{3}$. While for each of the queues $V_{ij}^{k}\left(i\right)$
and $V_{ij}^{k}\left(j\right)$ there is one inequality of the form
(\ref{eq:param1}), it turns out that these inequalities are identical.
Hence, for both queues $V_{ij}^{k}\left(i\right)$ and $V_{ij}^{k}\left(j\right)$
the following inequality holds 
\[
\epsilon^{3}(1-\epsilon)\cdot\phi_{i,j}^{j,i}\leq(1-\epsilon^{3})\cdot\phi_{ij,k}^{k,ij}+(1-\epsilon^{3})\cdot\phi_{ij}^{k}.
\]
We set $\phi_{ij}^{k}=0$, so that the previous inequality becomes
\begin{equation} \label{eq:level3-1}
\epsilon^{3}(1-\epsilon)\cdot\phi_{i,j}^{j,i}\leq(1-\epsilon^{3})\cdot\phi_{ij,k}^{k,ij}.
\end{equation}
Next, to satisfy (\ref{eq:level3-1}), we set 
\begin{equation} \label{eq:level3-1a}
\boxed{ \phi_{ij,k}^{k,ij}=\epsilon^{3}(1-\epsilon)\cdot\phi_{i,j}^{j,i}/(1-\epsilon^{3}) } ,
\end{equation}
where the second part of the inequality only depends on $\epsilon$
and $\lambda$, by substituting $\phi_{i,j}^{j,i}$ from (\ref{eq:level2-3}).
It follows that $\phi_{ij,k}^{k,ij}\geq 0$.
 
\item Possible incoming flows to $V_{i}^{jk}(i)$ are due to controls $I_{i},I_{i,j}^{j,i},I_{i,k}^{k,i},I_{i}^{j},I_{i}^{k},I_{ij,k}^{k,ij}$, 
$I_{ik,j}^{j,ik}$ and possible outgoing flows are due to controls $I_{jk,i}^{i,jk},I_{i,j,k}^{jk,ik,ij},I_{i,j}^{jk,ik},I_{i,k}^{jk,ij},I_{i}^{jk}$,
where $i,j,k\in\left\{ 1,\ldots,4\right\} $ with $i\neq j\neq k$.
For $V_{i}^{jk}(i)$, inequality (\ref{eq:param1}) becomes 
\begin{equation} \label{eq:level3-2}
\begin{array}{c}
\epsilon^{2}(1-\epsilon)^{2}\cdot\left(\phi_{i}+\phi_{i,j}^{j,i}+\phi_{i,k}^{k,i}\right)+\epsilon^{2}(1-\epsilon)\cdot
  \left(\phi_{i}^{j}+\phi_{i}^{k}+\phi^k_{ij}+\phi^j_{ik}+\phi_{ij,k}^{k,ij}+\phi_{ik,j}^{j,ik}\right)\\
  \leq (1-\epsilon^{2})\cdot\left(\phi_{i,j,k}^{jk,ik,ij}+\phi_{jk,i}^{i,jk}+\phi_{i,j}^{jk,ik}+\phi_{i,k}^{jk,ij}+\phi_{i}^{jk}\right)
\end{array}.
\end{equation}
For $V_{j}^{ik}(j)$ and $V_{k}^{ij}(k)$, inequality (\ref{eq:param1})
takes the form of (\ref{eq:level3-2}), with the appropriate exchange
of indices. Specifically, for $V_{j}^{ik}(j)$ and $V_{k}^{ij}(k)$,
we have the following inequalities, respectively 
\begin{equation} \label{eq:level3-3}
\begin{array}{c}
\epsilon^{2}(1-\epsilon)^{2}\cdot\left(\phi_{j}+\phi_{i,j}^{j,i}+\phi_{j,k}^{k,j}\right)+\epsilon^{2}(1-\epsilon)\cdot
  \left(\phi_{j}^{i}+\phi_{j}^{k}+\phi^k_{ij}+\phi^i_{jk}+\phi_{ij,k}^{k,ij}+\phi_{jk,i}^{i,jk}\right) \\
  \leq (1-\epsilon^{2})\cdot\left(\phi_{i,j,k}^{jk,ik,ij}+\phi_{ik,j}^{j,ik}+\phi_{i,j}^{jk,ik}+\phi_{j,k}^{ik,ij}+\phi_{j}^{ik}\right),
\end{array}
\end{equation}

\begin{equation} \label{eq:level3-4}
\begin{array}{c}
\epsilon^{2}(1-\epsilon)^{2}\cdot\left(\phi_{k}+\phi_{i,k}^{k,i}+\phi_{j,k}^{k,j}\right)+\epsilon^{2}(1-\epsilon)\cdot
  \left(\phi_{k}^{i}+\phi_{k}^{j}+\phi^j_{ik}+\phi^i_{jk}+\phi_{ik,j}^{j,ik}+\phi_{jk,i}^{i,jk}\right)\\
  \leq (1-\epsilon^{2})\cdot\left(\phi_{i,j,k}^{jk,ik,ij}+\phi_{ij,k}^{k,ij}+\phi_{i,k}^{jk,ij}+\phi_{j,k}^{ik,ij}+\phi_{k}^{ij}\right).
\end{array}
\end{equation}
All $\phi$ parameters in the LHS of inequalities (\ref{eq:level3-2}),
(\ref{eq:level3-3}) and (\ref{eq:level3-4}) have already been computed (or set to 0, by selection).
Therefore, the unknown parameters at this point are $\phi_{i,j,k}^{jk,ik,ij}$,
$\phi_{i,j}^{jk,ik}$, $\phi_{i,k}^{jk,ij}$, $\phi_{j,k}^{ik,ij}$,
$\phi_{i}^{jk}$, $\phi_{j}^{ik}$ and $\phi_{k}^{ij}$. We set all of these values to 0,
with the exception of $\phi^{jk,ik,ij}_{i,j,k}$, so that we can combine
(\ref{eq:level3-2})--(\ref{eq:level3-4}) to get the following equivalent expression 
(only the non-zero values are included)
\begin{equation} \begin{split} \label{eq:first_triple_lvl3}
\max \Bigg[ & \frac{\epsilon^2 (1-\epsilon)^2 \left( \phi_i+\phi^{j,i}_{i,j}+\phi^{k,i}_{i,k} \right)
  +\epsilon^2 (1-\epsilon) \left( \phi^{k,ij}_{ij,k}+\phi^{j,ik}_{ik,j} \right)}{1-\epsilon^2}
  -\phi^{i,jk}_{jk,i} , \\
& \frac{\epsilon^2 (1-\epsilon)^2 \left( \phi_j+\phi^{j,i}_{i,j}+\phi^{k,j}_{j,k} \right)
  +\epsilon^2 (1-\epsilon) \left( \phi^{k,ij}_{ij,k} + \phi^{i,jk}_{jk,i} \right)}{1-\epsilon^2}
   -\phi^{j,ik}_{ik,j} , \\
& \frac{\epsilon^2 (1-\epsilon)^2 \left( \phi_k+\phi^{k,i}_{i,k}+\phi^{k,j}_{j,k} \right)
  +\epsilon^2 (1-\epsilon) \left( \phi^{j,ik}_{ik,j}+\phi^{i,jk}_{jk,i} \right)}{1-\epsilon^2}
  -\phi^{k,ij}_{ij,k} \Bigg] \leq \phi^{jk,ik,ij}_{i,j,k} .
\end{split} \end{equation}

The formulas are getting very convoluted at this point but they are easily calculated as 
functions of the erasure probabilities and the arrival rates using symbolic computation packages. 
Using such a package (we used Maple 13.0), it is easy to see that, for $i<j<k$, the first
term in (\ref{eq:first_triple_lvl3}) is the maximum term and is also non-negative. Hence, we
select for all $i,j,k$, with $i<j<k$,
\begin{equation} \label{eq:level3-final}
\boxed{
\phi^{jk,ik,ij}_{i,j,k}= \frac{\epsilon^2 (1-\epsilon)^2 \left( \phi_i+\phi^{j,i}_{i,j}+\phi^{k,i}_{i,k} \right)
  +\epsilon^2 (1-\epsilon) \left( \phi^{k,ij}_{ij,k}+\phi^{j,ik}_{ik,j} \right)}{1-\epsilon^2}
  -\phi^{i,jk}_{jk,i} }  .
\end{equation}

\end{itemize}

\textbf{Level 4}: At level 4, there are 4 real queues of the form
$Q_{ijk}^{l}$ (which corresponds to virtual queues $V_{ijk}^{l}\left(i\right)$,
$V_{ijk}^{l}\left(j\right)$, $V_{ijk}^{l}\left(k\right)$), 6 real
queues of the form $Q_{ij}^{kl}$ (which corresponds to virtual queues
$V_{ij}^{kl}\left(i\right)$, $V_{ij}^{kl}\left(j\right)$) and 4 real queues
of the form $Q_{i}^{jkl}$ (corresponding to virtual queue $V_{i}^{jkl}\left(i\right)$).

\begin{itemize}
\item Incoming flows to the virtual queues corresponding to $Q_{ijk}^{l}$
are due to controls $I_{ij,k}^{k,ij}$,
$I_{ik,j}^{j,ik}$, $I_{jk,i}^{i,jk}$ and $I_{i,j,k}^{jk,ik,ij}$,
with probability $\epsilon^{3}(1-\epsilon)$, while outgoing flows
are due to controls $I_{ijk,l}^{l,ijk}$ and $I_{ijk}^{l}$ with probability
$1-\epsilon^{3}$. We set $\phi_{ijk}^{l}=0$ so that inequality (\ref{eq:param1})
becomes 
\begin{equation} \label{eq:level4-1}
\epsilon^{3}(1-\epsilon)\left(\phi_{ij,k}^{k,ij}+\phi_{ik,j}^{j,ik}+\phi_{jk,i}^{i,jk}+\phi_{i,j,k}^{jk,ik,ij}\right)
  \leq \left(1-\epsilon^{3}\right)\phi_{ijk,l}^{l,ijk}.
\end{equation}
To satisfy (\ref{eq:level4-1}), we set 
\begin{equation} \label{eq:level4-2}
\boxed{
\phi_{ijk,l}^{l,ijk}=\epsilon^{3}(1-\epsilon)\left(\phi_{ij,k}^{k,ij}+\phi_{ik,j}^{j,ik}+\phi_{jk,i}^{i,jk}
  +\phi_{i,j,k}^{jk,ik,ij}\right)/\left(1-\epsilon^{3}\right)  }  .
\end{equation}

\item Incoming flows to the virtual queues corresponding to $Q_{ij}^{kl}$
are due to controls $I_{i,j}^{j,i}$,
$I_{ij,k}^{k,ij}$, $I_{ij,l}^{l,ij}$, $I_{ik,j}^{j,ik}$, $I_{il,j}^{j,il}$,
$I_{jk,i}^{i,jk}$, $I_{jl,i}^{i,jl}$, $I_{i,j,k}^{jk,ik,ij}$, $I_{i,j,l}^{jl,il,ij}$,
with probability $\epsilon^{2}(1-\epsilon)^{2}$, and $I_{i,j}^{jk,ik}$,
$I_{i,j}^{jl,il}$, $I_{ijk,l}^{l,ijk}$, $I_{ijl,k}^{k,ijl}$ with
probability $\epsilon^{2}(1-\epsilon)$. Outgoing flows are due to
controls $I_{ij,kl}^{kl,ij}$, $I_{ij,k,l}^{kl,ijl,ijk}$ and $I_{ij}^{kl}$
with probability $1-\epsilon^{2}$. Therefore, inequality (\ref{eq:param1})
becomes 
\begin{equation} \label{eq:level4-3}
\begin{array}{c}
\epsilon^{2}(1-\epsilon)^{2}\left(\phi_{i,j}^{j,i}+\phi_{ij,k}^{k,ij}+\phi_{ij,l}^{l,ij}+\phi_{ik,j}^{j,ik}
  +\phi_{il,j}^{j,il}+\phi_{jk,i}^{i,jk}+\phi_{jl,i}^{i,jl}+\phi_{i,j,k}^{jk,ik,ij}+\phi_{i,j,l}^{jl,il,ij}\right) \\
  +\epsilon^{2}(1-\epsilon)\left(\phi_{i,j}^{jk,ik}+\phi_{i,j}^{jl,il}+\phi_{ijk,l}^{l,ijk}+\phi_{ijl,k}^{k,ijl}\right)
  \leq \left(1-\epsilon^{2}\right)\left(\phi_{ij,kl}^{kl,ij}+\phi_{ij,k,l}^{kl,ijl,ijk}+\phi_{ij}^{kl}\right).
\end{array}
\end{equation}

Similarly, for the virtual queues corresponding to $Q_{kl}^{ij}$, inequality (\ref{eq:param1}) becomes 
\begin{equation} \label{eq:level4-4}
\begin{array}{c}
\epsilon^{2}(1-\epsilon)^{2}\left(\phi_{k,l}^{l,k}+\phi_{kl,i}^{i,kl}+\phi_{kl,j}^{j,kl}+\phi_{ik,l}^{l,ik}
  +\phi_{jk,l}^{l,jk}+\phi_{il,k}^{k,il}+\phi_{jl,k}^{k,jl}+\phi_{i,k,l}^{kl,il,ik}+\phi_{j,k,l}^{kl,jl,jk}\right) \\
  +\epsilon^{2}(1-\epsilon)\left(\phi_{k,l}^{il,ik}+\phi_{k,l}^{jl,jk}+\phi_{ikl,j}^{j,ikl}+\phi_{jkl,i}^{i,jkl}\right)
  \leq \left(1-\epsilon^{2}\right)\left(\phi_{ij,kl}^{kl,ij}+\phi_{kl,i,j}^{ij,jkl,ikl} +\phi^{ij}_{kl} \right).
\end{array}
\end{equation}
All $\phi$ parameters in the LHS of inequalities (\ref{eq:level4-3})
and (\ref{eq:level4-4}) have already been computed (or set to 0, by selection). We now set all terms in the RHS
of (\ref{eq:level4-3}), (\ref{eq:level4-4}) to 0, with the exception of $\phi^{kl,ij}_{ij,kl}$. Without loss of generality,
we can also restrict our attention to the case $i=1$, $i<j$ and $k<l$, for distinct $i,j,k,l$. Similarly to 
the argument in level 3, we can combine (\ref{eq:level4-3}), (\ref{eq:level4-4}) to the equivalent expression
\begin{equation} \begin{split}
\frac{1}{1-\epsilon^2} \max \Bigg[ & \epsilon^2 (1-\epsilon)^2 \left( \phi^{j,i}_{i,j}+\phi^{k,ij}_{ij,k}+
  \phi^{l,ij}_{ij,l}+\phi^{j,ik}_{ik,j}+\phi^{j,il}_{il,j}+\phi^{i,jk}_{jk,i}+\phi^{i,jl}_{jl,i}
  +\phi^{jk,ik,ij}_{i,j,k}+\phi^{jl,il,ij}_{i,j,l} \right) \\
& +\epsilon^2 (1-\epsilon) \left( \phi^{l,ijk}_{ijk,l}+\phi^{k,ijl}_{ijl,k} \right) , \\
& \epsilon^2 (1-\epsilon)^2 \left( \phi^{l,k}_{k,l}+\phi^{i,kl}_{kl,i}+\phi^{j,kl}_{kl,j}+\phi^{l,ik}_{ik,l}+
  \phi^{l,jk}_{jk,l}+\phi^{k,il}_{il,k}+\phi^{k,jl}_{jl,k}+\phi^{kl,il,ik}_{i,k,l}+\phi^{kl,jl,jk}_{j,k,l} \right) \\
& +\epsilon^2 (1-\epsilon) \left( \phi^{j,ikl}_{ikl,j}+\phi^{i,jkl}_{jkl,i} \right) \Bigg] \leq \phi^{kl,ij}_{ij,kl} .
\end{split} \end{equation}

Again, symbolic manipulations show that the maximum is achieved for the first term (which is clearly non-negative)
so that we select
\begin{equation} \boxed{ \begin{split} \label{eq:level4-6}
\phi_{ij,kl}^{kl,ij}= \frac{1}{1-\epsilon^2} \Bigg[ & \epsilon^2 (1-\epsilon)^2 \left( \phi^{j,i}_{i,j}
  +\phi^{k,ij}_{ij,k}+\phi^{l,ij}_{ij,l}+\phi^{j,ik}_{ik,j}+\phi^{j,il}_{il,j}+\phi^{i,jk}_{jk,i}
  +\phi^{i,jl}_{jl,i}+\phi^{jk,ik,ij}_{i,j,k}+\phi^{jl,il,ij}_{i,j,l} \right) \\
& +\epsilon^2 (1-\epsilon) \left( \phi^{l,ijk}_{ijk,l}+\phi^{k,ijl}_{ijl,k} \right) \Bigg]  .
\end{split} }  \end{equation}

\item For the virtual queues corresponding to $Q_{i}^{jkl}$, incoming flows are due to controls
of the form $I_{\mc{D}_{1},\ldots,\mc{D}_{\nu}}^{\mc{L}_{1},\ldots,\mc{L}_{\nu}},\: i\in\mc{D}_{1},\:\left|\mc{D}_{j}\cup\mc{L}_{j}\right|
\leq 3,\: j\in\left\{ 1,\ldots,3\right\} $, as well as controls of the form $I_{\mc{D}_{1},\ldots,\mc{D}_{\nu}}^{\mc{L}_{1},\ldots,\mc{L}_{\nu}},
\: i\in\mc{D}_{1},\:\left|\mc{D}_{1}\right|\geq2,\:\left|\mc{D}_{j}\cup\mc{L}_{j}\right|=4,\: j\in\left\{ 1,\ldots,3\right\} $.
Outgoing flows are due to controls of the form $I_{\mc{D}_{1},\ldots,\mc{D}_{\nu}}^{\mc{L}_{1},\ldots,\mc{L}_{\nu}},\: i\in\mc{D}_{1},\:
\left|\mc{D}_{1}\right|=1,\:\left|\mc{D}_{j}\cup\mc{L}_{j}\right|=4,\: j\in\left\{ 1,\ldots,4\right\} $,
with probability $\left(1-\epsilon\right)$. Therefore, inequality (\ref{eq:param1}) becomes 
\begin{equation} \label{eq:level4-8}
\begin{array}{c}
\epsilon \left(1-\epsilon\right)^{3}\left(\phi_{i}+ \sum_{a\neq i} \phi_{i,a}^{a,i}+ \sum_{a,b\neq i} \phi^{i,ab}_{ab,i}
  +\sum_{a,b\neq i} \phi^{b,ia}_{ia,b} +\sum_{a,b\neq i} \phi^{ab,ib,ia}_{i,a,b} \right)\\
  +\epsilon \left(1-\epsilon\right)^{2} \left( \sum_{a\neq i} \phi^a_i + \sum_{a,b\neq i} \phi^{ab,ib}_{i,a}
  +\sum_{a,b,c\neq i} \phi^{c,iab}_{iab,c} \right)\\
  +\epsilon \left(1-\epsilon\right) \left( \sum_{a,b\neq i} \phi^{ab}_i +\sum_{a,b,c\neq i} \phi^{bc,ia}_{ia,bc}
  +\sum_{a,b,c\neq i} \phi^{bc,ibc,iac}_{ia,b,c} \right) \leq \\
  \left(1-\epsilon\right) \left( \phi_{i,j,k,l}^{jkl,ikl,ijl,ijk}+\phi_{jkl,i}^{i,jkl}+\phi_{jk,i,l}^{il,jkl,ijk}
  +\phi_{i,j,k}^{jkl,ikl,ijl} \right. \\
  +\left. \phi_{i,j,l}^{jkl,ikl,ijk}+\phi_{i,k,l}^{jkl,ijl,ijk}+\phi_{i,j}^{jkl,ikl}+\phi_{i,k}^{jkl,ijl}
  +\phi_{i,l}^{jkl,ikl}+\phi_{i}^{jkl} \right) ,
\end{array}
\end{equation}
where $a,b,c,d$ are distinct summation indices that take values in the set $\{i,j,k,l\}$.
Similar inequalities to (\ref{eq:level4-8}) can be formed for $Q_{j}^{ikl}$, $Q_{k}^{ijl}$
and $Q_{l}^{ijk}$. We now set
\begin{equation}
\begin{array}{c}
\phi_{a,b,c}^{bcd,acd,abd}=0, \: \forall\: a,b,c,d\in\left\{ i,j,k,l\right\} ,\\
\phi_{a,b}^{bcd,acd}=0, \: \forall \: a,b,c,d\in\left\{ i,j,k,l\right\} ,\\
\phi_{a}^{bcd}=0,\: \forall \: a,b,c,d\in\left\{ i,j,k,l\right\} .
\end{array}
\end{equation}
Therefore, when we write down (\ref{eq:level4-8}) for $i=1,\ldots,4$, only parameter
$\phi^{234,134,124,123}_{1,2,3,4}$ is unknown in the RHS while all LHS parameters in 
(\ref{eq:level4-8}) have been previously determined. Hence, (\ref{eq:level4-8}) as written
for $i=1,\ldots,4$ is equivalent to
\begin{equation} \begin{split}
\max_{i=1,\ldots,4} \; \Bigg[ & \epsilon (1-\epsilon)^2 \left( \phi_i +\sum_{a\neq i} \phi^{a,i}_{i,a} 
  +\sum_{a,b\neq i} \phi^{i,ab}_{ab,i} + \sum_{a,b\neq i} \phi^{b,ia}_{ia,b} + \sum_{a,b\neq i} 
  \phi^{ab,ib,ia}_{i,a,b} \right) \\
& +\epsilon (1-\epsilon) \sum_{a,b,c\neq i} \phi^{c,iab}_{iab,c} +\epsilon 
  \sum_{a,b,c\neq i} \phi^{bc,ia}_{ia,bc} -\phi^{i,jkl}_{jkl,i} \Bigg] \leq \phi^{234,134,124,123}_{1,2,3,4} ,
\end{split} \end{equation}
and some simple algebra reveals that the maximum term (which is also non-negative) is for $i=1$, so that we select
\begin{equation} \boxed{ \begin{split} \label{eq:all}
\phi^{234,134,124,123}_{1,2,3,4} = &\epsilon (1-\epsilon)^2 \left( \phi_1 +\sum_{a\neq 1} \phi^{a,1}_{1,a}
  +\sum_{a,b\neq 1} \phi^{1,ab}_{ab,1} + \sum_{a,b\neq 1} \phi^{b,1a}_{1a,b} +\sum_{a,b\neq 1} 
  \phi^{ab,1b,1a}_{1,a,b} \right) \\
& +\epsilon (1-\epsilon) \sum_{a,b,c\neq 1} \phi^{c,1ab}_{1ab,c} +\epsilon \sum_{a,b,c\neq 1} 
  \phi^{bc,1a}_{1a,bc}-\phi^{1,234}_{234,1} . 
\end{split} } \end{equation}

\end{itemize}

For the reader's convenience, the selected controls $\phi$ are given in closed form in Appendix~\ref{app:closed-form}.
Finally, to ensure that (\ref{eq:param6}) is satisfied, we calculate the sum of all flows, and find 
\[
\sum_{I\in\mc{I}}\phi_{I}=\sum_{i=1}^{4}\frac{\lambda_{i}}{1-\epsilon^{i}}.
\]
Since, by assumption, it holds $\sum_{i=1}^{4}\frac{\lambda_{i}}{1-\epsilon^{i}}\leq1,$
we conclude that $\sum_{I\in\mc{I}}\phi_{I}\leq1$, as desired.
Hence, we have proved the following result. 
\begin{theorem}
For the case of 4 users, and for i.i.d erasure events, the stability
region of the system is given by 
\[
\mc{R}_{\Pi}=\left\{ \boldsymbol{\lambda}:\:\max_{\sigma\in\mc{P}}\sum_{i=1}^{4}
  \frac{\lambda_{\sigma(i)}}{1-\epsilon^{i}}\leq1\right\} ,
\]
where $\mc{P}$ is the set of permutations
$\sigma$ on $\left\{ 1,\ldots,4\right\}$. Moreover, the policy
$\pi^{\ast}\in\Pi$ described in Section \ref{sub:Stabilizing-Policy}
using the XOR controls described in Table \ref{table1} is stabilizing.
The stability region coincides with the information theoretic capacity
region of the standard BEC with feedback, and is within one bit (actually,
$O(2^{-L})$ bits according to Theorem \ref{thm:outer_bound_ext})
from the capacity of the extended BEC with feedback. The latter is
equal to the stability region of the system under any coding strategy. 
\end{theorem}

\section{Implementation Issues}  \label{sec:Implementation-Issues}

\subsection{Packet overhead}

As mentioned in Section \ref{sec:Stability-region}, for the proposed
network coding scheme to work, every user must know the identities of
all native packets that constitute a composite (i.e.~XOR combination)
packet it receives. Having this information, a user is able to decode
the native packet destined for it. A simple mechanism that can be used
to provide users with this information is equipping every native
packet with a Packet ID, which consists of the packet's destination
and a sequence number. If a transmitted packet is composed of $\mu$
native packets, then it contains in its packet header the $\mu$ packet IDs.
Depending on the feedback from the users and in accordance to the
Rules for Packet Movement, either the transmitted packet
$p=\bigoplus_{k=1}^{\nu}p_{\mc{D}_{k}}^{\mc{L}_{k}}$ is moved as a
whole to a real queue, or some of the packets
$p_{\mc{D}_{1}}^{\mc{L}_{1}},\ldots,p_{\mc{D}_{\nu}}^{\mc{L}_{\nu}}$
are individually moved to real queues. More precisely, the following
Lemma follows immediately from the Rules for Packet Movement.
\begin{lemma}  \label{lem:A-packet-}
After transmission of a packet at slot $t$, let packet $q$ (not
necessarily the transmitted packet) be placed at a real queue of level
k.n. Then, either a) $q$ is a combination of packets that at the
beginning of slot $t$ were at queues of level less than $k$, or b) $q$
is a copy of a packet that at the beginning of slot $t$ was either at
level $r$, $r\in\left\{ 0,\ldots,k-1\right\} $, or at sublevel $k.l$,
$1\leq l\leq n-1$.
\end{lemma}
To compute the overhead bits needed to implement the above mechanism,
we need to find the maximum number of Packet IDs that may be included
in a packet that is placed in a real queue of a certain level.  This
is expressed in Lemma \ref{lem:The-maximum-number} below (all queues
and packets referred to in this lemma are real queues and packets,
respectively). In the following, when we say that a packet
\textit{comes from level} $k$ (or\textit{ exits level} $k$) we mean
that it is an XOR combination of packets placed in queues of levels
$1$ to $k$ (with at least one packet being in a level $k$ queue).
\begin{lemma}  \label{lem:The-maximum-number}
Under the coding scheme of Section \ref{sec:Network-coding-algorithms}, it holds
a) Any packet placed in queues at sublevel $k.n,\: n=1,2,\ldots,k-1,$
$k\geq2,$ contains at most $(k-1)!$ packet IDs.

b) Any packet exiting level $k\geq2$ contains at most $k!$ packet
IDs.
\end{lemma}

\begin{IEEEproof}
We use induction on $k$ to prove the Lemma. For $k=2,$ the Lemma
follows immediately from the Rules for Packet Movement in Section
\ref{sec:Network-coding-algorithms}. We now assume that the Lemma
holds for levels $2$ up to $k-1$ and show that it also holds for
level $k.$ We first prove part a) of the Lemma by induction on $n$.

Part a): If a packet $p$ is placed in a queue at the lowest sublevel
of level $k,$ i.e.~$k.1$, then according to Lemma \ref{lem:A-packet-},
$p$ comes from levels $l\leq k-1$. Hence, according to part b) of
the inductive hypothesis, it contains at most $(k-1)!$ packet IDs,
so that part a) holds for $n=1.$ Assume next that part a) holds for
all packets $p$ placed at any sublevel from $k.1$ up to $k.n$ with
$2\leq n<k-1$, i.e.~assume that all packets $p$ in sublevels from
$k.1$ up to $k.n$ contain at most $(k-1)!$ packet IDs. We shall
prove that any packet in sublevel $k.\left(n+1\right)$ also contains
at most $(k-1)!$ packet IDs. According to Lemma \ref{lem:A-packet-}
for a packet $p$ at sublevel $k.\left(n+1\right)$, one of the following
two cases holds.

\begin{enumerate}
\item Packet $p$ comes from level $l$, where $2\leq l\leq k-1$. Then,
according to part b) of the inductive hypothesis, $p$ contains at
most $(k-1)!$ packet IDs.

\item Packet $p$ was placed before the current slot transmission at
a queue in a lower sublevel of the same level, i.e.~a sublevel from
$k.1$ up to $k.n$. According to the inductive hypothesis on $n,$
packets in these sublevels contain at most $(k-1)!$ packet IDs. Since
Lemma \ref{lem:A-packet-} states that packets from lower sublevels
are merely copied to higher sublevels, it follows that the maximum
number of packet IDs they contain remains the same, so packet $p$
at sublevel $k.\left(n+1\right)$ will also contain at most $(k-1)!$
packet IDs. Therefore, packets at all sublevels $k.n,\: n=1,2,\ldots,k-1,$
$k\geq2,$ contain at most $(k-1)!$ packet IDs. This completes the
proof of part a) of the Lemma.
\end{enumerate}

To prove part b) of the Lemma, consider a packet $p$ exiting level
$k.$ This packet is of the form $p=p_{\mc{D}_{1}}^{\mc{L}_{1}}\oplus\ldots\oplus p_{\mc{D}_{\nu}}^{\mc{L}_{\nu}}$,
where each $p_{\mc{D}_{r}}^{\mc{L}_{r}}$ belongs to a queue
of at most level $k$, hence the maximum number of packet IDs $p$
may contain is the sum of the packet IDs contained in packets $p_{\mc{D}_{r}}^{\mc{L}_{r}},\: r\in\left\{ 1,\ldots,\nu\right\}$,
which is at most $\nu(k-1)!$ due to part a). From Lemma \ref{lem:Consider-packet-,},
it holds $\nu\leq k$, therefore any packet exiting level $k$ contains
at most $k(k-1)!=k!$ packet IDs. 
\end{IEEEproof}
Up to level 4, the maximum number of Packet IDs that may need to be
included in a packet is $4!=24$. Assuming a packet ID of 20 bits
and packet length of 1500 bytes, i.e.~12000
bits, the overhead is approximately $4\%$. Hence,
for $N=4$ receivers, since only queues up to level 4 may be formed,
the overhead of the proposed algorithm is fairly acceptable. It can
be seen that the maximum number of Packet IDs needed increases dramatically
with the number of users $N$ and it is very important to address
this matter as $N$ increases. Various suboptimal policies that reduce
the necessary number of Packet IDs can be investigated. For example,
the transmitter may choose not to send packet combinations if the
resulting packet header exceeds a certain number of Packet IDs. Another
policy towards this direction could involve coding of packets only
until a certain level. Specifically, for $N$ users, only the real
queues until level $l$ could be created, where $l<N$. In case a
packet is received by more than $l$ users, additional receivers would
be ignored and the packet would be placed in one of the level $l$
queues. The detailed study of these possibilities
and the performance of the resulting algorithm is a subject of future
work.

\subsection{Queue stability at the receivers}

As mentioned in Section \ref{sec:Stability-region}, another problem
that may arise is possible instability of queues at the receivers,
where all packets received by a certain user are stored. A simple
way to avert this possibility is to take advantage of the fact that
when the queue sizes at the base station become empty, all packets
formed during previous transmissions are not needed at the receivers.
Therefore, we can let the base station inform all receivers when its
queues become empty, by, for example, leaving a slot empty after a
series of transmissions taking place when the queues are nonempty.
Under this modification, using standard results from regenerative
theory, it can be shown that the system is stable if and only if the
total queue size at the base station is stable.

\section{Conclusions} \label{sec:Conclusions}

In this work, we presented a network coding scheme for the broadcast
erasure channel with $N$ multiple unicast sessions based on the coding
scheme we proposed in \cite{AGG}. In this scheme, only XOR operations
are allowed. Also, instant decodability, i.e.~the ability of any user
that receives a coded packet to instantly decode its own native packet,
is ensured. 

Furthermore, we assumed random packet arrivals and presented a stabilizing
policy based on this coding scheme. We then derived an upper bound
on the stability region of the system under examination.
For the case of $4$ users and i.i.d. erasure events, we proved that
the stability region of the system is identical to the capacity outer
bound of the BEC channel with feedback.

Finally, implementation issues were examined, such as the increase
of packet overhead as the number of users increases, which is due
to the number of packet addresses needed to completely describe a
coded packet. The maximum number of addresses needed in the general
case of $N$ users was found to be $N!$. Future work could be aimed
towards the development of suboptimal variations of the proposed policy
that will require a smaller number of packet addresses, thus reducing
packet overhead.

\appendix
{}

\subsection{Proof of Lemma \ref{lem:movok}} \label{sub:Proof-of-Lemma}

Let the transmitted packet $p$ at slot $t$ have the form
$p=\bigoplus_{k=1}^{\nu}p_{\mc{D}_{k}}^{\mc{L}_{k}}$ where
$\mc{L}_{k},\mc{D}_{k}$ satisfy BCR. The proof is easier if we assume
that any exogenous arrivals of native packets for user $i\in \mc{N}$
at slot $t$ enter the network (and are stored in queue $Q_{i}$ while a
corresponding token is stored in virtual queue $V_i(i)$ and $K_i(i)$
is also increased by 1) \textit{after} the transmission of $p$ takes
place, i.e.~any exogenous packet stays in the network for at least one
slot. However, it should be emphasized that this assumption is only
made to simplify the subsequent proof; Lemma \ref{lem:movok} still
holds regardless of this assumption.  For brevity, we hereafter write
``BP at $t$'' to mean the BP properties being true at the beginning of
slot $t$ (which is the assumption in Lemma \ref{lem:movok}) and ``BP
at $t+1$'' as the BP properties being true at the end of slot $t$, or
beginning of slot $t+1$ (which is the result we wish to
prove). Clearly, BP\ref{enu:xor_native} at $t+1$ follows immediately
from BP\ref{enu:xor_native} at $t$, so we concentrate on proving
BP\ref{enu:LD}--BP\ref{enu:nolost} at $t+1$. Notice that the exogenous
arrivals that enter at slot $t$ automatically satisfy
BP\ref{enu:LD}--BP\ref{enu:nolost} at $t+1$.  Since BP at $t+1$ is
trivially true if $\mc{S}=\emptyset$ (i.e.~$p$ is erased by all users,
so that the slot effectively ``never happened''), we hereafter assume
$\mc{S}\neq\emptyset$. In the following, we only examine the case
$\nu>1$ in detail, since $\nu=1$ can be handled as a special case.

\begin{figure}[t]
\centering
\centering
\psfrag{QL_D}[][][0.6]{all other $Q^{\mc{L}}_{\mc{D}}$}
\psfrag{QL1_D1}[][][0.8]{$Q^{\mc{L}_1}_{\mc{D}_1}$} 
\psfrag{QLm_Dm}[][][0.8]{$Q^{\mc{L}_\nu}_{\mc{D}_\nu}$} 
\psfrag{QL1_D1(1)}[][][0.8]{$V^{\mc{L}_1}_{\mc{D}_1}(i_{1,1})$} 
\psfrag{QL1_D1(n)}[][][0.8]{$V^{\mc{L}_1}_{\mc{D}_1}(i_{1,n_1})$} 
\psfrag{QLm_Dm(1)}[][][0.8]{$V^{\mc{L}_\nu}_{\mc{D}_\nu}(i_{\nu,1})$} 
\psfrag{QLm_Dm(n)}[][][0.8]{$V^{\mc{L}_\nu}_{\mc{D}_\nu}(i_{\nu,n_\nu})$} 
\psfrag{pL1_D1}[][][0.65]{$p^{\mc{L}_1}_{\mc{D}_1}$} 
\psfrag{pLm_Dm}[][][0.65]{$p^{\mc{L}_\nu}_{\mc{D}_\nu}$} 
\psfrag{phL1_D11}[][][0.65]{$\hat{p}^{\mc{L}_1}_{\mc{D}_1}[1]$} 
\psfrag{phLm_Dm1}[][][0.65]{$\hat{p}^{\mc{L}_\nu}_{\mc{D}_\nu}[1]$} 
\psfrag{phL1_D12}[][][0.65]{$\hat{p}^{\mc{L}_1}_{\mc{D}_1}[2]$} 
\psfrag{phLm_Dm2}[][][0.65]{$\hat{p}^{\mc{L}_\nu}_{\mc{D}_\nu}[2]$} 
\psfrag{pL1_D1_1}[][][0.65]{$p^{\mc{L}_1}_{\mc{D}_1}(i_{1,1})$} 
\psfrag{phL1_D1_1}[][][0.65]{$\hat{p}^{\mc{L}_1}_{\mc{D}_1}[1](i_{1,1})$} 
\psfrag{pL1_D1_n}[][][0.65]{$p^{\mc{L}_1}_{\mc{D}_1}(i_{1,n_1})$} 
\psfrag{phL1_D1_2}[][][0.65]{$\hat{p}^{\mc{L}_1}_{\mc{D}_1}[2](i_{1,1})$} 
\psfrag{phL1_D1_n1}[][][0.65]{$\hat{p}^{\mc{L}_1}_{\mc{D}_1}[1](i_{1,n_1})$} 
\psfrag{phL1_D1_n2}[][][0.65]{$\hat{p}^{\mc{L}_1}_{\mc{D}_1}[2](i_{1,n_1})$} 
\psfrag{pLm_Dm_1}[][][0.65]{$p^{\mc{L}_\nu}_{\mc{D}_\nu}(i_{\nu,1})$} 
\psfrag{phLm_Dm_1}[][][0.65]{$\hat{p}^{\mc{L}_\nu}_{\mc{D}_\nu}[1](i_{\nu,1})$} 
\psfrag{phLm_Dm_2}[][][0.65]{$\hat{p}^{\mc{L}_\nu}_{\mc{D}_\nu}[2](i_{\nu,1})$} 
\psfrag{pLm_Dm_n}[][][0.65]{$p^{\mc{L}_\nu}_{\mc{D}_\nu}(i_{\nu,n_\nu})$} 
\psfrag{phLm_Dm_n1}[][][0.65]{$\hat{p}^{\mc{L}_\nu}_{\mc{D}_\nu}[1](i_{\nu,n_\nu})$} 
\psfrag{phLm_Dm_n2}[][][0.65]{$\hat{p}^{\mc{L}_\nu}_{\mc{D}_\nu}[2](i_{\nu,n_\nu})$} 
\psfrag{native packets for QL1_D1}[][][0.7]{unknown native packets in $Q^{\mc{L}_1}_{\mc{D}_1}$} 
\psfrag{native packets for QLm_Dm}[][][0.7]{unknown native packets in $Q^{\mc{L}_\nu}_{\mc{D}_\nu}$} 
\psfrag{QL_D(i)}[][][0.6]{all other $V^{\mc{L}}_{\mc{D}}(i)$}
\includegraphics[scale=0.5]{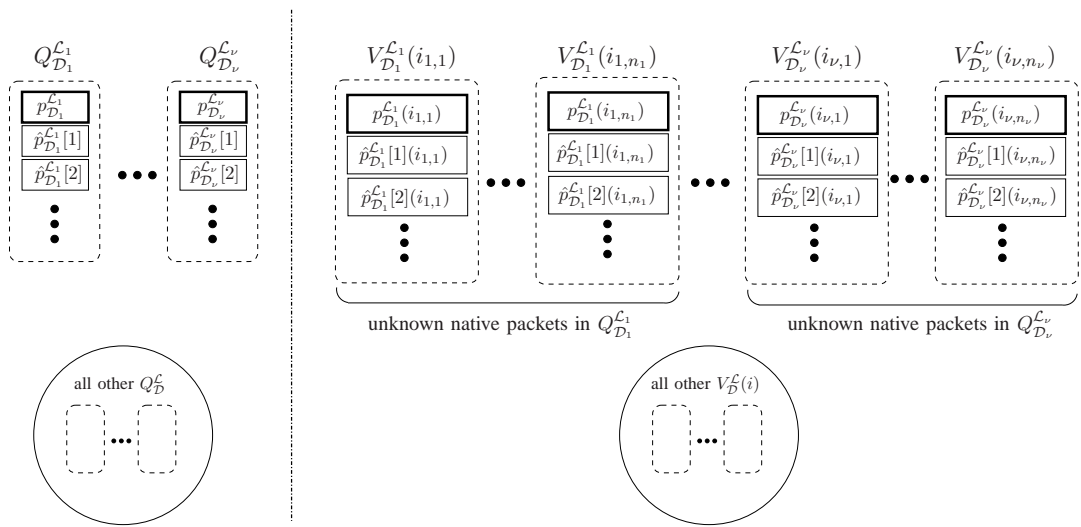}
\caption{Real (LHS) and virtual (RHS) queue contents at beginning of slot $t$. Only the real queues are actually stored at the transmitter.}
\label{fig:begint}
\end{figure}

We examine each case of the Rules for Packet Movement (RPM)
separately.  It will also be useful to have a graphical representation
for the queue contents at $t$, as shown in Fig.~\ref{fig:begint}. The
following notation is introduced to illustrate Fig.~\ref{fig:begint}:
we denote $n_{k}=\left|\mc{D}_{k}\right|$ so that each set
$\mc{D}_{k}$ can be written w.l.o.g. as $\mc{D}_{k}=\left\{
i_{k,1},i_{k,2},\ldots,i_{k,n_{k}}\right\} $, for each
$k=1,\ldots,\nu$. The real queues are shown in the LHS of
Fig.~\ref{fig:begint}, where the rectangles denote packets and the
topmost packets (shown in bold edges) in queues
$Q_{\mc{D}_{k}}^{\mc{L}_{k}}$, for $k=1,\ldots,\nu$, are the ones that
comprise the transmitted packet $p$ according to the BCR. All other
packets (including the ones contained in the queues
$Q_{\mc{D}}^{\mc{L}}$, with
$(\mc{L},\mc{D})\neq(\mc{L}_{k},\mc{D}_{k})$, shown in the circles at
the bottom of Fig.~\ref{fig:begint}) are non-bold. All packets denoted
with $\hat{p}$ in Fig.~\ref{fig:begint} are not included in $p$ and
are therefore unaffected by the RPM (the $[\cdot]$ notation is used
only for indexing purposes to visually distinguish the packets in the
same queue).

The virtual queues are shown in the RHS of Fig.~\ref{fig:begint},
where the bold edges denote the tokens for the unknown native packets
contained in the packets that comprise $p$. The tokens for the unknown
native packets contained in $p_{\mc{D}_{k}}^{\mc{L}_{k}}$ are denoted
as $p_{\mc{D}_{k}}^{\mc{L}_{k}}(i_{k,1}), \ldots,
p_{\mc{D}_{k}}^{\mc{L}_{k}}(i_{k,n_{k}})$ while those contained in
$\hat{p}_{\mc{D}_{k}}^{\mc{L}_{k}}[l]$ are denoted as
$\hat{p}_{\mc{D}_{k}}^{\mc{L}_{k}}[l](i_{k,1}), \ldots,
\hat{p}_{\mc{D}_{k}}^{\mc{L}_{k}}[l](i_{k,n_{k}})$. The duality
between a token and its corresponding native packet will be
consistently used below.

A careful examination of the RPM leads to the following observation:
in \textit{all} cases, the non-bold-edged real and virtual packets
in Fig.~\ref{fig:begint} are not affected by the RPM. Specifically,
these packets possess the following properties.

\textbf{Properties of non-bold-edged packets (PNB)}:
\begin{enumerate}
\item non-bold-edged real and virtual packets (tokens) are not moved
  from the queues they are stored at $t$ and the XOR decomposition of
  the non-bold-edged real packets remains the same between $t$ and
  $t+1$.

\item none of the unknown native packets corresponding to
  non-bold-edged tokens in the virtual queues at $t$ are decoded at
  $t+1$ (i.e.~these packets remain unknown at $t+1$).
\end{enumerate}
The second item in the above list follows from the fact that, by Corollary
\ref{cor:Dforall} and Fact \ref{instant-decoding}, only the users
in $\mc{S}\cap\left(\cup_{k=1}^{\nu}\mc{D}_{k}\right)$
actually decode unknown native packets (i.e.~the bold-edged native
packets in Fig.~\ref{fig:begint}) contained in the $p_{\mc{D}_{k}}^{\mc{L}_{k}}$
that comprise the transmitted packet $p$. Since, by BP\ref{enu:nolost}
at $t$, each unknown native packet is contained in exactly one real
packet, it follows that no (non-bold-edged) native packet contained
in a non-bold-edged real packet is decoded at $t+1$.

We now use the above observations to show that all non-bold-edged real
packets in Fig.~\ref{fig:begint} (which, by assumption, satisfy BP at
$t$) satisfy BP at $t+1$. Specifically, consider any non-bold-edged
real packet $p_{\mc{D}}^{\mc{L}}$ (this packet must either be stored
in a queue contained in the left circle of Fig.~\ref{fig:begint}, or
in queue $Q_{\mc{D}_{k}}^{\mc{L}_{k}}$,
i.e.~$p_{\mc{D}}^{\mc{L}}=\hat{p}_{\mc{D}_{k}}^{\mc{L}_{k}}[\cdot]$).
Any $i\in\mc{D}$ is, by BP\ref{enu:LD} at $t$, a Destination for
$p_{\mc{D}}^{\mc{L}}$ and, since the native unknown packet contained
in $p_{\mc{D}}^{\mc{L}}$ is still unknown at $t+1$ (second item in
PNB) and $p_{\mc{D}}^{\mc{L}}$ retained its XOR decomposition (first
item in PNB), we conclude that $i$ is also a Destination for
$p_{\mc{D}}^{\mc{L}}$ at $t+1.$ Also, any user $i$ that is a
Destination of $p_{\mc{D}}^{\mc{L}}$ at $t+1$ is also a Destination of
the same packet at $t$, since the XOR decomposition of
$p_{\mc{D}}^{\mc{L}}$ did not change during slot $t$. Hence, by
BP\ref{enu:LD} at $t$, it follows that $i\in\mc{D}$. The absorbing
property of Listeners for $p_{\mc{D}}^{\mc{L}}$ now implies
BP\ref{enu:LD} at $t+1$.

Furthermore, any unknown native packet for some user $i$ contained
in $p_{\mc{D}}^{\mc{L}}$ at $t+1$ is also unknown at $t$
(again, due to PNB) so that by BP\ref{enu:onlyD}
at $t$, user $i$ is a Destination for $p_{\mc{D}}^{\mc{L}}$
at $p$ and $i\in\mc{D}$ (by BP\ref{enu:LD} at $t$). Hence,
BP\ref{enu:LD} at $t+1$ (which was proved in the previous paragraph)
implies that $i$ is a Destination for $p_{\mc{D}}^{\mc{L}}$
at $t+1$, which also proves BP\ref{enu:onlyD} at $t+1$. Finally,
BP\ref{enu:nolost} at $t+1$ follows immediately from Fig.~\ref{fig:begint},
since any unknown native packet is either a new exogenous arrival
at $t$ for some user $i$ (and, by the scheme's construction, it
is contained in exactly one packet in $Q_{i}$) or it was already
in the network at $t$ and, by BP\ref{enu:nolost} at $t$, was stored
in exactly one non-bold-edged packet $p_{\mc{D}}^{\mc{L}}$
for some $\mc{L},\mc{D}$. 

Since the above arguments show that BP\ref{enu:LD}--BP\ref{enu:nolost}
at $t+1$ is true for all non-bold-edged real packets, it suffices
to only examine whether the packets moved between different queues
in the network according to RPM satisfy BP\ref{enu:LD}--BP\ref{enu:nolost}
at $t+1$. This is performed next.

\textbf{Case 2.1}: it holds
$\cup_{k=1}^{\nu}\mc{D}_{k}\subseteq\mc{S}$ (equivalently,
$\cup_{k=1}^{\nu}\mc{D}_{k}-\mc{S}=\emptyset$), so that all users in
$\cup_{k=1}^{\nu}\mc{D}_{k}$ decode their unknown native packets. By
the RPM in this case, all packets and tokens shown with bold edges in
Fig.~\ref{fig:begint} leave the network at $t+1$, whereas all other
packets remain in their queues (recall that the network actually
consists of the real queues only; virtual queues are
conceptual). Hence, the network representation at $t+1$ is the same as
in Fig.~\ref{fig:begint} minus the bold-edged packets and tokens (and
the possible addition of exogenous arrivals, which we have already
shown to satisfy BP\ref{enu:LD}--BP\ref{enu:onlyD} at $t+1$) so that
no packets/tokens are moved between queues in the network and no
further examination is necessary. BP\ref{enu:nolost} at $t+1$ also
follows trivially from BP\ref{enu:nolost} at $t$.

\textbf{Case 2.2.1}: it holds
$\cup_{k=1}^{\nu}\mc{D}_{k}-\mc{S}\neq\emptyset$ and
$\mc{S}\subseteq\cup_{k=1}^{\nu}\left(\mc{L}_{k}\cup\mc{D}_{k}\right)$
so that
$\hat{\mc{S}}=\mc{S}-\cup_{k=1}^{\nu}\left(\mc{L}_{k}\cup\mc{D}_{k}\right)=\emptyset$.
Again, all users in
$\mc{S}\cap\left(\cup_{k=1}^{\nu}\mc{D}_{k}\right)$ decode the unknown
native packets contained in the $p_{\mc{D}_{k}}^{\mc{L}_{k}}$ that
comprise $p$. Applying the RPM for this case to the network in
Fig.~\ref{fig:begint}, for each $k=1,\ldots,\nu$, bold-edged
packet $p_{\mc{D}_{k}}^{\mc{L}_{k}}$ is
moved to $Q_{\mc{D}_{k}-\mc{S}}^{\mc{L}_{k}\cup\left(\mc{D}_{k}
  \cap\mc{S}\right)\cup\tilde{\mc{S}}}$ and, for each
$i\in\mc{D}_{k}-\mc{S}$, bold-edged token
$p_{\mc{D}_{k}}^{\mc{L}_{k}}(i)$ is ``virtually'' moved to
$V_{\mc{D}_{k}-\mc{S}}^{\mc{L}_{k}\cup\left(\mc{D}_{k}
  \cap\mc{S}\right)\cup\tilde{\mc{S}}}(i)$ (which
  is captured by the fact that $K^{\mc{L}_k}_{\mc{D}_k}(i)$,
  $K^{\mc{L}_k\cup (\mc{D}_k \cap \mc{S}) \cup
    \tilde{\mc{S}}}_{\mc{D}_k-\mc{S}}(i)$ are reduced and increased by
  1, respectively), so that the queue contents at $t+1$ are
pictorially shown in Fig.~\ref{fig:endt_move1}. Recall also the
convention mentioned in Section \ref{sub:Packet-movement} that a
packet actually leaves the network if
$\mc{D}_{k}-\mc{S}=\emptyset$. Hence, to prove
BP\ref{enu:LD}--BP\ref{enu:nolost} for the moved packets, we can
assume w.l.o.g. that $\mc{D}_{k}-\mc{S}\neq\emptyset$ and we need to
show the following:

\begin{figure}[t]
\centering
\psfrag{QL_D}[][][0.6]{all other $Q^{\mc{L}}_{\mc{D}}$}
\psfrag{QL1_D1}[][][0.7]{$Q^{\mc{L}_1}_{\mc{D}_1}$} 
\psfrag{QLm_Dm}[][][0.7]{$Q^{\mc{L}_\nu}_{\mc{D}_\nu}$} 
\psfrag{QL1_D1(1)}[][][0.7]{$V^{\mc{L}_1}_{\mc{D}_1}(i_{1,1})$} 
\psfrag{QL1_D1(n)}[][][0.7]{$V^{\mc{L}_1}_{\mc{D}_1}(i_{1,n_1})$} 
\psfrag{QLm_Dm(1)}[][][0.7]{$V^{\mc{L}_\nu}_{\mc{D}_\nu}(i_{\nu,1})$} 
\psfrag{QLm_Dm(n)}[][][0.7]{$V^{\mc{L}_\nu}_{\mc{D}_\nu}(i_{\nu,n_\nu})$} 
\psfrag{pL1_D1}[][][0.65]{$p^{\mc{L}_1}_{\mc{D}_1}$} 
\psfrag{pLm_Dm}[][][0.65]{$p^{\mc{L}_\nu}_{\mc{D}_\nu}$} 
\psfrag{phL1_D11}[][][0.65]{$\hat{p}^{\mc{L}_1}_{\mc{D}_1}[1]$} 
\psfrag{phLm_Dm1}[][][0.65]{$\hat{p}^{\mc{L}_\nu}_{\mc{D}_\nu}[1]$} 
\psfrag{phL1_D12}[][][0.65]{$\hat{p}^{\mc{L}_1}_{\mc{D}_1}[2]$} 
\psfrag{phLm_Dm2}[][][0.65]{$\hat{p}^{\mc{L}_\nu}_{\mc{D}_\nu}[2]$} 
\psfrag{pL1_D1_1}[][][0.65]{$p^{\mc{L}_1}_{\mc{D}_1}(i_{1,1})$} 
\psfrag{phL1_D1_1}[][][0.65]{$\hat{p}^{\mc{L}_1}_{\mc{D}_1}[1](i_{1,1})$} 
\psfrag{pL1_D1_n}[][][0.65]{$p^{\mc{L}_1}_{\mc{D}_1}(i_{1,n_1})$} 
\psfrag{phL1_D1_2}[][][0.65]{$\hat{p}^{\mc{L}_1}_{\mc{D}_1}[2](i_{1,1})$} 
\psfrag{phL1_D1_n1}[][][0.65]{$\hat{p}^{\mc{L}_1}_{\mc{D}_1}[1](i_{1,n_1})$} 
\psfrag{phL1_D1_n2}[][][0.65]{$\hat{p}^{\mc{L}_1}_{\mc{D}_1}[2](i_{1,n_1})$} 
\psfrag{pLm_Dm_1}[][][0.65]{$p^{\mc{L}_\nu}_{\mc{D}_\nu}(i_{\nu,1})$} 
\psfrag{phLm_Dm_1}[][][0.65]{$\hat{p}^{\mc{L}_\nu}_{\mc{D}_\nu}[1](i_{\nu,1})$} 
\psfrag{phLm_Dm_2}[][][0.65]{$\hat{p}^{\mc{L}_\nu}_{\mc{D}_\nu}[2](i_{\nu,1})$} 
\psfrag{pLm_Dm_n}[][][0.65]{$p^{\mc{L}_\nu}_{\mc{D}_\nu}(i_{\nu,n_\nu})$} 
\psfrag{phLm_Dm_n1}[][][0.65]{$\hat{p}^{\mc{L}_\nu}_{\mc{D}_\nu}[1](i_{\nu,n_\nu})$} 
\psfrag{phLm_Dm_n2}[][][0.65]{$\hat{p}^{\mc{L}_\nu}_{\mc{D}_\nu}[2](i_{\nu,n_\nu})$} 
\psfrag{native packets for QL1_D1}[][][0.7]{unknown native packets in $Q^{\mc{L}_1}_{\mc{D}_1}$} 
\psfrag{native packets for QLm_Dm}[][][0.7]{unknown native packets in $Q^{\mc{L}_\nu}_{\mc{D}_\nu}$} 
\psfrag{QL_D(i)}[][][0.6]{all other $V^{\mc{L}}_{\mc{D}}(i)$} 
\psfrag{QL1_D1e}[][][0.65]{$Q^{\mc{L}_1\cup (\mc{D}_1\cap \mc{S})\cup \tilde{\mc{S}}}_{\mc{D}_1-\mc{S}}$} 
\psfrag{QLm_Dme}[][][0.65]{$Q^{\mc{L}_\nu\cup (\mc{D}_\nu\cap \mc{S}) \cup \tilde{\mc{S}}}_{\mc{D}_\nu-\mc{S}}$} 
\psfrag{contents}[][][0.65]{contents} 
\psfrag{begin}[][][0.65]{@ start} 
\psfrag{slot}[][][0.65]{of slot $t$} 
\psfrag{QL1_D1e(1)}[][][0.65]{$V^{\mc{L}_1\cup (\mc{D}_1-\mc{S}) \cup \tilde{\mc{S}}}_{\mc{D}_1-\mc{S}}(i_{1,1})$} 
\psfrag{QL1_D1e(n)}[][][0.65]{$V^{\mc{L}_1\cup (\mc{D}_1-\mc{S}) \cup \tilde{\mc{S}}}_{\mc{D}_1-\mc{S}}(i_{1,n_1})$} 
\psfrag{QLm_Dme(1)}[][][0.65]{$V^{\mc{L}_\nu\cup (\mc{D}_\nu-\mc{S}) \cup \tilde{\mc{S}}}_{\mc{D}_\nu-\mc{S}}(i_{\nu,1})$} 
\psfrag{QLm_Dme(n)}[][][0.65]{$V^{\mc{L}_\nu\cup (\mc{D}_\nu-\mc{S}) \cup \tilde{\mc{S}}}_{\mc{D}_\nu-\mc{S}}(i_{\nu,n_\nu})$}
\includegraphics[scale=0.5]{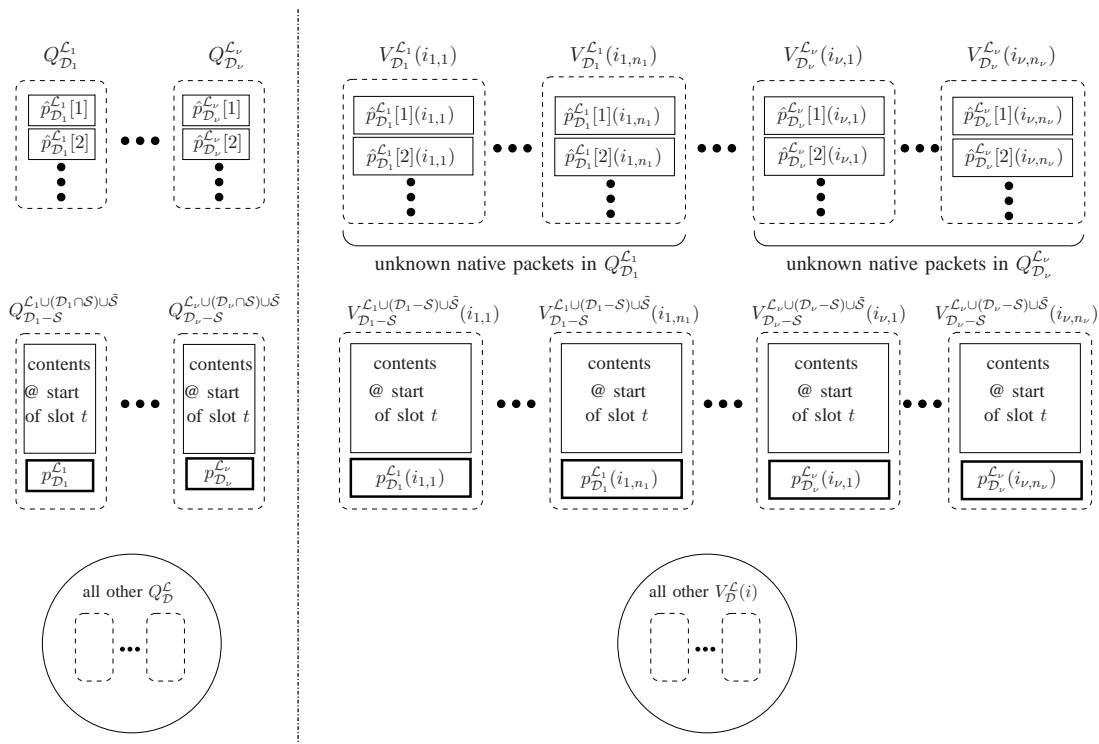}
\caption{Real (LHS) and virtual (RHS) queue contents at end of slot $t$ for case 2.2.1.}
\label{fig:endt_move1}
\end{figure}

$\bullet$ \textbf{BP\ref{enu:LD} at $t+1$: for each bold-edged
$p_{\mc{D}_{k}}^{\mc{L}_{k}}$ moved to $Q_{\mc{D}_{k}-\mc{S}}^{\mc{L}_{k}\cup(\mc{D}_{k}\cap\mc{S})\cup\tilde{\mc{S}}}$,
the set of Destinations for this packet is $\mc{D}_{k}-\mc{S}$
and all users in $Q_{\mc{D}_{k}-\mc{S}}^{\mc{L}_{k}\cup(\mc{D}_{k}\cap\mc{S})\cup\tilde{\mc{S}}}$
are Listeners:} we start with the Listener part. Notice that, since
it holds $\tilde{\mc{S}}\subseteq\mc{S}$, we can write
$\mc{L}_{k}\cup(\mc{D}_{k}\cap\mc{S})\cup\tilde{\mc{S}}=\mc{L}_{k}\cup(\mc{D}_{k}\cap\mc{S})\cup\left[\tilde{\mc{S}}
\cap\mc{L}_{k}^{c}\cap\mc{D}_{k}^{c}\right]$ so that we can examine each of the three sets separately. Any user
$i\in\mc{L}_{k}$ is, by BP\ref{enu:LD} at $t$, a Listener
for $p_{\mc{D}_{k}}^{\mc{L}_{k}}$ and this property also
holds at $t+1$, due to the absorbing property of Listener. Also,
as previously described, any user $i\in\mc{D}_{k}\cap\mc{S}$
decodes at $t+1$ its unknown native packet $q$ contained in $p$.
Since any $i\in\mc{D}_{k}\cap\mc{S}$ also satisfies $i\in\mc{D}_{k}$,
BP\ref{enu:LD} at $t$ implies that $i$ is a Destination for $p_{\mc{D}_{k}}^{\mc{L}_{k}}$,
so that $q$ is contained in $p_{\mc{D}_{k}}^{\mc{L}_{k}}$
and it holds $p_{\mc{D}_{k}}^{\mc{L}_{k}}=q\oplus c$, where
$i$ is Listener for $c$. Since $i$ decodes $q$ at $t+1$, it follows
that $i$ becomes a Listener for $p_{\mc{D}_{k}}^{\mc{L}_{k}}$
at $t+1$. Finally, by definition of $\tilde{\mc{S}},$ any user
$i\in\tilde{\mc{S}}\cap\mc{L}_{k}^{c}\cap\mc{D}_{k}^{c}$
must belong to all $\mc{L}_{r}$ for $r\neq k$ (since $i\in\tilde{\mc{S}}$
implies that $i$ received $p$ and belongs to at least $\nu-1$ of
the Listener sets) so that $i\in\mc{D}_{r}^{c}$ for all $r\neq k$.
Hence, we write $p_{\mc{D}_{k}}^{\mc{L}_{k}}=p\oplus\bigoplus_{r\neq k}p_{\mc{D}_{r}}^{\mc{L}_{r}}$,
where $i\not\in\cup_{r=1}^{\nu}\mc{D}_{r}$, and note that, by
Corollary \ref{cor:Dforall}, $p$ contains no unknown native packets
for any $i\not\in\cup_{r=1}^{\nu}\mc{D}_{r}$. Since $i$ knows
the value of $p$ and is a Listener of $p_{\mc{D}_{r}}^{\mc{L}_{r}}$
(BP\ref{enu:LD} at $t$), we conclude that $i$ is also a Listener
for $p_{\mc{D}_{k}}^{\mc{L}_{k}}$ at $t+1$.

For the Destination part, consider any $i\in\mc{D}_{k}-\mc{S}$,
which implies $i\in\mc{D}_{k}$. By BP\ref{enu:LD} at $t$,
$i$ is a Destination for $p_{\mc{D}_{k}}^{\mc{L}_{k}}$
and the unknown at $t$ packet $q$ for $i$ is still unknown at $t+1$
(since only users in $\mc{S}\cap\left(\cup_{k=1}^{\nu}\mc{D}_{k}\right)$
can decode packets at $t$). Hence, $i$ is still a Destination for
$p_{\mc{D}_{k}}^{\mc{L}_{k}}$ at $t+1$. Conversely, consider
any user $i$ that is a Destination of $p_{\mc{D}_{k}}^{\mc{L}_{k}}$
at $t+1$. This implies that $i\not\in\mc{D}_{k}\cap\mc{S}$
(otherwise, $i$ would have decoded its unknown native packet contained
in $p_{\mc{D}_{k}}^{\mc{L}_{k}}$ and would be a Listener
for it). Additionally, since the XOR decomposition of $p_{\mc{D}_{k}}^{\mc{L}_{k}}$
did not change between $t$ and $t+1$, it follows that $i$ is also
a Destination for $p_{\mc{D}_{k}}^{\mc{L}_{k}}$ at $t$,
so that BP\ref{enu:LD} at $t$ implies that $i\in\mc{D}_{k}$.
Hence, $i\in\mc{D}_{k}\cap\left(\mc{D}_{k}\cap\mc{S}\right)^{c}=\mc{D}_{k}-\mc{S}$,
which is the desired result. 

$\bullet$\textbf{ BP\ref{enu:onlyD} at $t+1$: if the non-bold-edged
packet $p_{\mc{D}_{k}}^{\mc{L}_{k}}$ stored in $Q_{\mc{D}_{k}-\mc{S}}^{\mc{L}_{k}\cup(\mc{D}_{k}\cap\mc{S})\cup\tilde{\mc{S}}}$
contains an unknown native packet for some user $i$, then $i$ is
a Destination for $p_{\mc{D}_{k}}^{\mc{L}_{k}}$:} let $p_{\mc{D}_{k}}^{\mc{L}_{k}}$
contain an unknown native packet $q$ for some user $i$ at $t+1$.
Then, since the XOR decomposition of $p_{\mc{D}_{k}}^{\mc{L}_{k}}$
did not change between $t$ and $t+1$, we conclude that $q$ was
also unknown at $t$, so that BP\ref{enu:onlyD} at $t$ implies that
$i$ was a Destination for $p_{\mc{D}_{k}}^{\mc{L}_{k}}$
at $t$ and (by BP\ref{enu:LD} at $t$) $i\in\mc{D}_{k}$. Also,
it holds $i\not\in\mc{D}_{k}\cap\mc{S}$ (otherwise $q$
would be decoded by $i$ at $t+1$, due to Corollary \ref{cor:Dforall}),
so that $i\in\mc{D}_{k}-\mc{S}$. Hence, by the previously
proved BP\ref{enu:LD} at $t+1$, $i$ is a Destination for $p_{\mc{D}_{k}}^{\mc{L}_{k}}$.

Finally, BP\ref{enu:nolost} at $t+1$ follows immediately from
BP\ref{enu:nolost} at $t$, since any unknown native packet at $t$ is
contained in exactly one XOR packet $p^{\mc{L}}_{\mc{D}}$ (stored in a
real queue $Q^{\mc{L}}_{\mc{D}}$) and, under the RPM,
$p^{\mc{L}}_{\mc{D}}$ either exits the real network or is moved
(\textit{not} copied) to another real queue
$Q^{\mc{L}^\prime}_{\mc{D}^\prime}$ at $t+1$.

\textbf{Case 2.2.2A}: it holds $\cup_{k=1}^{\nu}\mc{D}_{k}-\mc{S}\neq\emptyset$,
$\hat{\mc{S}}=\mc{S}-\cup_{k=1}^{\nu}\left(\mc{L}_{k}\cup\mc{D}_{k}\right)\neq\emptyset$
and $\left|\left(\cap_{k=1}^{\nu}\mc{L}_{k}\cup\mc{S}\right)\cup\left(\cup_{k=1}^{\nu}\mc{D}_{k}-\mc{S}\right)\right|>
\max_{k=1,\ldots,\nu}\left|\mc{L}_{k}\cup\mc{D}_{k}\right|$.
As in the previous case, all users in $\mc{S}\cap\left(\cup_{k=1}^{\nu}\mc{D}_{k}\right)$
decode their unknown native packets. RPM now requires that all bold-edged
packets $p_{\mc{D}_{k}}^{\mc{L}_{k}}$ in Fig.~\ref{fig:begint}
exit the network and the transmitted packet $p$ is moved to queue
$Q_{\cup_{k=1}^{\nu}\mc{D}_{k}-\mc{S}}^{\cap_{k=1}^{\nu}\mc{L}_{k}\cup\mc{S}}$. 
Also, for $k=1,\ldots,\nu$ and $i\in\mc{D}_{k}-\mc{S}$,
all bold-edged native tokens $p_{\mc{D}_{k}}^{\mc{L}_{k}}(i)$
are moved to $V_{\cup_{k=1}^{\nu}\mc{D}_{k}-\mc{S}}^{\cap_{k=1}^{\nu}\mc{L}_{k}\cup\mc{S}}(i)$ 
(this is captured by the fact that $K^{\mc{L}_k}_{\mc{D}_k}(i)$,
$K^{\cap_{k=1}^\nu \mc{L}_k \cup \mc{S}}_{\cup_{k=1}^\nu \mc{D}_k-\mc{S}}(i)$ are
reduced and increased by 1, respectively).
Hence, the network status at $t+1$ is shown in Fig.~\ref{fig:endt_move2}.
We now need to show the following:

\begin{figure}[t]
\centering
\psfrag{QL_D}[][][0.6]{all other $Q^{\mc{L}}_{\mc{D}}$}
\psfrag{QL1_D1}[][][0.7]{$Q^{\mc{L}_1}_{\mc{D}_1}$}
\psfrag{QLm_Dm}[][][0.7]{$Q^{\mc{L}_\nu}_{\mc{D}_\nu}$} 
\psfrag{QL1_D1(1)}[][][0.7]{$V^{\mc{L}_1}_{\mc{D}_1}(i_{1,1})$} 
\psfrag{QL1_D1(n)}[][][0.7]{$V^{\mc{L}_1}_{\mc{D}_1}(i_{1,n_1})$} 
\psfrag{QLm_Dm(1)}[][][0.7]{$V^{\mc{L}_\nu}_{\mc{D}_\nu}(i_{\nu,1})$} 
\psfrag{QLm_Dm(n)}[][][0.7]{$V^{\mc{L}_\nu}_{\mc{D}_\nu}(i_{\nu,n_\nu})$}
\psfrag{pL1_D1}[][][0.65]{$p$} 
\psfrag{phL1_D11}[][][0.65]{$\hat{p}^{\mc{L}_1}_{\mc{D}_1}[1]$} 
\psfrag{phLm_Dm1}[][][0.65]{$\hat{p}^{\mc{L}_\nu}_{\mc{D}_\nu}[1]$} 
\psfrag{phL1_D12}[][][0.65]{$\hat{p}^{\mc{L}_1}_{\mc{D}_1}[2]$}
\psfrag{phLm_Dm2}[][][0.65]{$\hat{p}^{\mc{L}_\nu}_{\mc{D}_\nu}[2]$}
\psfrag{pL1_D1_1}[][][0.65]{$p^{\mc{L}_1}_{\mc{D}_1}(i_{1,1})$} 
\psfrag{phL1_D1_1}[][][0.65]{$\hat{p}^{\mc{L}_1}_{\mc{D}_1}[1](i_{1,1})$} 
\psfrag{pL1_D1_n}[][][0.65]{$p^{\mc{L}_1}_{\mc{D}_1}(i_{1,n_1})$} 
\psfrag{phL1_D1_2}[][][0.65]{$\hat{p}^{\mc{L}_1}_{\mc{D}_1}[2](i_{1,1})$} 
\psfrag{phL1_D1_n1}[][][0.65]{$\hat{p}^{\mc{L}_1}_{\mc{D}_1}[1](i_{1,n_1})$} 
\psfrag{phL1_D1_n2}[][][0.65]{$\hat{p}^{\mc{L}_1}_{\mc{D}_1}[2](i_{1,n_1})$} 
\psfrag{pLm_Dm_1}[][][0.65]{$p^{\mc{L}_\nu}_{\mc{D}_\nu}(i_{\nu,1})$} 
\psfrag{phLm_Dm_1}[][][0.65]{$\hat{p}^{\mc{L}_\nu}_{\mc{D}_\nu}[1](i_{\nu,1})$} 
\psfrag{phLm_Dm_2}[][][0.65]{$\hat{p}^{\mc{L}_\nu}_{\mc{D}_\nu}[2](i_{\nu,1})$} 
\psfrag{pLm_Dm_n}[][][0.65]{$p^{\mc{L}_\nu}_{\mc{D}_\nu}(i_{\nu,n_\nu})$} 
\psfrag{phLm_Dm_n1}[][][0.65]{$\hat{p}^{\mc{L}_\nu}_{\mc{D}_\nu}[1](i_{\nu,n_\nu})$} 
\psfrag{phLm_Dm_n2}[][][0.65]{$\hat{p}^{\mc{L}_\nu}_{\mc{D}_\nu}[2](i_{\nu,n_\nu})$} 
\psfrag{native packets for QL1_D1}[][][0.7]{unknown native packets in $Q^{\mc{L}_1}_{\mc{D}_1}$} 
\psfrag{native packets for QLm_Dm}[][][0.7]{unknown native packets in $Q^{\mc{L}_\nu}_{\mc{D}_\nu}$} 
\psfrag{QL_D(i)}[][][0.6]{all other $V^{\mc{L}}_{\mc{D}}(i)$}
\psfrag{QL1_D1e}[][][0.65]{$Q^{\mc{L}_1\cup (\mc{D}_1\cap \mc{S})\cup \tilde{\mc{S}}}_{\mc{D}_1-\mc{S}}$} 
\psfrag{QLm_Dme}[][][0.65]{$Q^{\mc{L}_\nu\cup (\mc{D}_\nu\cap \mc{S}) \cup \tilde{\mc{S}}}_{\mc{D}_\nu-\mc{S}}$} 
\psfrag{contents}[][][0.65]{contents} 
\psfrag{begin}[][][0.65]{@ start} 
\psfrag{slot}[][][0.65]{of slot $t$} 
\psfrag{QL1_D1e}[][][0.65]{$Q^{\cap_{k=1}^\nu \mc{L}_k\cup \mc{S}}_{\cup_{k=1}^\nu \mc{D}_k -\mc{S}}$}
\psfrag{QL1_D1e(1)}[][][0.65]{$V^{\cap_{k=1}^\nu \mc{L}_k\cup \mc{S}}_{\cup_{k=1}^\nu \mc{D}_k -\mc{S}}(i_{1,1})$}
\psfrag{QL1_D1e(n)}[][][0.65]{$V^{\cap_{k=1}^\nu \mc{L}_k\cup \mc{S}}_{\cup_{k=1}^\nu \mc{D}_k -\mc{S}}(i_{1,n_1})$}
\psfrag{QLm_Dme(1)}[][][0.65]{$V^{\cap_{k=1}^\nu \mc{L}_k\cup \mc{S}}_{\cup_{k=1}^\nu \mc{D}_k -\mc{S}}(i_{\nu,1})$}
\psfrag{QLm_Dme(n)}[][][0.65]{$V^{\cap_{k=1}^\nu \mc{L}_k\cup \mc{S}}_{\cup_{k=1}^\nu \mc{D}_k -\mc{S}}(i_{\nu,n_\nu})$}
\includegraphics[scale=0.5]{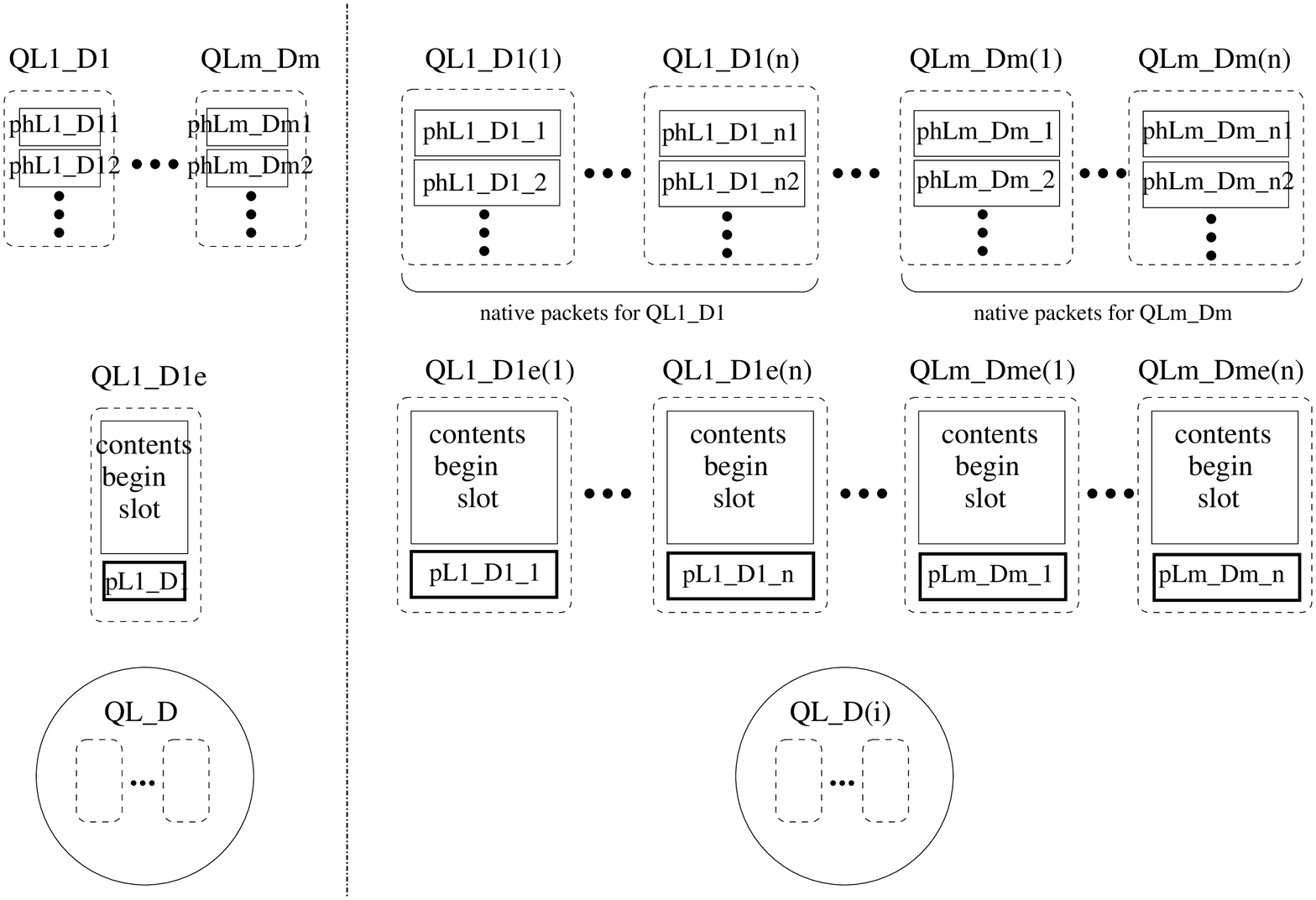}
\caption{Real (LHS) and virtual (RHS) queue contents at end of slot $t$ for case 2.2.2A.}
\label{fig:endt_move2}
\end{figure}

$\bullet$ \textbf{BP\ref{enu:LD} at $t+1$: for the packet $p$
moved to $Q_{\cup_{k=1}^{\nu}\mc{D}_{k}-\mc{S}}^{\cap_{k=1}^{\nu}\mc{L}_{k}\cup\mc{S}}$,
the set of Destinations for this packet is $\cup_{k=1}^{\nu}\mc{D}_{k}-\mc{S}$
and all users in $\cap_{k=1}^{\nu}\mc{L}_{k}\cup\mc{S}$
are Listeners: }for the Destination part, consider any $i\in\cup_{k=1}^{\nu}\mc{D}_{k}-\mc{S}$.
Then, there exists some $k^{\ast}\in\left\{ 1,\ldots,\nu\right\} $
such that $i\in\mc{D}_{k^{\ast}}-\mc{S}$ and, by the BCR,
$i\in\mc{L}_{r}$ for all $r\neq k^{\ast}$. By BP\ref{enu:LD}
at $t$, $i$ is a Destination for $p_{\mc{D}_{k^{\ast}}}^{\mc{L}_{k^{\ast}}}$
and Listener for all $p_{\mc{D}_{r}}^{\mc{L}_{r}}$, $r\neq k^{\ast}$.
Hence, we can write $p_{\mc{D}_{k^{\ast}}}^{\mc{L}_{k^{\ast}}}=q\oplus c$,
where $q$ is an unknown native packet for $i$
at $t$ and $i$ is Listener for $c$, so that $p=q\oplus c\oplus\bigoplus_{r\neq k^{\ast}}p_{\mc{D}_{r}}^{\mc{L}_{r}}$.
Due to Corollary \ref{cor:Dforall}, $q$ is not decoded by $i$ so
that it is still unknown at $t+1$, which implies that $i$ is a Destination
for $p$ at $t+1$ . Conversely, let $i$ be a Destination of $p$
at $t+1$, so that $p$ contains an unknown native packet $q$ for
$i$ at $t+1$. Obviously, $q$ is contained in some packet $p_{\mc{D}_{k^{\ast}}}^{\mc{L}_{k^{\ast}}}$
and, by BP\ref{enu:onlyD} at $t$, $i$ is Destination for $p_{\mc{D}_{k^{\ast}}}^{\mc{L}_{k^{\ast}}}$
so that $i\in\mc{D}_{k^{\ast}}$ (by BP\ref{enu:LD} at $t$).
It must then hold $i\not\in\mc{S}$ (otherwise, $i$ would be
able to decode $q$ by Corollary \ref{cor:Dforall}) so that $i\in\mc{D}_{k^{\ast}}-\mc{S}$,
which implies $i\in\cup_{k=1}^{\nu}\mc{D}_{k}-\mc{S}$.

For the Listener part, consider any $i\in\cap_{k=1}^{\nu}\mc{L}_{k}\cup\mc{S}$.
If $i\in\cap_{k=1}^{\nu}\mc{L}_{k}$, then by BP\ref{enu:LD}
at $t$, $i$ is a Listener for each packet $p_{\mc{D}_{k}}^{\mc{L}_{k}}$
so that it is also a Listener for $p$ at $t$. The absorbing property
of Listener then implies that $i$ is a Listener for $p$ at $t+1$.
If $i\in\mc{S}-\cap_{k=1}^{\nu}\mc{L}_{k}=\mc{S}\cap\left(\cup_{k=1}^{\nu}\mc{L}_{k}^{c}\right)$,
then it suffices to show that $p$ contains no unknown native packet
for this $i$ at $t+1$ (which immediately implies that $i$ is a
Listener for $p$). Since $i\in\mc{S}\cap\left(\cup_{k=1}^{\nu}\mc{L}_{k}^{c}\right)$,
there exists some $k^{\ast}=1,\ldots,\nu$ such that $i\in\mc{S}\cap\mc{L}_{k^{\ast}}^{c}$
and the BCR implies, for all $r\neq k^{\ast}$,
$\mc{L}_{k^{\ast}}\supseteq\mc{D}_{r}\Rightarrow\mc{L}_{k^{\ast}}^{c}\subseteq\mc{D}_{r}^{c}$
so that $i\in\mc{S}\cap\mc{L}_{k^{\ast}}^{c}\cap\mc{D}_{r}^{c}$
for all $r\neq k^{\ast}$. By BP\ref{enu:onlyD}, BP\ref{enu:LD}
at $t$, each $p_{\mc{D}_{r}}^{\mc{L}_{r}}$, for $r\neq k^{\ast}$,
contains no unknown native packet for this $i$ at $t$, and therefore
at $t+1$ as well. We now distinguish two cases: a) it holds $i\not\in\mc{D}_{k^{\ast}}$
so that, by BP\ref{enu:onlyD}, BP\ref{enu:LD} at $t$, $p_{\mc{D}_{k^{\ast}}}^{\mc{L}_{k^{\ast}}}$
contains no unknown native packet for $i$ at $t$, as well as at
$t+1$. Hence, $p$ contains no unknown native packet for $i$, which
is the desired result b) if $i\in\mc{D}_{k^{\ast}}$, then since
it also holds $i\in\mc{S}$, Corollary \ref{cor:Dforall} implies
that $i$ decodes its unknown native packet contained in $p_{\mc{D}_{k^{\ast}}}^{\mc{L}_{k^{\ast}}}$
at $t+1$. Hence, $p$ again contains no unknown native packet for
$i$ at $t+1$ and the Listener part is complete.

$\bullet$ \textbf{BP\ref{enu:onlyD} at $t+1$: if $p$ stored in
$Q_{\cup_{k=1}^{\nu}\mc{D}_{k}-\mc{S}}^{\cap_{k=1}^{\nu}\mc{L}_{k}\cup\mc{S}}$
contains an unknown native packet for some user $i$, then $i$ is
a Destination for $p$: }let $p$ contain an unknown native packet
$q$ for user $i$ at $t+1$. Clearly, $q$ is contained in one of
the $p_{\mc{D}_{k}}^{\mc{L}_{k}}$ that comprise $p$ and
was also unknown at $t$. BP\ref{enu:onlyD} at $t$ now implies that
$i$ is a Destination for $p_{\mc{D}_{k}}^{\mc{L}_{k}}$
and $i\in\mc{D}_{k}$ (by BP\ref{enu:LD} at $t$). Since $q$
is unknown at $t+1$ and $i\in\mc{D}_{k}$, Corollary \ref{cor:Dforall}
now implies that $i\not\in\mc{S}$, whence we conclude that $i\in\cup_{k=1}^{\nu}\mc{D}_{k}-\mc{S}$.
Since $p$ is stored in $Q_{\cup_{k=1}^{\nu}\mc{D}_{k}-\mc{S}}^{\cap_{k=1}^{\nu}\mc{L}_{k}\cup\mc{S}}$
at $t+1$, BP\ref{enu:LD} at $t+1$ now implies that $i$ is a Destination
for $p$.

As in Case 2.2.1, BP\ref{enu:nolost} at $t+1$ follows
from BP\ref{enu:nolost} at $t$ and the fact that no packet copying
is performed.

\textbf{Case 2.2.2B}: it holds $\cup_{k=1}^{\nu}\mc{D}_{k}-\mc{S}\neq\emptyset$,
$\hat{\mc{S}}=\mc{S}-\cup_{k=1}^{\nu}\left(\mc{L}_{k}\cup\mc{D}_{k}\right)\neq\emptyset$
and $\left|\left(\cap_{k=1}^{\nu}\mc{L}_{k}\cup\mc{S}\right)\cup\left(\cup_{k=1}^{\nu}\mc{D}_{k}-\mc{S}\right)\right|
\leq\max_{k=1,\ldots,\nu}\left|\mc{L}_{k}\cup\mc{D}_{k}\right|$. We further distinguish two subcases:
\begin{itemize}
\item if $\mc{S}\cap\left(\cup_{k=1}^{\nu}(\mc{L}_{k}\cup\mc{D}_{k})\right)=\emptyset$,
which implies $\mc{S}\cap\left(\cup_{k=1}^{\nu}\mc{D}_{k}\right)=\emptyset$,
then no native packets are decoded at $t+1$ (due to Corollary \ref{cor:Dforall})
and no packet movement takes place under the RPM. Hence, the network
status at $t+1$ is exactly the same as in $t$ so that BP holds trivially
at $t+1$.

\item if $\mc{S}\cap\left(\cup_{k=1}^{\nu}(\mc{L}_{k}\cup\mc{D}_{k})\right)\neq\emptyset$,
we set $\mc{S}\leftarrow\mc{S}\cap\left(\cup_{k=1}^{\nu}(\mc{L}_{k}\cup\mc{D}_{k})\right)$
and RPM reverts to Case 2.2.1, which has already
been shown to satisfy BP at $t+1$.
\end{itemize}
Since all possible cases under RPM have been examined and shown to
satisfy BP at $t+1$, the proof is complete.

\subsection{2-user stability region through Fourier-Motzkin elimination} \label{sec:Proof-FM}

The Fourier-Motzkin algorithm for eliminating a variable in a set of inequalities, consists of
splitting the set of inequalities into 3 sets $\mc{K}_{FM}$, $\mc{L}_{FM}$, $\mc{U}_{FM}$, where
the first set has inequalities which do not contain the variable to be eliminated, and the second 
and third sets have inequalities which provide, respectively, lower and upper bounds for the 
variable to be eliminated. We then combine the equations in $\mc{L}_{FM}$, $\mc{U}_{FM}$ to get
a new set of inequalities. This can be repeated for each variable to be eliminated. We provide
below a step-by-step application of Fourier-Motzkin to eliminate $\phi^{2,1}_{1,2},\phi^1_2,\phi^2_1,
\phi_2,\phi_1$ in this order.

Initial set of equations:
\beq \begin{split}
\mc{K}_{FM} &= \left\{ \lambda_1 \leq (1-\epsilon_{12}) \phi_1, \, \lambda_2 \leq (1-\epsilon_{12}) \phi_2,
  \, \phi_1\geq 0, \, \phi_2 \geq 0, \, \phi^1_2 \geq 0, \, \phi^2_1 \geq 0 \right\} , \\
\mc{L}_{FM} &= \left\{ 0\leq \phi^{2,1}_{1,2}, \, 
  \frac{\epsilon_1-\epsilon_{12}}{1-\epsilon_1} \phi_1-\phi^2_1 \leq \phi^{2,1}_{1,2}, \, 
  \frac{\epsilon_2-\epsilon_{12}}{1-\epsilon_2}\phi_2-\phi^1_2 \leq \phi^{2,1}_{1,2} \right\} , \\
\mc{U}_{FM} &= \left\{ \phi^{2,1}_{1,2} \leq 1-\phi_1-\phi_2 -\phi^1_2-\phi^2_1 \right\} .
\end{split} \eeq

New inequalities after elimination of $\phi^{2,1}_{1,2}$: 
\beq \begin{split}
& 0\leq 1-\phi_1-\phi_2 -\phi^1_2-\phi^2_1, \\
& \frac{\epsilon_1-\epsilon_{12}}{1-\epsilon_1} \phi_1-\phi^2_1 \leq 1-\phi_1-\phi_2 -\phi^1_2-\phi^2_1, \\
& \frac{\epsilon_2-\epsilon_{12}}{1-\epsilon_2} \phi_2-\phi^1_2 \leq 1-\phi_1-\phi_2 -\phi^1_2-\phi^2_1 ,
\end{split} \eeq
so that we proceed to recast the equations in terms of $\phi^1_2$ to get
\beq \begin{split}
\mc{K}_{FM} &= \left\{ \lambda_1 \leq (1-\epsilon_{12}) \phi_1, \, \lambda_2 \leq (1-\epsilon_{12}) \phi_2,
  \, \phi^2_1 \leq 1-\phi_1 -\frac{1-\epsilon_{12}}{1-\epsilon_2} \phi_2, \, \phi_1\geq 0, \, \phi_2 \geq 0, \, 
  \phi^2_1 \geq 0 \right\} , \\
\mc{L}_{FM} &= \left\{ 0\leq \phi^1_2 \right\} , \\  
\mc{U}_{FM} &= \left\{ \phi^1_2 \leq 1-\phi_2 -\frac{1-\epsilon_{12}}{1-\epsilon_1} \phi_1, \,
  \phi^1_2 \leq 1-\phi_1-\phi_2-\phi^2_1 \right\} ,
\end{split} \eeq
and get the new equations
\beq \begin{split}
& 0 \leq 1-\phi_2 -\frac{1-\epsilon_{12}}{1-\epsilon_1} \phi_1 , \\
& 0 \leq 1-\phi_1-\phi_2-\phi^2_1 ,
\end{split} \eeq
and we can recast these in terms of $\phi^2_1$ to get
\beq \begin{split}
\mc{K}_{FM} &= \left\{ \lambda_1 \leq (1-\epsilon_{12}) \phi_1, \, \lambda_2 \leq (1-\epsilon_{12}) \phi_2,
  \, \phi_1 \geq 0, \, \phi_2 \geq 0, \, 0 \leq 1-\phi_2 -\frac{1-\epsilon_{12}}{1-\epsilon_1} \phi_1 \right\} , \\
  \mc{L}_{FM} &= \left\{ 0\leq \phi^1_2 \right\} , \\
\mc{U}_{FM} &= \left\{ \phi^2_1 \leq 1-\phi_1-\frac{1-\epsilon_{12}}{1-\epsilon_2}\phi_2 , \,
  \phi^2_1 \leq 1-\phi_1-\phi_2 \right\} .
\end{split} \eeq

Eliminating $\phi^1_2$ yields the new equations
\beq \begin{split}
& 0 \leq 1-\phi_1-\frac{1-\epsilon_{12}}{1-\epsilon_2}\phi_2 , \\
& 0 \leq 1-\phi_1-\phi_2 ,
\end{split} \eeq
and recasting the equations in terms of $\phi_2$ yields
\beq \begin{split}
\mc{K}_{FM} &= \left\{ \lambda_1 \leq (1-\epsilon_{12}) \phi_1, \, \phi_1 \geq 0 \right\} , \\
\mc{L}_{FM} &= \left\{ \frac{\lambda_2}{1-\epsilon_{12}} \leq \phi_2 \right\} , \\
\mc{U}_{FM} &= \left\{ \phi_2 \leq \frac{1-\epsilon_2}{1-\epsilon_{12}} (1-\phi_1), \, \phi_2 \leq 1-\phi_1,
  \, \phi_2 \leq 1-\frac{1-\epsilon_{12}}{1-\epsilon_1} \phi_1 \right\} ,
\end{split} \eeq
whence we get the new equations
\beq \begin{split}
& \frac{\lambda_2}{1-\epsilon_{12}} \leq \frac{1-\epsilon_2}{1-\epsilon_{12}} (1-\phi_1) , \\
& \frac{\lambda_2}{1-\epsilon_{12}} \leq 1-\phi_1 , \\
& \frac{\lambda_2}{1-\epsilon_{12}} \leq 1-\frac{1-\epsilon_{12}}{1-\epsilon_1} \phi_1 .
\end{split} \eeq
We now recast in terms of the remaining $\phi_1$ to get
\beq \begin{split}
\mc{L}_{FM} &= \left\{ \frac{\lambda_1}{1-\epsilon_{12}} \leq \phi_1 \right\} , \\
\mc{U}_{FM} &= \left\{ \phi_1 \leq 1-\frac{\lambda_2}{1-\epsilon_2}, \, 
  \phi_1 \leq 1-\frac{\lambda_2}{1-\epsilon_{12}}, \, \phi_1 \leq \frac{1-\epsilon_1}{1-\epsilon_{12}}
  \left( 1-\frac{\lambda_2}{1-\epsilon_{12}} \right) \right\} ,
\end{split} \eeq
and applying the last step yields
\beq \begin{split}
& \frac{\lambda_1}{1-\epsilon_{12}} \leq 1-\frac{\lambda_2}{1-\epsilon_2} \Leftrightarrow 
\frac{\lambda_1}{1-\epsilon_{12}} +\frac{\lambda_2}{1-\epsilon_2} \leq 1 , \\
& \frac{\lambda_1}{1-\epsilon_{12}} \leq 1 -\frac{\lambda_2}{1-\epsilon_{12}} \Leftrightarrow
  \lambda_1+\lambda_2 \leq 1-\epsilon_{12} , \\
& \frac{\lambda_1}{1-\epsilon_{12}} \leq \frac{1-\epsilon_1}{1-\epsilon_{12}} \left(
  1-\frac{\lambda_2}{1-\epsilon_{12}} \right) \Leftrightarrow \frac{\lambda_1}{1-\epsilon_1}+
  \frac{\lambda_2}{1-\epsilon_{12}} \leq 1 .
\end{split} \eeq
Since the middle inequality is dominated by the first and third one, it can be removed and the final result
is the stability region in \cite{Leo_2user}.

\subsection{Proof Of Theorem \ref{thm:UpperBound}}  \label{sec:Proof-Of-Theorem}

We need some preliminary definitions. Define the sets $\mc{N}_{1}=\emptyset$
and $\mc{N}_{i}=\left\{ 1,2,\ldots,i-1\right\} $, for $i\in\mc{N}$
with $i\geq 2$, as well as 
\begin{align*}
\mc{M}_{i} & =\left\{ V_{\mc{D}}^{\mc{L}}\left(i\right):\: i\in\mc{D},\mbox{ and }\mc{L},\mc{D}-\left\{ i\right\} \subseteq\mc{N}_{i}\right\} ,\\
\mc{I}_{i} & =\left\{ I_{\mc{D},\mc{D}_{2},\ldots,\mc{D}_{\nu}}^{\mc{L},\mc{L}_{2},\ldots,\mc{L}_{\nu}}:\: i\in\mc{D},
  \mbox{ and }\mc{L},\mc{D}-\left\{ i\right\} \subseteq\mc{N}_{i}\right\} .
\end{align*}
Notice that $\mc{I}_{i}\cap\mc{I}_{j}=\emptyset$ for $i\neq j$.
This is due to the fact that the existence of a control $I\in\mc{I}_{i}\cap\mc{I}_{j}$
would imply that $i\in\mc{N}_{j}$ as well as $j\in\mc{N}_{i}$,
which is impossible. We also define the set $\mc{M}_{i}^{\mc{N}}$
in the subnetwork consisting of queues (i.e.~each node is a queue,
as described in Section \ref{sec:Stability-region}) as follows: 
\[
\mc{M}_{i}^{\mc{N}}=\left\{ V_{\mc{D}}^{\mc{L}}\left(i\right):\: i\in\mc{D},\mbox{ and }\mc{D},\mc{L}\subseteq\mc{N}\right\} \cup\{d\},
\]
Denote with $\mc{C}_{out}(\mc{M}_{i})$ the set of all outgoing
links in the cut $\left[\mc{M}_{i},\mc{M}_{i}^{\mc{N}}-\mc{M}_{i}\right]$,
i.e. 
\[
\mc{C}_{out}\left(\mc{M}_{i}\right)=\left\{ e=(m,l)\in\mc{E}:\: m\in\mc{M}_{i},\: l\in\mc{M}_{i}^{\mc{N}}-\mc{M}_{i}\right\} ,
\]
while the set $\mc{C}_{in}(\mc{M}_i)$ of incoming links to the cut is
\begin{equation} \label{eq:InCut14}
\mc{C}_{in}\left(\mc{M}_{i}\right)=\left\{ e=(m,l)\in\mc{E}:\: m\in\mc{M}_{i}^{\mc{N}}-\mc{M}_{i},\: l\in\mc{M}_{i}\right\} ,
\end{equation}

To prove Theorem \ref{thm:UpperBound}, it suffices to show that,
under (\ref{eq:param0}) and (\ref{eq:param2})--(\ref{eq:param6}), it holds 
\begin{equation} \label{eq:ToProve}
\sum_{i\in\mc{N}}\frac{\lambda_{i}}{1-\epsilon_{\mc{N}-\mc{N}_{i}}}\leq 1,
\end{equation}
which corresponds to the permutation $\sigma(i)=N-i+1$ in \eq{eq:capacityReg}.
The same argument can then be repeated verbatim for any permutation
$\sigma(i),\: i\in\mc{N}$. Summing (\ref{eq:param0}) over all
$m\in\mc{M}_{i}$ and using (\ref{eq:InRates8}) yields 
\[
\sum_{I\in\mc{I}_{i}}\phi_{I}\sum_{m\in\mc{M}_{i}}\sum_{e=(l,m)\in\mc{E}_{in}^{m}}\mu_{l}\left(I\right)p_{e}^{l}(I)
  +\lambda_{i}\leq\sum_{I\in\mc{I}_{i}}\sum_{m\in\mc{M}_{i}}\sum_{e\in\mc{E}_{out}^{m}}p_{e}^{m}(I)\mu_{m}\left(I\right)
  \phi_{I},\:\forall\, i\in\mc{N},
\]
or, rearranging the terms, 
\begin{equation} \label{eq:IneqFin19}
\lambda_{i}\leq\sum_{I\in\mc{I}_{i}}\sum_{m\in\mc{M}_{i}}\left(\sum_{e\in\mc{E}_{out}^{m}}p_{e}^{m}(I)\mu_{m}\left(I\right)
  -\sum_{e=(l,m)\in\mc{E}_{in}^{m}}\mu_{l}\left(I\right)p_{e}^{l}(I)\right)\phi_{I}.
\end{equation}
But (\ref{eq:InCut14}) and the construction of $\mc{C}_{out}(\mc{M}_{i}),\,\mc{C}_{in}(\mc{M}_{i})$ imply 
\begin{align*}
 & \sum_{m\in\mc{M}_{i}}\left(\sum_{e\in\mc{E}_{out}^{m}}p_{e}^{m}(I)\mu_{m}\left(I\right)-\sum_{e=(l,m)\in\mc{E}_{in}^{m}}
   \mu_{l}\left(I\right)p_{e}^{l}(I)\right)\\
 & =\sum_{e=(l,m)\in\mc{C}_{out}\left(\mc{M}_{i}\right)}\mu_{l}\left(I\right)p_{e}^{l}(I)-\sum_{e=(l,m)\in\mc{C}_{in}
   \left(\mc{M}_{i}\right)}\mu_{l}\left(I\right)p_{e}^{l}(I) \leq \sum_{e=(l,m)\in\mc{C}_{out}\left(\mc{M}_{i}\right)}
   \mu_{l}\left(I\right)p_{e}^{l}(I).
\end{align*}

Also, any control $I^{\mc{L},\mc{L}_2,\ldots,\mc{L}_\nu}_{\mc{D},\mc{D}_2,\ldots,\mc{D}_\nu}\in \mathcal{I}_i$ affects
only one real queue in $\mc{M}_i$ (namely, $V^{\mc{L}}_{\mc{D}}(i)$, since $i\in \mc{D}$ amd BCR is applied) that 
contains packets for $i$. Hence, when $I^{\mc{L},\mc{L}_2,\ldots,\mc{L}_\nu}_{\mc{D},\mc{D}_2,\ldots,\mc{D}_\nu}$
is applied, it holds $\mu_l(I)=1$ for $l=V^{\mc{L}}_{\mc{D}}(i)$ and $\mu_l(I)=0$ for all other queues in $\mc{M}_i$,
which implies
\begin{equation} \label{eq:sumR}
\sum_{e=(l,m)\in\mc{C}_{out}\left(\mc{M}_{i}\right)}\mu_{l}\left(I\right)p_{e}^{l}(I) \leq 1-\epsilon_{\mc{N}-\mc{N}_{i}}.
\end{equation}
This follows from the fact that, under $I_{\mc{D},\mc{D}_{2},\ldots,\mc{D}_{\nu}}^{\mc{L},\mc{L}_{2},\ldots,\mc{L}_{\nu}}$, 
whenever a native packet for user $i$ is transferred from $V^{\mc{L}}_{\mc{D}}(i)$ to one of the queues in queues in 
$\mc{M}^{\mc{N}}_i-\mc{M}_i$, the transmitted packet must have been received by at least one user in $\mc{N}-\mc{N}_i$, 
which occurs with probability $1-\epsilon_{\mc{N}-\mc{N}_i}$.

Hence, (\ref{eq:IneqFin19}) yields through (\ref{eq:sumR}) 
\[
\lambda_{i}\leq\sum_{I\in\mc{I}_{i}}\left(1-\epsilon_{\mc{N}-\mc{N}_{i}}\right)\phi_{I},
\]
and, summing over all $i\in\mc{N}$, we conclude that 
\[
\sum_{i\in\mc{N}}\frac{\lambda_{i}}{1-\epsilon_{\mc{N}-\mc{N}_{i}}}\leq\sum_{i\in\mc{N}}\sum_{I\in\mc{I}_{i}}\phi_{I}.
\]
However, since $\mc{I}_{i}\cap\mc{I}_{j}=\emptyset$ for
all $i\neq j$, it holds $\sum_{i\in\mc{N}}\sum_{I\in\mc{I}_{i}}\phi_{I}\leq\sum_{I\in\mc{I}}\phi_{I}\leq1$
and (\ref{eq:ToProve}) is proved.

\subsection{Closed form expressions for controls $\phi$ for 4 users and iid erasures} \label{app:closed-form}

Performing the algebra in (\ref{eq:level2-3}), (\ref{eq:level3-1a}), (\ref{eq:level3-final}),
(\ref{eq:level4-2}), (\ref{eq:level4-6}), (\ref{eq:all}) through Maple yields
\begin{equation} \label{eq:first}
\phi^{j,i}_{i,j}= \frac{\epsilon^3 (1-\epsilon)}{(1-\epsilon^3)(1-\epsilon^4)} \, \lambda_i \quad \mbox{for } i<j ,
\end{equation}

\begin{equation}
\phi^{k,ij}_{ij,k}= \frac{\epsilon^6 (1-\epsilon)^2}{(1-\epsilon^3)^2 (1-\epsilon^4)} \, \lambda_i \quad \mbox{for } i<j<k ,
\end{equation}

\begin{equation}
\phi^{jk,ik,ij}_{i,j,k}= \frac{\epsilon^2}{(1-\epsilon^4) (1+\epsilon+\epsilon^2)^2} \, \left(
  (1-\epsilon^4) \lambda_i + \epsilon^2 \lambda_i -\epsilon^4 \lambda_j \right) \quad \mbox{for } i<j<k ,
\end{equation}

\begin{equation}
\phi^{l,ijk}_{ijk,l}= \frac{\epsilon^5 (1-\epsilon+\epsilon^2)}{(1-\epsilon^4) (1+\epsilon+\epsilon^2)^2} \, 
  \lambda_i \quad \mbox{for } i<j<k<l ,
\end{equation}

\begin{equation} \label{eq:double}
\phi^{kl,ij}_{ij,kl}= \frac{\epsilon^4 (2-\epsilon+2\epsilon^2-\epsilon^4)}{(1-\epsilon^4)(1+\epsilon)(1+\epsilon+\epsilon^2)^2} 
  \, \lambda_i  \quad \mbox{for } i=1, \; j>1, \; k<l ,
\end{equation}

\begin{equation} \begin{split} \label{eq:toprove}
\phi^{234,134,124,123}_{1,2,3,4}= & \left[ \lambda_1 (1+\epsilon+3\epsilon^2-2\epsilon^3+4\epsilon^4-
  3\epsilon^5+\epsilon^6+\epsilon^7) -\lambda_2 (\epsilon^4+\epsilon^7) \right] \\
& \times \frac{\epsilon}{(1-\epsilon^4) (1+\epsilon) (1+\epsilon+\epsilon^2)^2} .
\end{split} \end{equation}

The non-negativity of (\ref{eq:first})--(\ref{eq:double}) is obvious (since $i<j$ implies $\lambda_i\geq \lambda_j$) 
while for (\ref{eq:toprove}) we observe that the coefficient of $\lambda_1$ in the RHS of (\ref{eq:toprove}) is 
non-negative, which implies
\begin{equation} \begin{split}
& \lambda_1 (1+\epsilon+3\epsilon^2-2\epsilon^3+4\epsilon^4-3\epsilon^5+\epsilon^6+\epsilon^7) -\lambda_2 (\epsilon^4+\epsilon^7) \\
& \geq \lambda_2 (1+\epsilon+3\epsilon^2-2\epsilon^3+4\epsilon^4-3\epsilon^5+\epsilon^6+\epsilon^7) -\lambda_2 (\epsilon^4+\epsilon^7) \\
& \geq \lambda_2 (1+\epsilon+3\epsilon^2-2\epsilon^3+3\epsilon^4-3\epsilon^5+\epsilon^6) \geq 0 ,
\end{split} \end{equation}
whence the non-negativity of (\ref{eq:toprove}) follows immediately.

\subsection{Proof Of Theorem \ref{thm:outer_bound_ext}}

We first need to establish some notation and prove a few intermediate
results. We consider the ``extended'' broadcast erasure channel
(BEC), where the transmitter has the option of not transmitting in
a given slot (as opposed to the ``standard'' BEC that appears
in the literature). This is equivalent to considering that the transmitter
sends in this slot a special (null) symbol, denoted as $\nulls$.
Hence, in information theoretic terms, given a standard point-to-point
BEC with an input alphabet of $\mc{X}$ and output alphabet of
$\mc{Y}=\mc{X}\cup\{*\}$, where $*$ denotes an erasure,
the extended point-to-point BEC has input alphabet $\mc{X}^{\prime}=\mc{X}\cup\{\nulls\}$
and output alphabet $\mc{Y}^{\prime}=\mc{X}^{\prime}\cup\{*\}=\mc{X}\cup\{*,\nulls\}$.
Since we consider feedback, we assume that, if the transmitter sends
symbol $\nulls$, \textit{all} users send $\nulls$ as feedback back
to the transmitter. Hence, at slot $l$, each user can send feedback
$Z\in\{ACK,NACK,\nulls\}$ to the transmitter, where $ACK$ (resp.~$NACK$)
denotes a successful reception (resp.~erasure) of a non-null symbol,
while $\nulls$ denotes a null symbol transmission (and reception).

The $N$ user version of the extended BEC follows from a simple ``vectorization''
procedure. Specifically, let $\mc{N}=\{1,\ldots,N\}$ be the
set of $N$ users and denote with $W_{i}$ the message for user $i\in\mc{N}$.
The transmitted symbol at slot $l$ is denoted as $X(l)$ (with $X(l)\in\mc{X}^{\prime}$)
and we also introduce the shortcut notation $X^{l}\stackrel{\vartriangle}{=}(X(1),\ldots,X(l))$.
Furthermore, let $Y_{i}(l)\in\mc{Y}^{\prime}$ be the symbol
received by user $i$ at slot $l$, while $Z_{i}(l)\in\{ACK,NACK,\nulls\}$
is the feedback sent by user $i$ to the transmitter at slot $l$.
We can also define an auxiliary random variable $\hat{Z}_{i}(l)\in\{ACK,NACK\}$
that is independent of $X(l)$ and all previously generated random
variables (up to slot $l$) so that it holds 
\[
Z_{i}(l)=\left\{ \begin{array}{l@{\quad}l}
\hat{Z}_{i}(l) & \mbox{if }X(l)\neq\nulls,\\
\nulls & \mbox{if }X(l)=\nulls,
\end{array}\right.
\]
Notice that, for any $z\neq\nulls$, the events $\{Z_{i}(l)=z\}$
and $\{\hat{Z}_{i}(l)=z,F(l)=1\}$ are identical. We now introduce
the following ``vectorized'' entities 
\begin{gather*}
W_{[1,j]}=(W_{1},\ldots,W_{j}),\\
Y_{i}^{l}=(Y_{i}(1),\ldots,Y_{i}(l)),\\
\vec{Y}_{[1,j]}(l)=(Y_{1}(l),\ldots,Y_{j}(l)),\quad\vec{Y}_{[1,j]}^{l}=(\vec{Y}_{[1,j]}(1),\ldots,\vec{Y}_{[1,j]}(l)),\\
\vec{Z}_{[1,j]}(l)=(Z_{1}(l),\ldots,Z_{j}(l)),\quad\vec{Z}_{[1,j]}^{l}=(\vec{Z}_{[1,j]}(1),\ldots,\vec{Z}_{[1,j]}(l)),\\
\hat{\vec{Z}}_{[1,j]}(l)=(\hat{\vec{Z}}_{1}(l),\ldots,\hat{\vec{Z}}_{j}(l)),
\end{gather*}
and use the shortcut $\vec{Y}=\vec{Y}_{[1,N]}$, $\vec{Y}^{l}=\vec{Y}_{[1,N]}^{l}$
(with similar interpretation for $\vec{Z}$, $\vec{Z}^{l}$).

The subsequent analysis closely follows the approach in \cite{LaCh},
with some necessary variations due to the fact that $\vec{Z}(l)$
are $X(l)$ are not independent. The following Lemma can be proved
by straightforward manipulations of information measures. 
\begin{lemma}  \label{lem:prelim}
Let $A,B,C,D$ be discrete random variables. The following identities hold. 
\begin{enumerate}
\item Conditioning can be added to either part of mutual information: \label{it1}
\[
I(A;B|C,D)=I(A,C;B|C,D)=I(A;B,C|C,D)=I(A,C;B,C|C,D)
\]

\item Let $B$ be independent of the joint ensemble $(C,D)$. It then holds
$I(A,B;C|D)=I(A;C|B,D)$. 

\item Let $D$ be independent of the joint ensemble $(A,B,C)$. It then
holds $I(A;B|C,D)=I(A;B|C)$. 

\item Conditioning can be augmented by redundant condition, i.e.~if the
event $\{B=b\}$ implies $\{C=c_{b}\}$, it then holds $H(A|B,D)=H(A|B,C,D)$. 

\item It holds $I(A;B|C)=I(A;B|C,D)+I(A;D|C)-I(A;D|B,C)$. \label{it5} 
\end{enumerate}
\end{lemma}
We now consider an arbitrary code $\frak{C}$ for the extended BEC
with feedback (see \cite{GPTL} for a detailed description of encoding
and decoding functions of $\frak{C}$) and denote $\pi(l)=\Pr(X(l)\neq\nulls)$
and $F(l)=\mathbb{I}[X(l)\neq\nulls]$. The following results, whose
proofs can be found, respectively, in sections \ref{sub:Proof-of-Lem1},
\ref{sub:Proof-of-Lem2} of the Appendix, will be used. 

\begin{lemma} \label{lem:lem1}
For any rate $\vec{R}=(R_{1},\ldots,R_{N})$ that
is achievable under $\frak{C}$, and for any $j\in\mc{N}$, it
holds
\[
n\sum_{k=1}^{j}R_{k}\leq\sum_{l=1}^{n}\left[h(\pi(l))+(1-\pi(l))(1-\epsilon_{\{1,\ldots,j\}})I(W_{[1,j]};X(l)|
  \vec{Y}_{[1,j]}^{l-1},\vec{Z}^{l-1},F(l)=1)\right]+o(n),
\]
where $h(\cdot)$ is Shannon's entropy function. 
\end{lemma}
 
\begin{lemma}
For any rate $\vec{R}=(R_{1},\ldots,R_{N})$ that is achievable under
$\frak{C}$, and for any $j\in\mc{N}$, it holds \label{lem:lem2}
\[
n\sum_{k=1}^{j}R_{k}\geq(1-\epsilon_{\{1,\ldots,j+1\}})\sum_{l=1}^{n}(1-\pi(l))I(W_{[1,j]};X(l)|\vec{Y}_{[1,j+1]}^{l-1},\vec{Z}^{l-1},F(l)=1).
\]
\end{lemma}
Applying Lemma~\ref{lem:lem1} for $j-1$ yields 
\begin{equation} \label{eq:third}
\frac{n\sum_{k=1}^{j-1}R_{k}}{1-\epsilon_{\{1,\ldots,j-1\}}}\leq o(n)+\sum_{l=1}^{n}\frac{h(\pi(l))}{1-\epsilon_{\{1,\ldots,j-1\}}}
  +\sum_{l=1}^{n}(1-\pi(l))I(W_{[1,j-1]};X(l)|\vec{Y}_{[1,j-1]}^{l-1},\vec{Z}^{l-1},F(l)=1),
\end{equation}
where the second line was produced by using the inequality 
\begin{equation}
\begin{split} & (1-\pi(l))I(W_{[1,j-1]};X(l)|\vec{Y}_{[1,j-1]}^{l-1},\vec{Z}^{l-1},F(l)=1)=I(W_{[1,j-1]};X(l)|\vec{Y}_{[1,j-1]}^{l-1},\vec{Z}^{l-1},F(l))\\
 & \stackrel{it.\ref{it5}}{\leq}I(W_{[1,j-1]};X(l)|\vec{Y}_{[1,j]}^{l-1},\vec{Z}^{l-1},F(l))+I(Y_{j}^{l-1};X(l)|\vec{Y}_{[1,j-1]}^{l-1},\vec{Z}^{l-1},F(l))\\
 & =(1-\pi(l))\left[I(W_{[1,j-1]};X(l)|\vec{Y}_{[1,j]}^{l-1},\vec{Z}^{l-1},F(l)=1)+I(Y_{j}^{l-1};X(l)|\vec{Y}_{[1,j-1]}^{l-1},\vec{Z}^{l-1},F(l)=1)\right],
\end{split}  \label{eq:t3}
\end{equation}
and applying Lemma~\ref{lem:lem2}, for $j-1$, to the first term
in the last line of (\ref{eq:t3}). Hence, we arrive at 
\begin{equation} \label{eq:overj}
\begin{split}n\sum_{k=1}^{j-1}R_{k}\left(\frac{1}{1-\epsilon_{\{1,\ldots,j-1\}}}-\frac{1}{1-\epsilon_{\{1,\ldots,j\}}}\right) & 
  \leq o(n)+\frac{1}{1-\epsilon_{\{1,\ldots,j-1\}}}\sum_{l=1}^{n}h(\pi(l))\\
 & +\sum_{l=1}^{n}(1-\pi(l))I(Y_{j}^{l-1};X(l)|\vec{Y}_{[1,j-1]}^{l-1},\vec{Z}^{l-1},F(l)=1).
\end{split}  
\end{equation}

We are now ready to prove Theorem \ref{thm:outer_bound_ext}. We only
consider the identity permutation (i.e.~$\sigma(i)=i$), since all
other permutations are handled similarly. Summing (\ref{eq:overj})
for $j=2,\ldots,N$, applying Lemma~\ref{lem:lem1} for $j=N$ and
summing the results yields after some manipulations (which involve
a change of order summation between $j$ and $k$) 
\begin{equation} \label{eq:almost}
\begin{split}n\sum_{k=1}^{N}\frac{R_{k}}{1-\epsilon_{\{1,\ldots,k\}}} & \leq\left(\sum_{j=1}^{N}\frac{1}{1-\epsilon_{\{1,\ldots,j\}}}\right)
  \sum_{l=1}^{n}h(\pi(l))+\sum_{l=1}^{n}(1-\pi(l))\sum_{j=2}^{N}I(Y_{j}^{l-1};X(l)|\vec{Y}_{[1,j-1]}^{l-1},\vec{Z}^{l-1},F(l)=1)\\
 & +\sum_{l=1}^{n}(1-\pi(l))I(W_{[1,N]};X(l)|\vec{Y}^{l-1},\vec{Z}^{l-1},F(l)=1)+o(n).
\end{split}
\end{equation}
For notational compactness, we hereafter denote $A=\sum_{j=1}^{N}\frac{1}{1-\epsilon_{\{1,\ldots,j\}}}$.
It also holds 
\begin{equation} \label{eq:boundX}
\begin{split} & L\geq H(X(l)|F(l)=1)=I(X(l);\vec{Y}_{[1,N]}^{l-1},\vec{Z}^{l-1}|F(l)=1)+H(X(l)|\vec{Y}^{l-1},\vec{Z}^{l-1},F(l)=1)\\
 & =\sum_{j=2}^{N}I(Y_{j}^{l-1};X(l)|\vec{Y}_{[1,j-1]}^{l-1},\vec{Z}^{l-1},F(l)=1)+H(X(l)|\vec{Y}^{l-1},\vec{Z}^{l-1},F(l)=1)\\
 & \geq\sum_{j=2}^{N}I(Y_{j}^{l-1};X(l)|\vec{Y}_{[1,j-1]}^{l-1},\vec{Z}^{l-1},F(l)=1)+H(W_{[1,N]};X(l)|\vec{Y}^{l-1},\vec{Z}^{l-1},F(l)=1),
\end{split}
\end{equation}
where the second line is derived by applying the chain rule over $j$.
Inserting (\ref{eq:boundX}) into (\ref{eq:almost}) yields 
\begin{equation} \label{eq:last}
n\sum_{k=1}^{N}\frac{R_{k}}{1-\epsilon_{\{1,\ldots,k\}}}\leq\sum_{l=1}^{n}\left[Ah(\pi(l))+(1-\pi(l))L\right]+o(n).
\end{equation}
The RHS of (\ref{eq:last}) is separable in terms of $\pi(l)$ and
its maximum can be computed via standard derivative arguments. In
fact, the maximum in the RHS of (\ref{eq:last}) is achieved for $\pi(l))=\frac{1}{1+2^{L/A}}$
for $l=1,\ldots,n$ which yields 
\begin{equation}
n\sum_{k=1}^{N}\frac{R_{k}}{1-\epsilon_{\{1,\ldots,k\}}}\leq nA\log_{2}(1+2^{L/A})+o(n)=nL+nA\log_{2}(1+2^{-L/A})+o(n).
\end{equation}
Dividing by $n$, taking a limit as $n\to\infty$ and using the inequality
$\ln(1+x)\leq x$, for any $x>0$, yields 
\begin{equation} \label{eq:perm}
\sum_{k=1}^{N}\frac{R_{k}}{1-\epsilon_{\{1,\ldots,k\}}}\leq L+2^{-L/A}A.
\end{equation}

Repeating the above procedure for an arbitrary permutation $\sigma$
on $\mc{N}$ produces 
\[
\sum_{k=1}^{N}\frac{R_{\sigma(k)}}{1-\epsilon_{\{1,\ldots,k\}}}\leq L+2^{-L/A_{\sigma}}A_{\sigma},
\]
where $A_{\sigma}=\sum_{k=1}^{N}\,\frac{1}{1-\epsilon_{\{\sigma(1),\ldots,\sigma(k)\}}}$
and since the last inequality must be true for all permutations $\sigma$,
the proof is complete.

\subsection{Proof of Lemma \ref{lem:lem1}} \label{sub:Proof-of-Lem1}

Fano's inequality implies 
\begin{equation} \label{eq:ent1}
n\sum_{k=1}^{j}R_{k}=H(W_{[1,j]})=I(W_{[1,j]};\vec{Y}_{[1,j]}^{n},\vec{Z}^{n})+o(n),
\end{equation}
with 
\begin{equation} \begin{split} \label{eq:messj}
& I(W_{[1,j]};\vec{Y}_{[1,j]}^{n},\vec{Z}^{n})=\sum_{l=1}^{n}I(W_{[1,j]};\vec{Y}_{[1,j]}(l),\vec{Z}(l)|\vec{Y}_{[1,j]}^{l-1},\vec{Z}^{l-1})\\
 & =\sum_{l=1}^{n}\left[I(W_{[1,j]};\vec{Z}(l)|\vec{Y}_{[1,j]}^{l-1},\vec{Z}^{l-1})+I(W_{[1,j]};\vec{Y}_{[1,j]}(l)|\vec{Y}_{[1,j]}^{l-1},\vec{Z}^{l-1},\vec{Z}_{[1,j]}(l)\right]\\
 & \stackrel{it.\ref{it1}}{=}\sum_{l=1}^{n}\left[I(W_{[1,j]};\vec{Z}(l)|\vec{Y}_{[1,j]}^{l-1},\vec{Z}^{l-1})+I(W_{[1,j]};\vec{Y}_{[1,j]}(l),\vec{Z}_{[1,j]}(l)|\vec{Y}_{[1,j]}^{l-1},\vec{Z}^{l-1},\vec{Z}_{[1,j]}(l)\right].
\end{split} \end{equation}
Applying the chain rule twice with different order yields 
\begin{equation} \begin{split} \label{eq:ch2}
I(W_{[1,j]};\vec{Z}(l),X(l)|\vec{Y}_{[1,j]}^{l-1},\vec{Z}^{l-1}) & =I(W_{[1,j]};X(l)|\vec{Y}_{[1,j]}^{l-1},\vec{Z}^{l-1})+I(W_{[1,j]};\vec{Z}(l)|\vec{Y}_{[1,j]}^{l-1},\vec{Z}^{l-1},X(l))\\
 & =I(W_{[1,j]};\vec{Z}(l)|\vec{Y}_{[1,j]}^{l-1},\vec{Z}^{l-1})+I(W_{[1,j]};X(l)|\vec{Y}_{[1,j]}^{l-1},\vec{Z}^{l-1},\vec{Z}(l)),
\end{split} \end{equation}
and since $\vec{Z}(l)$ is independent of all previous random variables
\textit{given} $X(l)$, (\ref{eq:ch2}) yields 
\begin{equation} \label{eq:messj2}
I(W_{[1,j]};\vec{Z}(l)|\vec{Y}_{[1,j]}^{l-1},\vec{Z}^{l-1})=I(W_{[1,j]};X(l)|\vec{Y}_{[1,j]}^{l-1},\vec{Z}^{l-1})-I(W_{[1,j]};X(l)|\vec{Y}_{[1,j]}^{l-1},\vec{Z}^{l-1},\vec{Z}(l)).
\end{equation}
Furthermore, since knowledge of $X(l)$ implies knowledge of $F(l)$,
it holds 
\begin{equation} \begin{split} \label{eq:messj3}
& I(W_{[1,j]};X(l)|\vec{Y}_{[1,j]}^{l-1},\vec{Z}^{l-1})=I(W_{[1,j]};X(l),F(l)|\vec{Y}_{[1,j]}^{l-1},\vec{Z}^{l-1})\\
 & =I(W_{[1,j]};F(l)|\vec{Y}_{[1,j]}^{l-1},\vec{Z}^{l-1})+I(W_{[1,j]};X(l)|\vec{Y}_{[1,j]}^{l-1},\vec{Z}^{l-1},F(l)).
\end{split} \end{equation}

Combining (\ref{eq:messj2}), (\ref{eq:messj3}) yields 
\begin{equation} \begin{split} \label{eq:diff}
I(W_{[1,j]};\vec{Z}(l)|\vec{Y}_{[1,j]}^{l-1},\vec{Z}^{l-1}) & =I(W_{[1,j]};F(l)|\vec{Y}_{[1,j]}^{l-1},\vec{Z}^{l-1})+I(W_{[1,j]};X(l)|\vec{Y}_{[1,j]}^{l-1},\vec{Z}^{l-1},F(l))\\
 & -I(W_{[1,j]};X(l)|\vec{Y}_{[1,j]}^{l-1},\vec{Z}^{l-1},\vec{Z}(l)).
\end{split} \end{equation}
Defining the set $\mc{Z}_{[1,j]}=\left\{ \vec{Z}_{[1,j]}:\vec{Z}_{[1,j]}\neq(\nulls,\ldots,\nulls)\right\} $,
we can compute 
\begin{equation} \begin{split} \label{eq:reduce}
& I(W_{[1,j]};X(l)|\vec{Y}_{[1,j]}^{l-1},\vec{Z}^{l-1},\vec{Z}(l))=\sum_{\vec{z}\in\mc{Z}_{[1,j]}}I(W_{[1,j]};X(l)|\vec{Y}_{[1,j]}^{l-1},\vec{Z}^{l-1},\vec{Z}(l)=\vec{z})\Pr(\vec{Z}(l)=\vec{z})\\
 & =\sum_{\vec{z}\in\mc{Z}_{[1,j]}}I(W_{[1,j]};X(l)|\vec{Y}_{[1,j]}^{l-1},\vec{Z}^{l-1},\hat{\vec{Z}}(l)=\vec{z},F(l)=1)\Pr(\hat{\vec{Z}}(l)=\vec{z})\Pr(F(l)=1)\\
 & =\sum_{\vec{z}\in\mc{Z}_{[1,j]}}I(W_{[1,j]};X(l)|\vec{Y}_{[1,j]}^{l-1},\vec{Z}^{l-1},F(l)=1)\Pr(\hat{\vec{Z}}(l)=\vec{z})\Pr(F(l)=1)\\
 & =I(W_{[1,j]};X(l)|\vec{Y}_{[1,j]}^{l-1},\vec{Z}^{l-1},F(l)=1)\Pr(F(l)=1)\\
 & =I(W_{[1,j]};X(l)|\vec{Y}_{[1,j]}^{l-1},\vec{Z}^{l-1},F(l)),
\end{split} \end{equation}
where we exploited the independence of $\hat{\vec{Z}}(l)$ from all
variables up to slot $l$ and used the facts that $F(l)=0$ implies
$X(l)=\nulls$ and $\sum_{\vec{z}\in\mc{Z}_{[1,j]}}\Pr(\hat{\vec{Z}}(l)=\vec{z})=1$.

To manipulate the last term in (\ref{eq:messj}), we define the set
$\tilde{\mc{Z}}_{[1,j]}=\left\{ \vec{Z}_{[1,j]}:\vec{Z}_{[1,j]}\neq(\nulls,\ldots,\nulls),(*,\ldots,*)\right\} $.
In words, $\tilde{\mc{Z}}_{[1,j]}$ is the set of feedback vectors
in which at least one user in $\{1,\ldots,j\}$ successfully receives
the transmitted symbol and sends back $ACK$. Notice that, for any
$\vec{z}\not\in\tilde{\mc{Z}}_{[1,j]}$, the event $\{\vec{Z}_{[1,j]}(l)=\vec{z}\}$
implies full knowledge of $\vec{Y}_{[1,j]}(l)$. It now holds 
\begin{equation} \begin{split} \label{eq:messj4}
& I(W_{[1,j]};\vec{Y}_{[1,j]}(l),\vec{Z}_{[1,j]}(l)|\vec{Y}_{[1,j]}^{l-1},\vec{Z}^{l-1},\vec{Z}_{[1,j]}(l))\\
 & =\sum_{\vec{z}\in\tilde{\mc{Z}}_{[1,j]}}I(W_{[1,j]};\vec{Y}_{[1,j]}(l),\vec{Z}_{[1,j]}(l)|\vec{Y}_{[1,j]}^{l-1},\vec{Z}^{l-1},\vec{Z}_{[1,j]}(l)=\vec{z})\Pr(\vec{Z}_{[1,j]}(l)=\vec{z})\\
 & =\sum_{\vec{z}\in\tilde{\mc{Z}}_{[1,j]}}\left[H(W_{[1,j]}|\vec{Y}_{[1,j]}^{l-1},\vec{Z}^{l-1},\vec{Z}_{[1,j]}(l)=\vec{z})-H(W_{[1,j]}|\vec{Y}_{[1,j]}^{l-1},\vec{Z}^{l-1},\vec{Y}_{[1,j]}(l),\vec{Z}_{[1,j]}(l)=\vec{z})\right]\Pr(\vec{Z}_{[1,j]}(l)=\vec{z})\\
 & =\sum_{\vec{z}\in\tilde{\mc{Z}}_{[1,j]}}\left[H(W_{[1,j]}|\vec{Y}_{[1,j]}^{l-1},\vec{Z}^{l-1},F(l)=1,\hat{\vec{Z}}_{[1,j]}(l)=\vec{z})-H(W_{[1,j]}|\vec{Y}_{[1,j]}^{l-1},\vec{Z}^{l-1},\vec{Y}_{[1,j]}(l),F(l)=1,\hat{\vec{Z}}_{[1,j]}(l)=\vec{z})\right]\\
 & \times\Pr(\hat{\vec{Z}}_{[1,j]}(l)=\vec{z})\Pr(F(l)=1)\\
 & =\sum_{\vec{z}\in\tilde{\mc{Z}}_{[1,j]}}\left[H(W_{[1,j]}|\vec{Y}_{[1,j]}^{l-1},\vec{Z}^{l-1},F(l)=1)-H(W_{[1,j]}|\vec{Y}_{[1,j]}^{l-1},\vec{Z}^{l-1},F(l)=1,X(l))\right]\Pr(\hat{\vec{Z}}_{[1,j]}(l)=\vec{z})\Pr(F(l)=1)\\
 & =(1-\epsilon_{\{1,\ldots,j\}})(1-\pi(l))I(W_{[1,j]};X(l)|\vec{Y}_{[1,j]}^{l-1},\vec{Z}^{l-1},F(l)=1).
\end{split} \end{equation}
In the transition from the third to the fourth line of (\ref{eq:messj4}),
we used the event identity $\{\vec{Z}_{[1,j]}(l)=\vec{z}\}=\{\hat{\vec{Z}}_{[1,j]}(l)=\vec{z},F(l)=1\}$,
which is valid for any $\vec{z}\in\tilde{\mc{Z}}_{[1,j]}$, while
in the transition from the fourth to the fifth line we used the facts
that $\hat{\vec{Z}}_{[1,j]}$ is independent of all variables up to
$l$ (including $F(l)$, $X(l)$) and knowledge of $\hat{\vec{Y}}_{[1,j]}(l)$,
$\vec{Z}_{[1,j]}(l)=\vec{z}$ implies knowledge of $X(l)$ for any
$\vec{z}\in\tilde{\mc{Z}}_{[1,j]}$.

Inserting (\ref{eq:messj4}), (\ref{eq:reduce}), (\ref{eq:diff})
into (\ref{eq:ent1}), via (\ref{eq:messj}), and using item \ref{it5}
in Lemma~\ref{lem:prelim} produces 
\begin{equation} \begin{split}
n\sum_{k=1}^{k}R_{k} & \leq o(n)+\sum_{l=1}^{n}\left[I(W_{[1,j]};F(l)|\vec{Y}_{[1,j]}^{l-1},\vec{Z}^{l-1})+(1-\epsilon_{\{1,\ldots,j\}})(1-\pi(l))I(W_{[1,j]};X(l)|\vec{Y}_{[1,j]}^{l-1},\vec{Z}^{l-1},F(l)=1)\right]\\
 & \leq o(n)+\sum_{l=1}^{n}\left[h(\pi(l))+(1-\epsilon_{\{1,\ldots,j\}})(1-\pi(l))I(W_{[1,j]};X(l)|\vec{Y}_{[1,j]}^{l-1},\vec{Z}^{l-1},F(l)=1)\right],
\end{split} \end{equation}
where we used the inequality $I(W_{[1,j]};F(l)|\vec{Y}_{[1,j]}^{l-1},\vec{Z}^{l-1})\leq H(F(l))=h(\pi(l))$.

\subsection{Proof of Lemma~\ref{lem:lem2}} \label{sub:Proof-of-Lem2}

Performing similar manipulations as in the proof of Lemma~\ref{lem:lem1},
we can write 
\begin{equation} \begin{split}
n\sum_{k=1}^{j}R_{k} & =H(W_{[1,j]})\geq I(W_{[1,j]};\vec{Y}_{[1,j+1]}^{n},\vec{Z}^{n})=\sum_{l=1}^{n}I(W_{[1,j]};\vec{Y}_{[1,j+1]}(l),\vec{Z}(l)|\vec{Y}_{[1,j+1]}^{l-1},\vec{Z}^{l-1})\\
 & \geq\sum_{l=1}^{n}I(W_{[1,j]};\vec{Y}_{[1,j+1]}(l)|\vec{Y}_{[1,j+1]}^{l-1},\vec{Z}^{l-1},\vec{Z}(l))\\
 & =\sum_{l=1}^{n}\sum_{\vec{z}\in\tilde{\mc{Z}}_{[1,j+1]}}I(W_{[1,j]};\vec{Y}_{[1,j+1]}(l)|\vec{Y}_{[1,j+1]}^{l-1},\vec{Z}^{l-1},\vec{Z}(l)=\vec{z})\Pr(\vec{Z}(l)=\vec{z})\\
 & =\sum_{l=1}^{n}\sum_{\vec{z}\in\tilde{\mc{Z}}_{[1,j+1]}}I(W_{[1,j]};\vec{Y}_{[1,j+1]}(l)|\vec{Y}_{[1,j+1]}^{l-1},\vec{Z}^{l-1},\hat{\vec{Z}}(l)=\vec{z},F(l)=1)\Pr(F(l)=1)\Pr(\hat{\vec{Z}}(l)=\vec{z})\\
 & =\sum_{l=1}^{n}\sum_{\vec{z}\in\tilde{\mc{Z}}_{[1,j+1]}}\left[H(W_{[1,j]}|\vec{Y}_{[1,j+1]}^{l-1},\vec{Z}^{l-1},\hat{\vec{Z}}(l)=\vec{z},F(l)=1)\right.\\
 & \left.-H(W_{[1,j]}|\vec{Y}_{[1,j+1]}^{l-1},\vec{Z}^{l-1},\vec{Y}_{[1,j+1]}(l),\hat{\vec{Z}}(l)=\vec{z},F(l)=1)\right]\Pr(F(l)=1)\Pr(\hat{\vec{Z}}(l)=\vec{z})\\
 & =\sum_{l=1}^{n}\sum_{\vec{z}\in\tilde{\mc{Z}}_{[1,j+1]}}\left[H(W_{[1,j]}|\vec{Y}_{[1,j+1]}^{l-1},\vec{Z}^{l-1},\hat{\vec{Z}}(l)=\vec{z},F(l)=1)\right.\\
 & \left.-H(W_{[1,j]}|\vec{Y}_{[1,j+1]}^{l-1},\vec{Z}^{l-1},X(l),\hat{\vec{Z}}(l)=\vec{z},F(l)=1)\right]\Pr(F(l)=1)\Pr(\hat{\vec{Z}}(l)=\vec{z})\\
 & =(1-\epsilon_{\{1,\ldots,j+1\}})\sum_{l=1}^{n}(1-\pi(l))I(W_{[1,j]};X(l)|\vec{Y}_{[1,j+1]}^{l-1},\vec{Z}^{l-1},F(l)=1),
\end{split} \end{equation}
where we used again the independence of $\hat{\vec{Z}}(l)$ from all other variables.

\bibliographystyle{IEEEtran}
\bibliography{IEEEabrv,conf-4-users-L}

\end{document}